\documentclass[11pt,sort&compress]{elsarticle}
\usepackage[paper=letterpaper,top=24mm, bottom=26mm, left=26mm, right=26mm]{geometry}
\makeatletter
\def\ps@pprintTitle{%
 \let\@oddhead\@empty
 \let\@evenhead\@empty
 \def\@oddfoot{\centerline{\thepage}}%
 \let\@evenfoot\@oddfoot}
\makeatother
\usepackage{amsmath}
\usepackage{amsfonts}
\usepackage{accents}
\usepackage{amssymb}
\usepackage{mathrsfs}
\usepackage{mathtools}
\usepackage[LGR,T1]{fontenc}
\usepackage[latin1]{inputenc}
\let\oldbibliography\thebibliography
\renewcommand{\thebibliography}[1]{%
  \oldbibliography{#1}%
  \setlength{\itemsep}{1.4pt}%
}
\usepackage[cal=boondox,calscaled=1]{mathalfa}
\DeclareMathAlphabet{\bbvar}{U}{BOONDOX-ds}{m}{n}
\DeclareMathAlphabet{\bbgreek}{U}{bbold}{m}{n}
\usepackage{psfrag}
\usepackage{pstool}
\usepackage{caption}
\usepackage{tabularx}
\usepackage{multirow}
\usepackage{hhline}
\usepackage{ltablex}
\usepackage{pbox}
\usepackage{graphicx}%

\newcommand{\hook}{\text{\large{$\lrcorner$}}}

\usepackage[colorlinks=true,
            linkcolor=blue,
            urlcolor=blue,
            citecolor=blue]{hyperref}
\usepackage[hyperref]{xcolor}
\definecolor{darkred}{rgb}{.95,.0,.0}

\newcommand{\qq}[1]{``#1''} 

\newcommand{\di}{\mathrm{d}}
\usepackage{tensor}\newcommand{\ou}[3]{{#1}{}^{#2}{}_{#3}}
\newcommand{\uo}[3]{{#1}{}_{#2}{}^{#3}}

\newcommand{\I}{\operatorname{i}} 
\newcommand{\E}{\mathrm{e}} 
\newcommand{\CC}{\mathrm{cc.}} 
\newcommand{\C}{\mathbb{C}}

\newcommand{\R}{\mathbb{R}}

\newenvironment{subalign}{\subequations\align}{\endalign\endsubequations}
\newcommand{\eref}[1]{(\ref{#1})}

\newcommand{\bbwedge}{\reflectbox{\rotatebox[origin=c]{180}{\fontsize{10pt}{10pt}$\hspace{0.7pt}\bbvar{V}\hspace{0.7pt}$}}}

\newcommand{\mtext}[1]{\text{\it #1}}
\usepackage{tikz-cd}

\newcommand\vpm{\mathbin{\vcenter{\hbox{
  \oalign{\hfil$\scriptstyle+$\hfil\cr
          \noalign{\kern-.3ex}
          $\scriptscriptstyle({-})$\cr}}}}}
\DeclareMathAlphabet{\sfit}{OT1}{fos}{sb}{it}
\DeclareMathAlphabet{\mathsf}{OT1}{fos}{sb}{n}

\definecolor{darkgreen}{rgb}{0.01, 0.75, 0.24}
\definecolor{darkblue}{rgb}{0.01, 0.24, 0.75}

\usepackage[multiple, flushmargin]{footmisc}

\let\originalleft\left
\let\originalright\right
\renewcommand{\left}{\mathopen{}\mathclose\bgroup\originalleft}
\renewcommand{\right}{\aftergroup\egroup\originalright}

\begin{document}

\begin{abstract}
\noindent 
When a system emits gravitational radiation, the Bondi mass decreases. If the Bondi energy is Hamiltonian, it can thus only be a time-dependent Hamiltonian. In this paper, we show that the Bondi energy can be understood as a time-dependent Hamiltonian on the covariant phase space. Our derivation starts from the Hamiltonian formulation in domains with boundaries that are null. We introduce the most general boundary conditions on a generic such null boundary, and compute quasi-local charges for boosts, energy and angular momentum. Initially, these domains are at finite distance, such that there is a natural IR regulator.  To remove the IR regulator, we introduce a double null foliation together with an adapted Newman\,--\,Penrose null tetrad. Both null directions are surface orthogonal. We study the falloff conditions for such specific null foliations and take the limit to null infinity. At null infinity, we recover the Bondi mass and the usual covariant phase space for the two radiative modes at the full non-perturbative level. Apart from technical results, the framework gives two important physical insights. First of all, it explains the physical significance of the corner term that is added in the Wald\,--\,Zoupas framework to render the quasi-conserved charges integrable. The term to be added is simply the derivative of the Hamiltonian with respect to the background fields that drive the time-dependence of the Hamiltonian. Secondly, we propose a new interpretation of the Bondi mass as the thermodynamical free energy of gravitational edge modes at future null infinity. The Bondi mass law is then simply the statement that the free energy always decreases on its way towards thermal equilibrium.
\end{abstract}%
\title{Null infinity as an open Hamiltonian system}
\author{Wolfgang Wieland}
\address{Institute for Quantum Optics and Quantum Information (IQOQI)\\Austrian Academy of Sciences\\Boltzmanngasse 3, 1090 Vienna, Austria\\{\vspace{0.5em}\normalfont 3 December 2020}
}
\maketitle
\vspace{-1.2em}
\hypersetup{
  linkcolor=black,
  urlcolor=black,
  citecolor=black
}
{\tableofcontents}
\hypersetup{
  linkcolor=black,
  urlcolor=darkred,
  citecolor=darkred,
}
\begin{center}{\noindent\rule{\linewidth}{0.4pt}}\end{center}\newpage
\section{Introduction}
\noindent Gravitational radiation carries energy. In an asymptotically flat spacetime, an asymptotic observer measures a flux of gravitational energy $F(u)$, which is sourced by the incoming radiation and satisfies the well-known Bondi mass law,
\begin{equation}
F(u)= \dot{M}_{\mathrm{B}}(u) = -\frac{1}{4\pi G}\oint_{S^2_u\subset\mathcal{I}^+}d^2\Omega\,|\dot{\sigma}|^2\leq 0,\label{fluxlaw}
\end{equation}
where\footnote{In the following, $(u,z,\bar{z})$ are asymptotic Bondi coordinates on $\mathcal{J}^+$ and $d^2\Omega=-2\I (1+|z|^2)^{-2}\di z\wedge \di\bar{z}$ is the two-dimensional fiducial area element.} $\sigma (u,z,\bar{z})$ is the asymptotic shear that characterises the strength and polarisation of the gravitational radiation  in addition to gravitational memory \cite{Bondi21,Sachs103,Horowitz:1981uw,AshtekarNullInfinity,Ashtekar:2014zsa}. The energy flux is the time derivative of the Bondi mass $M_{\mathrm{B}}(u)$, which is a two-dimensional integral at future null infinity. On the relativistic phase space, energy and time are canonically conjugate partial observables \cite{partobs,Dittrich:2005kc}, and the question arises whether the  Bondi energy is Hamiltonian\,---\,whether there is a phase space equipped with a symplectic two-form $\Omega$ such that the Hamilton equations are satisfied,
\begin{equation}
\delta [M_{\mathrm{B}}(u)] \stackrel{?}{=} -\Omega(\tfrac{\di}{\di u},\delta),
\end{equation}
with $\delta$ denoting a linearised solution of the field equations. The obvious trouble with such a point of view is that it seems to be incompatible with the existence of gravitational radiation, rendering it rather useless for any real-world applications. In fact, since the two-form $\Omega(X,Y)=-\Omega(Y,X)$ is anti-symmetric and the Bondi mass $M(u)$ is the supposed generator of asymptotic time translations, we would find immediately that the energy is conserved
\begin{equation}
\dot{M}_{\mathrm{B}}(u) = \{M_{\mathrm{B}}(u),M_{\mathrm{B}}(u)\} = -\Omega(\tfrac{\di}{\di u},\tfrac{\di}{\di u})=0.\label{puzzle}
\end{equation}
No gravitational radiation would be seen at null infinity. If the Bondi mass is Hamiltonian, it can thus only be a time-dependent Hamiltonian such that the Hamilton equations \eref{puzzle} must be modified by an explicitly time-dependent term.\footnote{Recall that the Hamilton equations for a time-dependent observable $O$ are $\frac{\di}{\di t}O=\{H,O\}+\partial_tO$ rather than $\frac{\di}{\di t}O=\{H,O\}$. Below, we will characterise  the partial time derivative $\partial_tH$ in terms of the radiative symplectic potential on the covariant phase space.} One of the main motivations for this paper is to understand such explicit time-dependence on the covariant phase space for general relativity \cite{Peierls,Ashtekar:1990gc,Lee:1990nz,Iyer:1994ys,Wald:1999wa}. Otherwise we cannot distinguish  background fields ($c$-numbers) that are responsible for the explicit time-dependence of the Bondi mass, from the actual phase space degrees of freedom ($q$-numbers) of the system to be studied. Our basic proposal is to \emph{remove}, in fact,  the incoming radiative modes from the covariant phase space and encode them into auxiliary background fields at null infinity. For given shear, the  resulting Hamiltonian generates the evolution on a reduced boundary phase space of gravitational edge modes alone. Incoming radiative modes are to be swapped, therefore, for background fields at null infinity. The situation is reminiscent of what happens in lower dimensions, where there are no radiative modes at all, and\,---\,apart from non-local moduli\,---\,all gravitational observables are the gravitational edge modes alone \cite{Balachandran:1994up,Strominger:1997eq,Banados:1998ta,carlipbook,Carlip:2005zn,Afshar:2016wfy,Compere:2017knf,Wieland:2018ymr,Namburi:2019qja,Wieland_2020}.

To set up the basic framework, we start out in \hyperref[sec2]{section 2} with a simple toy model, where we explain the covariant phase space approach in the presence of time-dependent background fields that drive the time-depedence of the Hamiltonian. 
\hyperref[sec3]{Section 3}  deals with the classical bulk plus boundary field theory. We introduce the symplectic potential in terms of a Newman\,--\,Penrose (NP) tetrad on a generic null boundary and identify the gauge symmetries and quasi-local observables on the light cone (\hyperref[sec4]{section 4}). The most technical part is \hyperref[sec5]{section 5}, where we consider the boundary and falloff conditions of the spin coefficients for a double null foliation. Our gauge conditions are different from what is usefully imposed in the NP framework, where only the outgoing (radial) null vector is surface orthogonal. However, a double null foliation is more appropriate in our context, because it allows us to first evaluate the null symplectic potential on some generic null surface $\mathcal{N}$ and obtain the radiative phase space in the limit where $\mathcal{N}$ is sent to $\mathcal{I}^+$.  Finally, we integrate the Hamilton equations and compute the energy and angular momentum from the radial $1/r$ expansion of quasi-local charges (\hyperref[sec5]{section 5} and \hyperref[sec6]{section 6}).\vspace{-0.5em}

\paragraph{- Notation} In the following, we will work in the first-order spin-connection representation of general relativity \cite{penroserindler, newmanpenrose, newvariables}. All  configuration variables are spinor-valued differential $p$-forms, i.e.\ sections $\psi_{AB\dots A'B'\dots ab\dots}$ of the tensor product bundle consisting of the spinor bundle $\mathcal{S}_{AB\dots A'B'\dots}(\mathcal{M})$ and the $p$-th exterior power of the cotangent bundle $T^\ast\mathcal{M}$. Indices $A,B,C,\dots$ and $A',B',C',\dots$ transform under the spin $(1/2,0)$ and complex conjugate spin $(0,1/2)$ representation of local $SL(2,\C)$ Lorentz  transformations. Indices $a,b,c,\dots$ are abstract tensor indices.\footnote{We will not distinguish between four-dimensional (bulk) and three-dimensional (boundary) indices. If there is a chance of confusion, we will use a prefix to distinguish four-dimensional tensors (vectors) from fields that are defined intrinsically on the null boundary: if e.g.\ $m_a$ is a one-form on the boundary, ${}^{4}m_a$ will denote a possible extension into the bulk.} The spinor indices are raised and lowered using the skew-symmetric Levi-Civita spinor $\epsilon_{AB}=-\epsilon_{BA}$, i.e.\ $\xi_A=\xi^B\epsilon_{BA}$ and $\bar{\xi}^{A'}=\bar{\epsilon}^{A'B'}\bar{\xi}_{B'}$. Falloff conditions for the double null foliation and the dictionary between our conventions and the Newman\,--\,Penrose formalism are summarised in a short \hyperref[appdx]{appendix} at the very end of this paper.
\section{Covariant phase space with background fields}\label{sec2}

\noindent The covariant phase space approach \cite{Peierls,Lee:1990nz,Iyer:1994ys,Wald:1999wa,Ashtekar:1990gc} is frequently used in relativistic field theories. It provides a straight-forward method to get the Poisson algebra of  conserved charges  without introducing an auxiliary foliation of spacetime, which is otherwise needed c.f.\ \cite{Arnowitt:2008aa,ADMEnergy}. The formalism is manifestly covariant. However, there is an important drawback. It is somewhat cumbersome to impose boundary conditions, and distinguish external sources, which are kept fixed at infinity\,---\,background fields that Poisson commute with the dynamical fields\,---\,, from physical phase space degrees of freedom, see e.g.\ \cite{Harlow:2020aa}. The goal of this introductory section is to study the problem in a very simplified setting. The gravitational case is considered below (\hyperref[sec3]{section 3} and the rest of the paper).

Let us start with a system of $N$ degrees of freedom on the real line with a time-dependent Hamiltonian, whose time-dependence appears, however, only through an external background field $\omega(t)$. In gravity, $\omega(t)$ will play the role of fixed boundary data at infinity. A simple example for such a system is a harmonic oscillator with a time-dependent frequency that can be tuned freely by the experimenter, i.e.
\begin{equation}
H[\vec{p},\vec{q}\,|\omega(t)] = \frac{1}{2}\big(|\vec{p}|^2+\omega^2(t)|\vec{q}|^2\big).
\end{equation}
The   action on phase space for such an $N$-dimensional system on an interval $I=(0,1)\subset \R$ is 
\begin{equation}
S[\vec{p},\vec{q}|\omega]=\int_I\Big(\sum_{i=1}^Np_i\di{q}^i-\di t\,H\big[\vec{p}(t),\vec{q}(t)\big|\omega(t)\big]\Big),
\end{equation}
where the Hamiltonian depends parametrically on the background field $\omega(t)$. Thus the action is a functional on the infinite-dimensional \emph{space of kinematical histories},
\begin{equation}
\mathcal{H}_{\mtext{kin}}=\big\{\gamma\in\mathcal{C}^1(I:\R^N\times\R^N\times \R):\gamma(t)=(\vec{p}(t),\vec{q}(t),\omega(t))\big\},
\end{equation}
where $\gamma(t)$ is a trajectory in the extended configuration space $\R^N\times\R^N\times\R\ni(\vec{p},\vec{q},\omega)$. To compute the variation of the action, we introduce the  \emph{pre-symplectic} two-form\footnote{The symbol \qq{$\bbvar{d}$} is the exterior derivative on the infinite-dimensional space of kinematical histories $\mathcal{H}_{\mtext{kin}}$, and \qq{{$\reflectbox{\rotatebox[origin=c]{180}{\fontsize{9pt}{9pt}$\hspace{0.7pt}\bbvar{V}\hspace{0.7pt}$}}$}} denotes the exterior product between $p$-forms in $\bigwedge^pT^\ast\mathcal{H}_{\mtext{kin}}$.} on the infinite-dimensional space of kinematical histories
\begin{equation}
\Omega_{\mtext{bulk}}\big|_t := \sum_{i=1}^N \bbvar{d}p_i(t)\bbwedge\bbvar{d}q^i(t)=\bbvar{d}\Theta_{\mtext{bulk}}\big|_t\in\Omega^2(\mathcal{H}_{\mtext{kin}}:\R).
\end{equation}
In addition, we define the following functional one-forms, namely the pre-symplectic potential and the pre-symplectic flux,
\begin{align}
\Theta_{\mtext{bulk}}\big|_t&:=\sum_{i=1}^N p_i(t)\bbvar{d}q^i(t),\\
{\Theta}_{\mtext{flux}}\big|_t &:= -\frac{\partial H[\vec{p}(t),\vec{q}(t)| \omega(t)]}{\partial\omega}\bbvar{d} \omega(t).\label{Thfluxdef}
\end{align}
Consider then a general vector field $\delta\in T\mathcal{H}_{\mtext{kin}}$, and let us consider the derivative of the action under such an infinitesimal variation, i.e.\
\begin{equation}
\delta S = -\int_I \di t \,\Omega_{\mtext{bulk}}(\partial_t-X_H,\delta) + \Theta_{\mtext{bulk}}(\delta)\big|_{\partial I}+\int_I\di t\,{\Theta}_{\mtext{flux}}(\delta),
\end{equation}
where $\partial_t$ and $X_H$ are the vector fields,
\begin{align}
\partial_t & = \int_{I} \di t\Big(\dot{q}^i(t)\frac{\delta}{\delta q^i(t)} + \dot{p}^i(t)\frac{\delta}{\delta p^i(t)} \Big)\in T\mathcal{H}_{\mtext{kin}},\label{dtdef}\\
X_H & = \int_{I} \di t\Big(\frac{\partial H[\vec{p},\vec{q}|\omega]}{\partial p_i}\frac{\delta}{\delta q^i(t)} - \frac{\partial H[\vec{p},\vec{q}|\omega]}{\partial q^i}\frac{\delta}{\delta p^i(t)} \Big)\in T\mathcal{H}_{\mtext{kin}}.\label{XHdef}
\end{align}
The space of \emph{kinematical histories} is an unphysical auxiliary space. The space of \emph{physical histories} $\mathcal{H}_{\mtext{phys}}$ contains all trajectories that satisfy the Hamilton equations for all possible choices of  $\omega(t)$,
\begin{equation}
\mathcal{H}_{\mtext{phys}}=\big\{(\vec{p},\vec{q},\omega)(t)\in\mathcal{H}_{\mtext{kin}} : \dot{q}^i=\partial_{p_i}H,\, \dot{p}_i=-\partial_{q^i}H\big\}.
\end{equation}

Notice that the space of  \emph{physical histories} is infinite-dimensional, because we include all possible configurations of the background field $\omega(t)$. To define the physical phase space $\mathcal{P}_{\omega_o}$, which is finite-dimensional, we need to make a specific choice for the background field,
\begin{equation}
\mathcal{P}_{\omega_o}=\big\{(\vec{p},\vec{q},\omega)(t)\in\mathcal{H}_{\mtext{phys}} : \omega(t)=\omega_o(t)\,\forall t\in I\big\}.
\end{equation}
We thus have a \emph{triple of history spaces},
\begin{equation}
\mathcal{P}_{\omega}\lhook\joinrel\xrightarrow{\hspace{0.6em}\varphi_\omega\hspace{0.6em}} \mathcal{H}_{\mtext{phys}} \lhook\joinrel\xrightarrow{\;\varphi_{\mtext{phys}}\;}\mathcal{H}_{\mtext{kin}}.
\end{equation}
For any given background field $\omega(t)$, the phase space $\mathcal{P}_\omega$ is equipped with a symplectic two-form $\Omega$, which is conserved and obtained from the pull-back
\begin{equation}
\Omega = (\varphi_{\mtext{phys}}\circ\varphi_\omega)^\ast\Omega_{\mtext{bulk}}.\label{Ompullback}
\end{equation}

Each phase space $\mathcal{P}_\omega$ is equipped with a corresponding Hamiltonian, which is a functional on the space of kinematical histories: for any instant $t\in I$, the Hamiltonian is a map $H_t:\mathcal{H}_{\mtext{kin}}\ni\gamma\mapsto  H[\vec{p}(t),\vec{q}(t)|\omega(t)]\in\R $. Given a vector field $\delta \in T\mathcal{H}_{\mtext{kin}}$, its variation satisfies the generalised Hamilton equations
\begin{equation}
\delta H_t = -(\Omega_{\mtext{bulk}})_t(X_H,\delta)-({\Theta}_{\mtext{flux}})_t(\delta).\label{Hvar}
\end{equation}

In relativity, situations occur, where the action blows up at infinity and it is not immediate to simply infer the Hamiltonian from a 3+1 split of the action.  In such a situation, the covariant phase space approach provides a simple method to infer the {on-shell} value of the Hamiltonian, i.e.\ the pull-back of the Hamiltonian to the space of physical histories $\mathcal{H}_{\mtext{phys}}$. This is possible, because on $\mathcal{H}_{\mtext{phys}}$ the Hamiltonian vector field $X_H$ coincides with the time translation $\partial_t$, see (\ref{dtdef}, \ref{XHdef}). If we restrict equation \eref{Hvar} to the space of physical histories, we can replace the Hamiltonian vector field $X_H$ by $\partial_t$. Integrating the 
 Hamilton equations on the space of physical histories, i.e.\ solving
\begin{equation}
\delta[H]\big|_{\mathcal{H}_{\mtext{phys}}} = -\Big((\varphi_{\mtext{phys}}^\ast\Omega_{\mtext{bulk}})(\partial_t,\delta)+(\varphi^\ast_{\mtext{phys}}{\Theta}_{\mtext{flux}})(\delta)\Big)\label{Hvar2}
\end{equation}
for all vector fields $\delta\in T\mathcal{H}_{\mtext{phys}}$, we will then obtain the \emph{on-shell} value of the Hamiltonian. In general, this Hamiltonian is implicitly time-dependent. In fact, by replacing the vector field $\delta\in T\mathcal{H}_{\mtext{phys}}$ by the infinitesimal time translation $\partial_t\in T\mathcal{H}_{\mtext{phys}}$, we obtain
\begin{equation}
\partial_t[{H}]\big|_{\mathcal{H}_{\mtext{phys}}} =-(\varphi^\ast_{\mtext{phys}}{\Theta}_{\mtext{flux}})(\partial_t).\label{Hdot}
\end{equation}
We will see in \hyperref[sec6]{section 6} how these equations show up in gravity.
\section{Null surface edge modes and quasi-local graviton}\label{sec3}
\subsection{Boundary conditions}
\noindent The action in the interior of the manifold is a functional of  the $SL(2,\C)$ spin connection $\ou{A}{A}{Ba}$ and the associate spin $(1/2,1/2)$ soldering form $e_{AA'a}$. For the metric to be real, the soldering form $e_{AA'a}$ satisfies the reality conditions\footnote{That the soldering form $e_{AA'a}$ is anti-hermitian is a result of our choice of $(-$$+$$+$$+)$ metric signature.}  $e_{AA'a}=-\bar{e}_{A'A}$. In addition, there are the Infeld\,--\,van der Waerden relations,
\begin{equation}
e_{AC'a}\tensor{e}{_B^{C'}_b}=\frac{1}{2}\epsilon_{AB}\,g_{ab}-\Sigma_{ABab},\label{Pauliident}
\end{equation}
that tell us that the self-dual Pleba\'{n}ski two-form $\Sigma_{ABab}=\Sigma_{BAab}=-\Sigma_{ABba}$ and the signature $(-$$+$$+$$+)$ Lorentzian  metric $g_{ab}$ are the irreducible spin $(0,0)$ and spin $(1,0)$ components of $e_{ACa} e_B{}^{C'}{}_b$. 

In terms of these variables, the action for vacuum general relativity with vanishing cosmological constant and vanishing Immirzi parameter is then given by the sum of the self-dual and anti-self-dual action,
\begin{equation}
S_{\mtext{bulk}}\big[e_{AA'a},\ou{A}{A}{Ba}\big] = \left[\frac{\I}{8\pi G}\int_{\mathcal{M}}\Sigma_{AB}\wedge F^{AB}\right]+\CC,\label{bulkactn}
\end{equation}
where $\ou{F}{A}{B}=\di\ou{A}{A}{B}+\ou{A}{A}{C}\wedge\ou{A}{C}{B}$ is the curvature of the self-dual connection and $\Sigma_{AB}$ is the self-dual Pleba\'{n}ski two-form \eref{Pauliident}.

In the following, we consider a manifold $\mathcal{M}$ that contains a light-like boundary. The limit, where $\mathcal{N}$ goes to $\mathcal{I}^+$, will be studied in \hyperref[sec5]{section 5}. The entire boundary $\partial\mathcal{M}=M_1\cup\mathcal{N}\cup M_o^{-1}$ of the manifold consists of the null surface $\mathcal{N}$ and two partial Cauchy surfaces $M_1$ and $M_o$ that have the topology of a three-dimensional disc. Each of these discs is anchored at $\mathcal{N}$, which is a three-dimensional null surface embedded into an abstract three-dimensional fibre bundle $P(S^2,\pi,\R)$. Every fibre $\gamma_z = \pi^{-1}(z)$ represents a light ray, and the canonical projection $\pi:P\rightarrow S^2$ maps every such light ray $\gamma_z$ into its base point $z\in S^2$.  
The fibre bundle $P$ is ruled by vertical vector fields,
\begin{equation}
\ell^a\in VP\Leftrightarrow \pi_\ast\ell^a =0,\label{nullvec1}
\end{equation}
where $\pi_\ast T^\ast P\rightarrow TS^2$ denotes the push-forward under the canonical projection. We call any two such vector fields equivalent, $\ell^a\sim{\ell'}^a$ and $[\ell^a]$ is the corresponding equivalence class. Notice that the null boundary $\mathcal{N}$ is only a portion (strip) of $P$. In fact, the boundary of $\mathcal{N}$ consists of two disconnected parts, $\partial\mathcal{N}=\mathcal{C}^{-1}_1\cup\mathcal{C}_o$, each of which is a section of $P$, i.e.\ $\pi(\mathcal{C}_o)=\pi(\mathcal{C}_1^{-1})= S^2$. The orientation of each of these sections is induced from the bulk, i.e.\ $\partial M_1 = \mathcal{C}_1$ and $\partial M_o = \mathcal{C}_o$.

To characterise the free radiative data at the null boundary $\mathcal{N}$, we need to introduce additional metrical structures that the boundary  inherits from the bulk. Consider first the pull-back\footnote{On a null surface $\mathcal{N}$, there is no canonical projector from $T\mathcal{M}$ into $T\mathcal{N}$, because any normal vector to $\mathcal{N}$ has zero length. On the other hand, for co-vectors there is a natural notion of projection, namely the pull-back $T^\ast \mathcal{M}\rightarrow T^\ast\mathcal{N}$.} of the soldering form $e_{AA'a}$. On the null hypersurface $\mathcal{N}$, we can always find a spinor dyad\footnote{Our notation may be a little confusing, because there is now the same pair of letters for different objects: $(k^a,\ell^a)$, $k_a\ell^a=-1$ is a pair of null vectors, and $(k^A,\ell^A)$ is an associate spin dyad, $k_A\ell^A=1$, such that $\I\ou{e}{AA'}{a}\ell_A\bar{\ell}_{A'}=\ell_a$ and $\I\ou{e}{AA'}{a}k_A\bar{k}_{A'}=k_a$. The notation is unambiguous, since $k^A$ (resp.\ $\ell^A$) and $k^a$ (resp.\ $\ell^a$) are ontologically different (spinors and vectors) and can never appear in the same spot.} $(k^A,\ell^A):k_A\ell^A = 1$, a one-form $k_a\in T^\ast \mathcal{N}$ and a complex dyad $(m_a,\bar{m}_a)$ in the complexified co-tangent space $T^\ast_\C \mathcal{N}$ such that we can parametrise the pull-back $\varphi^\ast_{\mathcal{N}}:T^\ast\mathcal{M}\rightarrow T^\ast P$ of the soldering form in terms of the triad $(k_a,m_a,\bar{m}_a)$ and the associate spin dyad $(k^A,\ell^A)$,
\begin{equation}
\varphi^\ast_{\mathcal{N}}e_{AA'} = -\I\ell_A\bar{\ell}_{A'} k + \I k^A\bar{\ell}^{A'}m + \I\ell^A\bar{k}^{A'}\bar{m},\label{nulltetra}
\end{equation}
where $k_a\ell^a=-1$ without loss of generality and the co-dyad $(m_a,\bar{m}_a)$ is always transversal to the null direction $\ell^a$, i.e.\
\begin{equation}
\ell^am_a=0.\label{nullvec2}
\end{equation}

Besides the soldering form $e_{AA'a}$, it is also useful to consider the self-dual Pleba\'{n}ski two-form $\Sigma_{ABab}$. For a given tetrad \eref{nulltetra}, the pull-back of the two-form $\Sigma_{ABab}$ to a null hypersurface can be parametrised by the \emph{null flag} $\ell^A$ (i.e.\ a section of $\mathcal{S}^A(\mathcal{N})$) and a spinor-valued two-form $\eta_{Aab}$ (i.e.\ a section of $\mathcal{S}_A(\mathcal{N})\otimes \bigwedge ^2 (T^\ast\mathcal{N})$), 
\begin{equation}
\varphi_{\mathcal{N}}^\ast\Sigma_{AB} = \eta_{(A}\ell_{B)}.\label{signull}
\end{equation}
If we decompose $\eta_A$ in terms of the triad $(k_a,m_a,\bar{m}_a)$ and the associate spin dyad $(k^A,\ell^A)$, we find
\begin{equation}
\eta_A = \big(\ell_A k - k_A m\big)\wedge \bar{m}.\label{etadef}
\end{equation}

The next step is to identify the appropriate boundary conditions and add the boundary terms to the action in the bulk \eref{bulkactn}.  The analysis simplifies considerably by disentangling the boundary fields $(\eta_{Aab},\ell^A)$ from the self-dual two-form  $\Sigma_{ABab}$ in the bulk. Accordingly, we introduce additional Lagrange multipliers $\omega_a$ (a complex-valued one-form on $\mathcal{N}$) and $\ou{N}{A}{ab}$ (a spinor-valued two-form on $\mathcal{N}$) to impose the \emph{gluing constraints} \eref{signull} and \eref{etadef} on the space of kinematical histories. The resulting bulk plus boundary action is a functional of the bulk fields $(e_{AA'a},\ou{A}{A}{Ba})$ and the boundary fields that consist of the boundary spinors $(\eta_{Aab},\ell^A)$, the co-dyad $(m_a,\bar{m}_a)$, a vertical vector field $\ell^a$, which lies tangential to the fibres $P\supset\mathcal{N}$, and an additional connection one-form $\varkappa_a\in\Omega^1(\mathcal{N})$ that encodes the non-affinity of $\ell^a\in T\mathcal{N}$. The coupled bulk plus boundary action is
\begin{align}\nonumber 
S\big[e_{AA'a},\ou{A}{A}{Ba}\big|&\eta_{Aab},\ell^A,\omega_a,\ou{N}{A}{ab},\varkappa_a,\ell^a,m_a\big]  
 = \frac{\I}{8\pi G}\left[\int_{\mathcal{M}}\Sigma_{AB}\wedge F^{AB}+\right.\\
   &\left.+\int_{\mathcal{N}}\Big(\eta_A\wedge\big(D - \omega - \tfrac{1}{2}\varkappa \big)\ell^A-\omega\wedge m\wedge\bar{m}+N^A\wedge\big(\ell\hook\eta_A+\bar{m}\ell_A\big)\Big)\right] + \CC,\label{actndef}
\end{align}
where $D$ denotes the induced $SL(2,\C)$ covariant derivative on the null hypersurface, i.e.\ $D=\varphi^\ast_{\mathcal{N}}\nabla$, and $\ell\hook\eta_A$ is the interior product\footnote{Using the abstract index notation, we have $(\ell\hook\eta_A)_a=\ell^b\eta_{Aba}$.} between the vertical vector field $\ell^a\in T\mathcal{N}$ and the two-form $\eta_{Aab}$, which is intrinsic to the boundary. In addition, $\ou{N}{A}{ab}$ and $\omega_a$ are Lagrange multipliers. The fixed boundary data on the null surface is a gauge equivalence class of the boundary fields $(\varkappa_a,\ell^a,m_a)$ that will be characterised below.

To obtain the equations of motion, we need to specify the boundary conditions for the action \eref{actndef} on the various parts of $\partial \mathcal{M}=M_1\cup\mathcal{N}\cup M_o^{-1}$. On the partial Cauchy hypersurfaces $M_o$ and $M_1$, we impose the usual Neumann\footnote{On shell, the imaginary part of the Ashtekar connection $\ou{A}{A}{Ba}=\ou{\omega}{AB}{a}+\I\ou{K}{A}{Ba}$ is the extrinsic curvature that represents the normal derivative of the metric to the hypersurface. The real part of the Ashtekar connection  contributes a boundary term to the symplectic two-form. On a closed manifold $\oint_M\Sigma_{AB}\wedge\bbvar{d}\omega^{AB}$ is exact $\oint_M\bbvar{d}\Sigma_{AB}\bbwedge\bbvar{d}\omega^{AB}=0$ and thus generates a symplectic transformation \cite{newvariables,Ashtekar:1987gu,Barberoparam}. } boundary conditions,
\begin{equation}
\varphi^\ast_{M_o}\delta\ou{A}{A}{B}=0,\quad\varphi^\ast_{M_1}\delta\ou{A}{A}{B}=0,\label{bndry1}
\end{equation}
where e.g.\ $\varphi^\ast_{M_o}:T^\ast\mathcal{M}\rightarrow T^\ast M_o$ is the pull-back from the bulk into the boundary. Consider then the boundary conditions on the null surface $\mathcal{N}$. In the interior of $\mathcal{N}$, the boundary fields $(\eta_{Aab},\ell^A,\omega_a,\ou{N}{A}{ab})$ are unconstrained and we will vary them in the boundary action to obtain the corresponding boundary field equations. Since the action contains also derivatives of the null flag $\ell^A$, we then also have to specify boundary conditions at the two consecutive endpoints of the null surface. At these corners, $\mathcal{C}_o=\partial M_o$ and $\mathcal{C}_1=\partial M_1$, we impose additional Dirichlet boundary conditions
\begin{equation}
\delta\ell^A\big|_{\mathcal{C}_o}=0,\quad\delta\ell^A\big|_{\mathcal{C}_1}=0.\label{bndry2}
\end{equation}
On the null surface itself, the boundary conditions constrain the variations of the triple $(\varkappa_a,\ell^a,m_a)$, such that a gauge equivalence class of boundary fields $(\varkappa_a,\ell^a,m_a)$ is kept fixed. A generic such boundary gauge transformation is a combination of vertical diffeomorphisms, dilations of the null normal, shifts of the abelian connection $\varkappa_a$ and complexified conformal transformations of the boundary fields. We will study each contribution below.

First of all, there are the fibre-preserving diffeomorphisms on $\mathcal{N}$. The light-like boundary $\mathcal{N}\supset\partial\mathcal{M}$ is part of a principle bundle $P(S^2,\pi,\R)$, which is ruled by the integral curves of the equivalence class $[\ell^a]$ of null generators (vertical vector fields). Let us denote by $\mathrm{Diff}_0(\mathcal{N})$  the group of (vertical) diffeomorphisms that preserve each individual fibre of $\mathcal{N}$,
\begin{equation}
\operatorname{Diff}_0(\mathcal{N}) = \big\{\varphi\in \operatorname{Diff}(\mathcal{N}) : \pi\circ\varphi\circ\pi^{-1}=\operatorname{id}_{S^2}\big\}.\label{diffN}
\end{equation}
Notice that any such diffeomorphism preserves the two ends of the null boundary, i.e.\ $\varphi|_{\mathcal{C_o}}=\operatorname{id}_{\mathcal{C_o}}$ and $\varphi|_{\mathcal{C_1}}=\operatorname{id}_{\mathcal{C_1}}$. We say any two such triples $(\varkappa_a,\ell^a,m_a)$ and $(\tilde{\varkappa}_a,\tilde{\ell}^a,\tilde{m}_a)$ are gauge equivalent, if they are related by a  vertical  diffeomorphism, 
\begin{equation}
\forall\varphi\in \operatorname{Diff}_0(\mathcal{N}) : \big((\varphi^\ast\varkappa)_a,\ell^a,(\varphi^\ast m)_a\big)\sim\big(\varkappa_a,(\varphi_\ast\ell)^a,m_a\big).\label{gauge1}
\end{equation}
Next, we introduce the dilations of the null generators,
\begin{equation}
\forall f:\mathcal{N}\rightarrow\R, f\big|_{\partial\mathcal{N}}=0: (\varkappa_a,\ell^a,m_a)\sim(\varkappa_a+\partial_a f,\E^f\ell^a,m_a).\label{gauge2}
\end{equation}
We then also have a shift symmetry that only affects the abelian connection $\varkappa_a$,
\begin{equation}
\forall \zeta:\mathcal{N}\rightarrow\C: (\varkappa_a,\ell^a,m_a)\sim(\varkappa_a+\bar{\zeta}m_a+{\zeta}\bar{m}_a,\ell^a,m_a).\label{gauge3}
\end{equation}
Finally, we also have the complexified conformal transformations,
\begin{equation}
\forall \lambda:\mathcal{N}\rightarrow\C: (\varkappa_a,\ell^a,m_a)\sim(\varkappa_a,\E^{\frac{1}{2}(\lambda+\bar{\lambda})}\ell^a,\E^{\lambda}m_a).\label{gauge4}
\end{equation}

\renewcommand{\arraystretch}{1.3}\setcounter{table}{-1}
\begin{table}{\small
\begin{tabularx}{\textwidth}[c]{p{0.9em} p{5em} p{14.3em} p{14.5em} X}\cline{1-5}
\multicolumn{1}{l}{}&\multicolumn{2}{l}{\raisebox{0.1em}{\it space of kinematical histories} $\mathcal{H}_{\mtext{kin}}$}&\raisebox{0.1em}{\it constraints on physical histories $\mathcal{H}_{\mtext{phys}}$}&\phantom{x}\\\hline
\parbox[t]{1em}{\multirow{2}{*}{\rotatebox[origin=c]{90}{\it bulk}}}  & $e_{AA'a}$ & {\it soldering form} & $\nabla e_{AA'}=0$\\[0.1em]
& $\ou{A}{A}{Ba}$ & {\it self-dual connection} & $\ou{F}{B}{A}\wedge e_{BA'} =0$ \\[0.1em]\hline
\parbox[t]{1em}{\multirow{5}{*}{\rotatebox[origin=c]{90}{\it null boundary}}} & $\ell^A$ & {\it null flag} & $D_a\ell^A = \big(\omega_a+\tfrac{1}{2}\varkappa_a\big)\ell^A + \ell^b\ou{N}{A}{ba}$\\
& $\eta_{Aab}$ & {\it spinor-valued two-form} &  $D \eta_A = - \big(\omega+\tfrac{1}{2}\varkappa\big)\wedge \eta_A -N_A\wedge \bar{m}$\\
& $(\omega_a,\ou{N}{A}{ab})$ & {\it $\C$-valued Lagrange multipliers} & $\mathfrak{Re}(\ell^a\omega_a)=0$\\
& $(\varkappa_a,\ell^a,m_a)$ & {\it non-affinity one-form, null generator, co-dyad: $\ell^am_a=0$} & $\varphi^\ast_{\mathcal{N}}\Sigma_{AB} =\eta_A\ell_B+\tfrac{1}{2}\epsilon_{AB}m\wedge\bar{m}$

$\ell\hook \eta_A = -\ell_A\bar{m}$& \\\hline
\end{tabularx}\vspace{1.2em}
}
\caption{On the space of kinematical histories $\mathcal{H}_{\mtext{kin}}$, there is no correlation between the boundary fields and the fields in the interior of the manifold. The correlation is established on the space of physical histories $\mathcal{H}_{\mtext{phys}}$, which consists of all solutions of the bulk plus boundary field equations: there are the Einstein equations and the torsionless condition in the bulk, but there are also additional constraints at the boundary (boundary field equations).}\label{tab1} 
\end{table}

The boundary data that needs to be fixed along the null hypersurface is the gauge equivalence class of the triple $(\varkappa_a,\ell^a,m_a)$ under the combination of these gauge symmetries 
\begin{equation}
\delta\mathcal{g}=0,\quad \mathcal{g} = [\varkappa_a,\ell^a,m_a]/_\sim.\label{gdef}
\end{equation}
Any such gauge equivalence class $\mathcal{g}$ characterises two degrees of freedom at the null boundary. Let us do the counting explicitly: since $\ell^a$ lies tangential to the fibres of $\mathcal{N}$ and $\xi^am_a=0$ for all $\xi\in[\ell^a]$, we see any given triple $(\varkappa_a,\ell^a,m_a)$ is characterised by $3+1+4$ numbers ($m_a$ is complex and all fields $(\varkappa_a,\ell^a,m_a)\in T^\ast\mathcal{N}\times T\mathcal{N}\times T^\ast_\C\mathcal{N}$ are intrinsic to $\mathcal{N}$). The fibre-preserving diffeomorphisms and the dilations remove one degree of freedom each, the shift symmetry removes two degrees of freedom from $\varkappa_a$ and the $U(1)\times\R_>$ complexified conformal transformations remove another two degrees of freedom. This leaves us with two physical degrees of freedom along the interior of $\mathcal{N}$, which are the two physical degrees of freedom of the quasi-local graviton $\mathcal{g}$.

Let us briefly summarise this section. We have defined the coupled bulk plus boundary action \eref{actndef} and specified the boundary conditions. On the partial Cauchy hypersurfaces $M_o$ and $M_1$, the pull-back of the self-dual connection is fixed. Along the null hypersurface $\mathcal{N}$, the boundary data is given by the quasi-local graviton, which is the gauge equivalence class \eref{gdef}. What is missing, is to demonstrate that the action is functionally differentiable for such  boundary conditions. This will be the purpose of the next section, where we compute the boundary field equations and introduce the symplectic potentials along the various portions of the boundary.

\subsection{Bulk plus boundary field equations}
\noindent The purpose of this section is to compute the saddle points of the bulk plus boundary action for the specified boundary conditions \eref{bndry1}, \eref{bndry2} and \eref{gdef}. In the interior of $\mathcal{M}$, the situation is straight forward. The combined variation of the self-dual connection $\ou{A}{A}{Ba}$ and the soldering-forms $e_{AA'a}$ yields the Einstein equations in the first-order formalism,
\begin{subequations}\begin{align}
\nabla e_{AA'}=0,\label{tors}\\
\ou{F}{B}{A}\wedge e_{BA'} = 0\label{curvt}, 
\end{align}\label{bulkEOM}\end{subequations}
where $\nabla$ is the $SL(2,\C)$ covariant exterior derivative for spinor-valued differential forms $\psi_{AB\dots A'B'}$, and $\ou{F}{A}{B}$ is the field strength of the self-dual connection. On the other hand, there are also the boundary fields. The variation of the boundary spinors $(\eta_{Aab},\ell^A)$ yields the boundary field equations along the null hypersurface,
\begin{subequations}\begin{align}
D_a\ell^A & = + \Big(\omega_a + \frac{1}{2}\varkappa_a\Big)\ell^A + \ell^b\ou{N}{A}{ba},\label{bndryeq1}\\
D\eta_A& = - \Big(\omega + \frac{1}{2}\varkappa\Big)\wedge\eta_A - N_A\wedge\bar{m},\label{bndryeq2}
\end{align}\label{bndryeq}\end{subequations}
where $D=\varphi^\ast_{\mathcal{N}}\nabla$ denotes the pull-back of the $SL(2,\C)$ covariant exterior derivative to the null boundary, and $\omega_a$ and $\ou{N}{A}{ab}$ are the Lagrange multipliers that appear in the bulk plus boundary action \eref{actndef}. It is easy to show that the boundary field equations \eref{bndryeq1} and \eref{bndryeq2} are unrestrictive. For any null surface $\mathcal{N}$, one can always find boundary fields $\omega_a$, $\varkappa_a$ and $\ou{N}{A}{ab}$ such that \eref{bndryeq1} and \eref{bndryeq2} are satisfied.\footnote{In addition, the shift symmetry \eref{gauge3} always allows us to achieve $\omega_a=-\bar{\omega}_a$ without loss of generality.}  
On shell, the boundary fields $(\ou{\eta}{A}{ab},\ell^A)$ are correlated to the fields in the bulk. The correlation is obtained from the condition that the coupled bulk plus boundary action be stationary under large variations of the connection,\footnote{The vector field $\delta_A\in T\mathcal{H}_{\mtext{kin}}$ annihilates all configuration variables on the space of kinematical histories except the connection, upon which it acts as $\delta_A[\ou{A}{A}{Ba}]=\delta \ou{A}{A}{Ba}$. The variation is \emph{large}, if $\delta\ou{A}{A}{Ba}$ does not vanish at the null boundary.}
\begin{align}\nonumber
\delta_A S  =  -\frac{\I}{8\pi G}\left[\int_{\mathcal{M}}(\nabla\Sigma_{AB})\wedge\delta A^{AB}-\right.&\int_{M_1\cup M_o^{-1}}\Sigma_{AB}\wedge\delta A^{AB}+\\
&\left.-\int_{\mathcal{N}}\big(\Sigma_{AB}-\eta_{(A}\ell_{B)}\big)\wedge\delta A^{AB}\right]+\CC\label{varAS}
\end{align}
The first term vanishes \emph{on-shell}, namely iff the connection is torsionless, see \eref{tors}. The second term vanishes provided the boundary conditions \eref{bndry1} are satisfied. The third term vanishes for given boundary conditions \eref{bndry1}, \eref{bndry2}, and \eref{gdef} provided the \emph{gluing conditions}, namely equation \eref{signull} is satisfied.
Finally, we also have to take the variation of the Lagrange multipliers $(\omega_a,\ou{N}{A}{ab})$ into account, which yield the additional algebraic constraints
\begin{subequations}\begin{align}
\eta_A\ell^A&=-m\wedge \bar{m},\label{glucond1}\\
\ell^b\ou{\eta}{A}{ba}&=-\ell_A\bar{m}_a.\label{glucond2}
\end{align}\label{glucond}\end{subequations}
Equation \eref{glucond2} aligns the kinematical ruling of $P(S^2,\pi,\R)$ with the causal structure in the interior. The  equation implies that the vector field $\ell^a\in T\mathcal{N}$ is null with respect to the metric in the interior and it also implies that the corresponding null flag is given by $\ell^A$, i.e.\ $\ell^a = \I\ell^A\bar{\ell}^{A'}\uo{e}{AA'}{a}$. Equation \eref{glucond1}, on the other hand, determines the area two-form $\varepsilon_{ab}$ in terms of the boundary spinors,
\begin{equation}
\varepsilon_{ab} = -2\I m_{[a}\bar{m}_{b]} = \I\eta_{Aab}\ell^A.\label{area2form}
\end{equation}

Finally, and most importantly, we also have to take into account  the variations of the triple $(\varkappa_a,\ell^a,m_a)$ for given boundary conditions \eref{gdef}. Any such variation $\delta[\cdot]$ is a sum of four contributions: it is a sum $\delta=\delta^{\mtext{diff}}+\delta^{\mtext{dilat}}+\delta^{\mtext{shift}}+\delta^{\mtext{con}}$ of an infinitesimal fibre-preserving diffeomorphism \eref{diffN}, a dilation \eref{gauge2},  an infinitesimal shift \eref{gauge3} and a complexified conformal transformations \eref{gauge4}. Let us briefly discuss each contribution separately, and show that it vanishes provided the bulk plus boundary equations of motion are satisfied, which are listed in \hyperref[tab1]{table 1}. 

\paragraph{- Fibre-preserving diffeomorphisms} The variation of $(\varkappa_a,\ell^a,m_a)$ along the orbits of the fibre-pre\-serv\-ing diffeomorphisms \eref{gauge1} does not give any further boundary equations of motion. This follows immediately from the invariance of the coupled bulk plus boundary action under vertical diffeomorphisms. In fact, any such diffeomorphism is generated by a vertical vector field $\xi^a\in[\ell^a]$ that vanishes at the boundary of $\mathcal{N}$. The components of the corresponding vector field $\delta_\xi^{\mtext{diff}}\in T\mathcal{H}_{\mtext{kin}}$ on the space of kinematical histories are given by the Lie derivative,
\begin{subequations}\begin{align}
\delta^{\mtext{diff}}_\xi[\varkappa]&:=\mathcal{L}_\xi \varkappa =\xi\hook(\di\varkappa) + \di(\xi\hook\varkappa),\label{diffvec1}\\
\delta_\xi^{\mtext{diff}}[m]&:=\mathcal{L}_\xi m =\xi\hook (\di m),\label{diffvec2}\\
\delta_\xi^{\mtext{diff}}[\ell^a]&:=\mathcal{L}_\xi \ell^a =[\xi,\ell]^a\label{diffvec3}.
\end{align}\label{dxi}\end{subequations}
In addition,  $\delta_\xi^{\mtext{diff}}\in T\mathcal{H}_{\mtext{kin}}$ only acts on $(\varkappa_a,\ell^a,m_a)$, and all other components vanish, i.e.\ $\delta_\xi^{\mtext{diff}}[\ou{A}{A}{Ba}]=0$, $\delta_\xi^{\mtext{diff}}[e_{AA'a}]=0$ etc. The action of the Lie derivative, on the other hand, is well-defined for all bulk plus boundary fields. If $\xi^a\in T\mathcal{M}$ is a vector field in $\mathcal{M}$, we have
\begin{subalign}
\mathcal{L}_\xi\ou{A}{A}{B} & = \xi\hook\ou{F}{A}{B},\\
\mathcal{L}_\xi e_{AA'} & = \xi\hook(\nabla e_{AA'})+\nabla (\xi\hook e_{AA'}).
\end{subalign}
If, in addition, the vector field $\xi^a$ happens to be tangential to the null boundary $\xi^a\big|_{\mathcal{N}}\in T\mathcal{N}$, the action of $\mathcal{L}_\xi$ can be naturally extended to the boundary fields as well,
\begin{subalign}
\mathcal{L}_\xi \ell^A & = \xi^aD_a\ell^A,\\
\mathcal{L}_\xi \eta_A & = \xi\hook (D \eta_A)+D (\xi\hook\eta_A),\\
\mathcal{L}_\xi N^A & =   \xi\hook (D N^A)+D (\xi\hook N^A),\\
\mathcal{L}_\xi\omega & = \xi\hook (\di \omega) + \di(\xi\hook\omega).
\end{subalign}
To show that the boundary conditions \eref{gdef} are satisfied, we need to show that the action is stationary under such $\delta_\xi^{\mtext{diff}}$-variations on the space of physical histories (i.e.\ \emph{on-shell}), i.e.\ $\delta_\xi^{\mtext{diff}}[S]\big|_{\mathcal{H}_{\mtext{phys}}}=0$. This can be seen as follows: let us first smoothly extend the vertical vector field $\xi^a\in [\ell^a]$ into the interior of the manifold in such a way that $\xi^a$ vanishes at the partial Cauchy hypersurfaces $M_o$ and $M_1$, which is possible since $\xi^a$ vanishes already at the endpoints of $\mathcal{N}$, see (\ref{diffN}, \ref{gauge1}). The resulting Lie derivative $\mathcal{L}_\xi\in T\mathcal{H}_{\mtext{kin}}$ preserves the boundary conditions \eref{bndry1}, \eref{bndry2} and \eref{gdef}. The bulk plus boundary action \eref{actndef} is invariant under any such fibre-preserving diffeomorphism, hence $\mathcal{L}_\xi[S]=0$ on $\mathcal{H}_{\mtext{kin}}$. Consider then the vector field $\delta_\xi^V:=\delta_\xi^{\mtext{diff}}-\mathcal{L}_\xi\in T\mathcal{H}_{\mtext{kin}}$. Such a vector field $\delta_\xi^V$ clearly satisfies the boundary conditions \eref{bndry1} and \eref{bndry2}. In addition, it also annihilates the triple $(\varkappa_a,\ell^a,m_a)$, i.e.\ $\delta^V_\xi[\varkappa_a]=0$, $\delta^V_\xi[\ell^a]=0$, $\delta^V_\xi[m_a]=0$. Therefore, all the boundary conditions are fulfilled. At the saddle points of the bulk plus boundary theory, the action is stationary with respect to any such variation that satisfies the boundary conditions, i.e.\ $\delta^V_\xi[S]\big|_{\mathcal{H}_{\mtext{phys}}}=0$.  On the other hand, $\mathcal{L}_\xi[S]=0$ anyways, since the action is invariant under the fibre preserving diffeomorphisms \eref{diffN}. Therefore, 
\begin{equation}\delta_\xi^{\mtext{diff}}[S]\big|_{\mathcal{H}_{\mtext{phys}}}=(\delta^V_\xi+\mathcal{L}_\xi)[S]\big|_{\mathcal{H}_{\mtext{phys}}}=0,\end{equation} such that  the action is invariant under fibre preserving diffeomorphisms \eref{diffN} of the boundary fields $(\varkappa_a,\ell^a,m_a)$ alone, provided the bulk plus boundary field equations \eref{signull}, \eref{bulkEOM},  \eref{bndryeq}, and \eref{glucond} are satisfied.

\paragraph{- Dilations} Next, we consider the dilatations \eref{gauge2} that act via $(\varkappa_a,\ell^a,m_a)\rightarrow (\varkappa_a+\partial_a f,\E^f\ell^a,m_a)$ for any $f\big|_{\partial{N}}=0$ onto the triple $(\varkappa_a,\ell^a,m_a)$ of boundary fields. The corresponding vector field $\delta_f^{\mtext{dilat}}\in T\mathcal{H}_{\mtext{kin}}$ generates the infinitesimal transformation
\begin{subequations}\begin{align}
\delta_f^{\mtext{dilat}}[\varkappa_a] =\partial_a f,\label{dilat1}\\
\delta_f^{\mtext{dilat}}[\ell^a] = f \ell^a,\label{dilat2}
\end{align}\label{dilat}\end{subequations}
and annihilates all other bulk and boundary configuration variables on the space of kinematical histories, e.g.\ $\delta_f[e_{AA'a}]=0$. The resulting variation of the bulk plus boundary action \eref{actndef} yields
\begin{align}
\nonumber \delta_f^{\mtext{dilat}}[S] & = \frac{\I}{8\pi G}\left[\int_{\mathcal{N}}\Big(-\tfrac{1}{2}\di f\wedge\eta_A\ell^A+f N^A\wedge\ell\hook\eta_A\Big)-\CC\right]=\\
&=\frac{\I}{8\pi G}\left[\int_{\mathcal{N}}\Big(\tfrac{1}{2} f\,\di\big(\eta_A\ell^A\big)+f N^A\wedge\ell\hook\eta_A\Big)-\CC\right].\label{dilat3a}
\end{align}
In going from the first to the second line, we used Stokes's theorem. There is no boundary term, since $f\big|_{\partial\mathcal{N}}=0$. Lets now simplify this expression. On shell, i.e. on the space of physical histories, the imaginary part of the $SL(2,\C)$ invariant singlet $\eta_{Aab}\ell^A$ equals the area two-form  $\varepsilon_{ab}\in\Omega^2(\mathcal{N})$, see \eref{area2form}. The exterior derivative of the area two-form defines the expansion of the null hypersurface,
\begin{equation}
\di\varepsilon = -\vartheta_{(\ell)} k\wedge \varepsilon,\label{thetadef1}
\end{equation}
where the one-form $k_a\in\Omega^1(\mathcal{N})$ is dual to $\ell^a$, i.e.\ $\ell^a k_a =-1$, as in e.g.\ \eref{nulltetra}. We then also know that the boundary spinors satisfy the boundary field equations \eref{bndryeq1} and \eref{bndryeq2}, which allow us to write the expansion of the null surface in terms of the Lagrange multiplier $\ou{N}{A}{ab}$. A short calculation gives
\begin{align}\nonumber
\di\varepsilon & = \I\di\left(\eta_A\ell^A\right) = \I (D\eta_A)\ell^A+\I\eta_A\wedge D\ell^A= -\I N_A\wedge \bar{m}\ell^A + \I \eta_A\wedge \ell\hook N^A=\\
& = -\I N_A\wedge \bar{m}\ell^A - \I (\ell\hook\eta_A)\wedge  N^A = - 2\I  (\ell\hook\eta_A)\wedge  N^A.\label{thetadef2}
\end{align}
Going back to the variation  of the action \eref{dilat3a}, we thus have
\begin{equation}
\delta_f^{\mtext{dilat}}[S]\big|_{\mathcal{H}_{\mtext{phys}}}=0.
\end{equation}
On the space of physical histories, the dilatations of the boundary fields $(\varkappa_a,\ell^a,m_a)\rightarrow (\varkappa_a+\partial_a f,\E^f\ell^a,m_a)$ for $f\big|_{\partial\mathcal{N}}=0$ preserve the action.

\paragraph{- Shifts of $\varkappa_a$} The case for the shift symmetry \eref{gauge3} is immediate. The corresponding vector field  acts as
\begin{equation}
\delta_\zeta^{\mtext{shift}}[\varkappa_a] = \zeta\bar{m}_a+\CC,\label{shiftdef1}
\end{equation}
and all other configuration variables are annihilated by the vector field $\delta_\zeta\in T\mathcal{H}_{\mtext{kin}}$, e.g.\ $\delta_\zeta[\ell^A]=0$. On-shell, the variation of the bulk plus boundary  action \eref{actndef} under such a shift  of $\varkappa_a$ is now simply given by
\begin{equation}
\delta_\zeta^{\mtext{shift}}[S]\big|_{\mathcal{H}_{\mtext{phys}}} = -\frac{\I}{16\pi G}\int_{\mathcal{N}}\big((\zeta\bar{m}+\bar\zeta m)\wedge \eta_A\ell^A-\CC\big)\Big|_{\mathcal{H}_{\mtext{phys}}}=
-\frac{1}{8\pi G}\int_{\mathcal{N}} (\zeta\bar{m}+\bar\zeta m)\wedge\varepsilon,\label{zetavarS}
\end{equation}
The last term of \eref{zetavarS} is identically zero, since the area two-form $\varepsilon_{ab}$ is given by $\varepsilon = -\I m\wedge\bar{m}$, such that e.g.\ $\bar\zeta m\wedge\varepsilon=0$. Therefore, the action is  invariant under the shift symmetry \eref{gauge3},
\begin{equation}
\delta_\zeta^{\mtext{shift}}[S]\big|_{\mathcal{H}_{\mtext{phys}}} = 0.
\end{equation}
\paragraph{- Complexified conformal transformations} Finally, let us consider the complexified conformal transformations \eref{gauge4}. The corresponding vector field $\delta^{\mtext{con}}_\lambda\in T\mathcal{H}_{\mtext{kin}}$ is defined by its components
\begin{subalign}
\delta^{\mtext{con}}_\lambda[\ell^a] & = \frac{1}{2}(\lambda+\bar{\lambda})\ell^a,\\
\delta^{\mtext{con}}_\lambda[m_a] & = \lambda m_a,
\end{subalign}
and all other bulk and boundary configuration variables are conserved by $\delta^{\mtext{con}}_\lambda$, e.g.\ $\delta_\lambda[e_{AA'a}]=0$. For any such a vector field $\delta^{\mtext{con}}_\lambda\in T\mathcal{H}_{\mtext{kin}}$, the corresponding  variation of the bulk plus boundary action is given by
\begin{equation}
\delta^{\mtext{con}}_\lambda[S] = \frac{\I}{8\pi G}\int_{\mathcal{N}}\Big(\operatorname{\mathfrak{Re}}(\lambda)\big(-2\omega\wedge m\wedge\bar{m}+N^A\wedge\big(\ell\hook\eta_A+\bar{m}\ell_A\big)\big)+\I\operatorname{\mathfrak{Im}}(\lambda)\,\big(N^A\wedge\bar{m}\ell_A\big)\Big)+\CC
\end{equation}
On shell, the second term vanishes thanks to the gluing condition \eref{glucond2}. The third term does not contribute either: the three-form $\tfrac{\I}{2} N^A\wedge\bar{m}\ell_A=\di\varepsilon$ is real ($\di\varepsilon=\di\bar{\varepsilon}$), such that the third term is cancelled against its complex conjugate. 
In other words,
\begin{equation}
\delta^{\mtext{con}}_\lambda[S] \stackrel{\text{(\ref{bndryeq},\ref{glucond})}}{=}\frac{\I}{8\pi G}\int_{\mathcal{N}}\Big(-2\operatorname{\mathfrak{Re}}(\lambda)\,\omega\wedge m\wedge\bar{m}+\tfrac{1}{2}\operatorname{\mathfrak{Im}}(\lambda)\,\di\varepsilon\Big)+\CC=\frac{1}{2\pi G}\int_{\mathcal{N}}\operatorname{\mathfrak{Re}}(\lambda)\,\operatorname{\mathfrak{Re}}(\omega)\wedge\varepsilon.
\end{equation}
For arbitrary $\lambda$ and $\varepsilon_{ab}\neq 0$, this variation vanishes iff
\begin{equation}
\operatorname{\mathfrak{Re}}(\ell^a\omega_a) = 0.\label{omega0}
\end{equation}
Therefore, the time component $\operatorname{\mathfrak{Re}}(\ell^a\omega_a)$ of the real part of the Lagrange multiplier $\omega_a$ must be set to zero on the space of physical histories $\mathcal{H}_{\mtext{phys}}$. This in turn implies that the one-form $\varkappa_a$ determines the non-affinity of $\ell^a$. In fact, if we go back to the boundary field equations \eref{bndryeq1}, and take into account that $\ell^a e_{AA'a} =\I\ell_A\bar{\ell}_{A'}$ and $\nabla e _{AA'}=0$ (vanishing of torsion), we find
\begin{equation}
\text{on $\mathcal{H}_{\mtext{phys}}$} : \ell^b\nabla_b\ell^a = \I\uo{e}{AA'}{a}\,\ell^bD_b(\ell^A\bar{\ell}^{A'}) = \ell^b\varkappa_b\,\ell^a \equiv \kappa\ell^a.
\end{equation}

\paragraph{- Summary} In this section, we identified the saddle points of the coupled bulk plus boundary action \eref{actndef} for fixed boundary conditions \eref{bndry1}, \eref{bndry2}, \eref{gdef}. The saddle points of the action are characterised by the Einstein equations in the bulk \eref{tors}, \eref{curvt}, and the boundary field equations \eref{bndryeq1}, \eref{bndryeq2} and \eref{omega0} that determine the evolution of the boundary fields $(\ou{\eta}{A}{ab},\ell^A)$ along the light-like boundary $\mathcal{N}$. In addition, there are the constraint equations \eref{signull} and \eref{glucond} that follow from the variation of the selfdual connection and the variation of the Lagrange multipliers $\ou{N}{A}{ab}$ and $\omega_a$. The free initial data along the lightlike portion of the boundary is given by the quasi-local graviton, i.e.\ the gauge equivalence class $\mathcal{g}$ that characterises the two radiative modes along the null boundary. \hyperref[tab1]{Table 1} summarises the field content of the bulk plus boundary field theory and the constraints that determine the space of physical histories. 

\subsection{Corner terms and symplectic structure}\label{sec3.3}
\noindent The variation of the action \eref{actndef} determines both the bulk plus boundary field equations as well as the symplectic potentials,
\begin{equation}
\delta[S] = - \Theta_{M_o}(\delta) + \Theta_{M_1}(\delta) + \Theta_{\mathcal{N}}(\delta) + \text{EOM}(\delta).\label{varS}
\end{equation}
In here, $\delta\in T\mathcal{H}_{\mtext{kin}}$ is a tangent vector on the space of kinematical histories, $\Theta_{M_o}$,  $\Theta_{M_o}$ and  $\Theta_{\mathcal{N}}\in T^\ast\mathcal{H}_{\mtext{kin}}$ are the pre-symplectic potentials for the various components of the boundary (recall that $\partial{\mathcal{M}}=M_1\cup\mathcal{N}\cup M_o^{-1}$), and $\text{EOM}\in T^\ast\mathcal{H}_{\mtext{kin}}$ denotes a one-form, whose pull-back to the space of physical histories vanishes (i.e.\ it determines the bulk plus boundary field equations). Clearly, there is no unique splitting  of the variation \eref{varS} into the various component parts. First of all, we can always add terms to the various pre-symplectic potentials that vanish on the space of physical histories (provided we add the appropriate counter-terms to the one-form $\text{EOM}$). In addition, there are  ambiguities at the corner \cite{Wieland:2017zkf,DePaoli:2018erh}. For example,  $\mathcal{N}$ has a boundary consisting of the two corners, $\partial\mathcal{N}=\mathcal{C}_o\cup\mathcal{C}_1^{-1}$, such that the $\ell^A$-variation of the boundary action \eref{actndef} generates a corner term,
\begin{equation}
\delta_{\ell^A}[S] = \frac{\I}{8\pi G}\left[\int_{\partial\mathcal{N}}\eta_A\delta\ell^A-\int_{\mathcal{N}}\delta\ell^A\underbrace{\Big(D\eta_A+(\omega+\tfrac{1}{2}\varkappa)\wedge\eta_A+N_A\wedge\bar{m})\Big)}_{=\,0\,\mtext{on-shell}}\right]+\CC,\label{varlS}
\end{equation}
where $\delta_{\ell^A}\in T\mathcal{H}_{\mtext{kin}}$ is the vector field on field space
\begin{equation}
\delta_{\ell^A} =\int_{\mathcal{N}}\delta\ell^A\frac{\delta}{\delta\ell^A}+\CC
\end{equation}
The second term of \eref{varlS} vanishes provided the boundary field equations \eref{bndryeq2} are satisfied. The first term is a sum of two corner terms. To make sure that large $SL(2,\C)$ transformations\footnote{Such gauge transformations are generated by vector fields $\delta_\Lambda\in T\mathcal{H}_{\mtext{kin}}$ for gauge parameter $\ou{\Lambda}{A}{B}:\mathcal{M}\rightarrow\mathfrak{sl}(2,\C)$ such that $\delta_\Lambda[\ou{A}{A}{Ba}]=-\nabla_a\ou{\Lambda}{A}{B}$, $\delta_\Lambda[\ou{\Sigma}{AB}{ab}]=2\ou{\Lambda}{(A}{C}\ou{\Sigma}{B)C}{ab}$, $\delta_\Lambda[\ou{\eta}{A}{ab}]=\ou{\Lambda}{A}{B}\ou{\eta}{A}{ab}$, $\delta_\Lambda[\ell^A]=\ou{\Lambda}{A}{B}\ell^B$ and $\ou{\Lambda}{A}{B}\big|_{\partial\mathcal{M}}\neq 0$. } of the bulk plus boundary fields represent unphysical  gauge transformations, we  include these corner terms into the pre-symplectic potentials $\Theta_{M_o}$ resp.\ $\Theta_{M_1}$ on the partial Cauchy surfaces $M_o$ and $M_1$, see \cite{Wieland:2017zkf} for details. Accordingly,
\begin{equation}
\Theta_M = \frac{\I}{8\pi G}\left[\int_M\Sigma_{AB}\wedge\bbvar{d}A^{AB}-\oint_{\partial{M}}\eta_A\bbvar{d}\ell^A\right]+\CC\label{thetamdef}
\end{equation}
The pre-symplectic potential $\Theta_{\mathcal{N}}$ on the null hypersurface $\mathcal{N}$ is then inferred from the variation of the action \eref{varS} for given boundary conditions \eref{gdef}. Taking into account the boundary field equations (see \hyperref[tab1]{table 1}), we find\footnote{The null boundary $\mathcal{N}$ is a part of the abstract fibre bundle $P(S^2,\pi,\R)$, whose standard fibres $\pi^{-1}(z)$ for $z\in S^2$ are light rays in $\mathcal{N}$. The null vector $\ell^a$ lies tangential to these fibres, and the differential of $\ell^a$ satisfies, therefore, $\bbvar{d}\ell^a\propto\ell^a$, $m_a\bbvar{d}\ell^a=0$.   }
\begin{align}\nonumber
\Theta_{\mathcal{N}} &= \frac{\I}{8\pi G}\int_{\mathcal{N}}\Big[-\tfrac{1}{2}\eta_A\ell^A\wedge\bbvar{d}\varkappa+N^A\wedge(\bbvar{d}\ell)\hook\eta_A+N^A\ell_A\wedge\bbvar{d}\bar{m}\Big]+\CC=\\
 &= \frac{\I}{8\pi G}\int_{\mathcal{N}}\Big[-\tfrac{1}{2}\eta_A\ell^A\wedge\bbvar{d}\varkappa-\big((\bbvar{d}\ell)\hook N^A\big)\wedge\eta_A-k\wedge (\ell\hook N^A)\ell_A\wedge\bbvar{d}\bar{m}\Big]+\CC,\label{NTheta}
\end{align}
where $k_a\in T^\ast\mathcal{N}$ is  a one-form dual to $\ell^a$ that satisfies $k_a\ell^a=-1$. For all practical calculations, it is now  useful to replace $\I\eta_A\ell^A$ by the area two-form $\varepsilon=-\I m\wedge\bar{m}$ and eliminate the Lagrange multiplier $\ou{N}{A}{ab}$ from \eref{NTheta} in terms of derivatives of $\ell^A$. Going back to the boundary field equation \eref{bndryeq1}, we have
\begin{equation}
\ell\hook N^A\ell_A = \ell_AD\ell^A,\label{lDl}
\end{equation}
We may now replace \eref{NTheta} by the following simplified expression\footnote{Modulo terms that vanish provided the bulk plus boundary field equations are satisfied.}
\begin{equation}
\Theta_{\mathcal{N}} = -\frac{1}{8\pi G}\int_{\mathcal{N}}\varepsilon\wedge\bbvar{d}\varkappa +
\frac{\I}{8\pi G}\int_{\mathcal{N}}\Big(\ell_AD\ell^A\wedge\bbvar{d}(k\wedge\bar{m})-\CC\Big).\label{thetanull}
\end{equation}
It is important to realise that the one-form $\Theta_{\mathcal{N}}$ depends only on the triple $(\varkappa_a,\ell^a,m_a)$. This is obvious for the first term in \eref{thetanull}. The second term, on the other hand, involves the one-form $\ell_AD\ell^A$, which is completely determined by shear and expansion of $\ell^a$. This can be seen as follows. Consider an adapted Newman\,--\,Penrose null co-tetrad\footnote{Given the boundary variables $(\ell^a,m_a,\bar{m}_a)$, such a null tetrad is unique modulo residual Lorentz transformations $(\tensor[^4]{k}{_a},\tensor[^4]{\ell}{_a},\tensor[^4]{m}{_a},\tensor[^4]{\bar{m}}{_a})\rightarrow(\tensor[^4]{k}{_a}+\bar{f}\,\tensor[^4]{m}{_a}+f\,\tensor[^4]{\bar{m}}{_a},\tensor[^4]{\ell}{_a},\tensor[^4]{m}{_a}+f\,\tensor[^4]{\ell}{_a},\tensor[^4]{\bar{m}}{_a}+\bar{f}\,\tensor[^4]{\ell}{_a})$ for $f:\mathcal{N}\rightarrow\C$.} $(\tensor[^4]{k}{_a},\tensor[^4]{\ell}{_a},\tensor[^4]{m}{_a},\tensor[^4]{\bar{m}}{_a})$, whose pull-back to $\mathcal{N}$ is given by $\varphi^\ast_{\mathcal{N}}(\tensor[^4]{k}{_a},\tensor[^4]{\ell}{_a},\tensor[^4]{m}{_a},\tensor[^4]{\bar{m}}{_a})=(k_a,0,m_a,\bar{m}_a)$, with $k_a\ell^a=-1$. Let $(k^A,\ell^A), k_A\ell^A=1$ be the associated spin dyad such that \eref{nulltetra} is satisfied. If we then impose the torsionless condition \eref{tors}, we can express shear and expansion of the null surface in terms of the one-form $\ell_AD\ell^A$,
\begin{subalign}
\sigma_{(\ell)} & =  [\tensor[^4]{m}{}]^a[\tensor[^4]{m}{}]^b\nabla_a\tensor[^4]{\ell}{_b}=-\bar{k}^{A'}\ell^A m^a D_a(\ell_A\bar{\ell}_{A'})=-\ell_Am^aD_a\ell^A,\label{sheardef}\\
\frac{1}{2}\vartheta_{(\ell)} & = [\tensor[^4]{m}{}]^{(a}[\tensor[^4]{\bar{m}}{}]^{b)} \nabla_a\tensor[^4]{\ell}{_b} =
[\tensor[^4]{m}{}]^{a}[\tensor[^4]{\bar{m}}{}]^{b} \nabla_a\tensor[^4]{\ell}{_b} =  - \bar{k}^{A'}\ell^A \bar{m}^a D_a(\ell_A\bar{\ell}_{A'})=- \ell_A\bar{m}^aD_a\ell^A,\label{exdef}
\end{subalign}
where $m^a$ is the  complexified tangent vector $m^a\in T\mathcal{N}_\C$ such that the push-forward satisfies $(\varphi_{\mathcal{N}})_\ast m^a = [\tensor[^4]{m}{}]^a$. Since $\ell_A \ell^aD_a\ell^A=0$, see \eref{bndryeq1}, we then also have that
\begin{equation}
\ell_A D\ell^A=-\frac{1}{2}\vartheta_{(\ell)}m-\sigma_{(\ell)}\bar{m}.\label{thetashear}
\end{equation}
Notice that shear and expansion can be easily computed without explicit knowledge of the spin connection. In fact, the ordinary Lie derivative of the boundary intrinsic dyads $m_a\in\Omega^1(\mathcal{N}:\C)$ along the null generators $\ell^a\in T\mathcal{N}$ determines both shear and expansion,\footnote{Notice that $\phi_{(\ell)}:\mathcal{N}\rightarrow\R$ transforms as the time component of a $U(1)$ connection under the $U(1)$ transformations $m_a\rightarrow\E^{\I \varphi}m_a$.}
\begin{equation}
\mathcal{L}_\ell m = +\frac{1}{2}\left(\vartheta_{(\ell)}+\I\phi_{(\ell)}\right)m +\sigma_{(\ell)}\bar{m}.\label{Liem}
\end{equation}
Therefore, shear and expansion are  intrinsic to the null surface. The only extrinsic spin coefficient that enters the pre-symplectic structure on the null surface $\mathcal{N}$ is the one-form $\varkappa_a$.

\subsection{Quasi-local radiative phase space}\label{sec3.4}
\noindent In the previous section, we identified the pre-symplectic potential \eref{thetanull} on a generic null boundary $\mathcal{N}$ in terms of an adapted Newman\,--\,Penrose null tetrad. The next step is to characterise the pull-back of the pre-symplectic structure to the space of radiative modes, which are encoded into a gauge equivalence class \eref{gdef} of boundary fields $(\varkappa_a,\ell^a,m_a)\in T^\ast\mathcal{N}\otimes V\mathcal{N}\otimes T^\ast \mathcal{N}_\C$. In here, $\ell^a$ is a vertical vector field, and the dyads $(m_a,\bar{m}_a)$, diagonalise the intrinsic signature $(0$$+$$+)$ metric $q_{ab}=2m_{(a}\bar{m}_{b)}$ on the null surface. The space of such boundary fields $(\varkappa_a,\ell^a,m_a)$ is equipped with a pre-symplectic two-form $\Omega_{\mathcal{N}}$, which is given by the exterior functional derivative of the pre-symplectic potential \eref{thetanull},
\begin{equation}
\Omega_{\mathcal{N}} = \bbvar{d}\Theta_{\mathcal{N}}.\label{OmN}
\end{equation}

At the kinematical level, there are eight degrees of freedom per point: since $\ell^a$ is a vertical vector field, it is characterised by only one degree of freedom (a lapse function). The co-dyad $(m_a,\bar{m}_a)$, on the other hand, is transversal to the null direction, i.e.\ $\forall\xi^a\in[\ell^a]$, $\xi^a\in[\ell^a]:\xi^am_a=0$, hence there are only $2\times 2=4$ degrees of freedom in $m_a$. The entire triple $(\varkappa_a,\ell^a,m_a)$ represents, therefore,  $3+1+2\times 2 = 8$ kinematical degrees of freedom. The gauge symmetries and constraints reduce them to two physical degrees of freedom per point.

\paragraph{- Fibre-preserving diffeomorphisms} Let us consider the fibre-preserving diffeomorphisms \eref{gauge1} first. Such diffeomorphisms are generated by vertical vector fields $\xi^a\in[\ell^a]$ that vanish at the boundary of $\mathcal{N}$, i.e.\ $\xi^a\big|_{\mathcal{C}_o}=\xi^a\big|_{\mathcal{C}_1}=0$. The Lie derivative $\mathcal{L}_\xi$, see \eref{diffvec1}, \eref{diffvec2}, \eref{diffvec3} lifts any such vector field on space time into an associated vector field $\delta^{\mtext{diff}}_\xi\in T\mathcal{H}_{\mtext{kin}}$ on field space. 

We now want to convince ourselves that such a vector field defines a degenerate null direction of the pre-symplectic two-form \eref{OmN} on the space of physical geometries. Consider thus a second linearly independent vector field $\delta\in T\mathcal{H}_{\mtext{kin}}$, and assume further (without loss of generality) that the commutator vanishes, i.e.\ $[\delta,\delta^{\mtext{diff}}_\xi]=0$. Going back to the definition of the pre-symplectic potential, we now immediately have
\begin{align}\nonumber
\Omega_{\mathcal{N}}(\delta^{\mtext{diff}}_\xi,\delta)&=-\frac{1}{8\pi G}\int_{\mathcal{N}}\Big(\mathcal{L}_\xi\varepsilon\wedge\delta[\varkappa]-\delta[\varepsilon]\wedge\mathcal{L}_\xi\varkappa\Big)+\\\nonumber
&\qquad+\frac{\I}{8\pi G}\int_{\mathcal{N}}\Big(\mathcal{L}_\xi(\ell_AD\ell^A)\wedge\delta[k\wedge\bar{m}]-\delta[\ell_AD\ell^A]\wedge\mathcal{L}_\xi(k\wedge\bar{m})-\CC\Big)=\\
&=-\delta[C_\xi].\label{Cxi0}
\end{align}
where we repeatedly used the identity $\int_{\mathcal{N}}\mathcal{L}_\xi X\wedge\delta{Y}=\int_{\partial{\mathcal{N}}}\xi\hook(X\wedge\delta(Y))-\int_{\mathcal{N}} X\wedge\delta(\mathcal{L}_\xi Y)$, i.e.\ a partial integration and the vanishing of the commutator $[\delta,\mathcal{L}_\xi]=0$. In addition, $C_\xi$ denotes the generator of fibre-preserving diffeomorphisms
\begin{equation}
C_\xi = \frac{1}{8\pi G}\int_{\mathcal{N}}\mathcal{L}_\xi\varepsilon\wedge\varkappa-\frac{\I}{8\pi G}\int_{\mathcal{N}}\Big(\mathcal{L}_\xi(\ell_A D\ell^A)\wedge (k\wedge \bar{m})-\CC\Big).\label{Cxi}
\end{equation}
To demonstrate that $C_\xi$ generates a gauge symmetry, we must show that it vanishes (as a constraint) on the space of physical histories.\footnote{Equations \eref{Lie1}, \eref{Lie2}, \eref{Lie3}, \eref{Cxi2} are satisfied only on-shell, i.e.\ provided the bulk plus boundary field equations are satisfied, see \hyperref[tab1]{table 1}.} The first step is to compute the Lie derivative of the one-form $\ell_AD\ell^A$. Taking into account the boundary field equation $\xi^aD_a\ell^A=\tfrac{1}{2}\xi^a(\varkappa_a+2\omega_a)\ell^A$, see \eref{bndryeq1}, we have 
\begin{align}\nonumber
\mathcal{L}_\xi(\ell_AD\ell^A) &= \di\underbrace{\big(\xi\hook\ell_AD\ell^A\big)}_{=0} + \xi\hook D(\ell_AD\ell^A) = \xi\hook (D\ell_A\wedge D\ell^A)-\xi\hook F_{AB}\ell^A\ell^B=\\\nonumber
& =  - 2\xi\hook (D\ell_A k^A\wedge \ell_BD\ell^B)-\xi\hook F_{AB}\ell^A\ell^B=\\
& =  + \xi\hook(\varkappa+2\omega)\ell_AD\ell^A-\xi\hook F_{AB}\ell^A\ell^B,\label{Lie1}
\end{align}
where $F_{AB}$ is the curvature two-form. Going from the first to the second line of \eref{Lie1}, we wrote the identity $\uo{\epsilon}{A}{B}$ in terms of the spin dyad, i.e.\ $\uo{\epsilon}{A}{B}=\ell^B k_A-k^B\ell_A$, and repeatedly used the on-shell identity $\ell_A\xi^aD_a\ell^A\propto\ell_A\ell^A=0$. Inserting \eref{Lie1} and the expression for $\ell_AD\ell^A$ in terms of shear and expansion, see \eref{thetashear}, back into \eref{Cxi}, we obtain
\begin{align}\nonumber
\mathcal{L}_\xi(\ell_A D\ell^A)\wedge (k\wedge \bar{m})-\CC & =  +\frac{1}{2}\xi\hook(\varkappa+2\omega)\vartheta_{(\ell)} k\wedge m\wedge\bar{m}-\xi\hook F_{AB}\ell^A\ell^B\wedge k\wedge \bar{m}-\CC=\\
& = -\frac{1}{2}(\xi\hook\varkappa)\di\varepsilon-\xi\hook F_{AB}\ell^A\ell^B\wedge k\wedge \bar{m}-\CC,\label{Lie2}
\end{align}
where the one-form $\omega_a$ falls out of the final result since the reality condition \eref{omega0} is satisfied (on-shell). If we now take the pull-back of the co-tetrad to the null boundary, recall \eref{nulltetra}, we may replace the one-form $\ell_A\ell_B\bar{m}$ by the pull-back $-\I\varphi^\ast_{\mathcal{N}}e_{AA'}\bar{\ell}^{A'}\equiv -\I{}^3e_{AA'}\bar{\ell}^{A'}$. This in turn allows us to rewrite the generator  solely in terms of the Einstein three-form \eref{curvt},
\begin{align}\nonumber
\xi\hook F_{AB}\ell^A\ell^B\wedge k\wedge \bar{m}&= +\I (\xi\hook F_{AB})\ell^A\wedge \tensor[^3]{e}{^B_{B'}}\bar{\ell}^{B'}\wedge k =-\I F_{AB}\ell^A\bar{\ell}^{B'}\wedge \xi\hook(\tensor[^3]{e}{^B_{B'}}\wedge k)\\
&= +\I F_{AB}\wedge\tensor[^3]{e}{^B_{B'}} \ell^A\bar{\ell}^{B'}\xi\hook k=
 -F_{AB}\wedge\tensor[^3]{e}{^B_{A'}}\xi^{AA'}\label{Lie3}
\end{align}
Substituting \eref{Lie2} and \eref{Lie3} back into \eref{Cxi}, we find
\begin{equation}
C_\xi = \frac{1}{8\pi G}\int_{\mathcal{N}}(\xi\hook\di\varepsilon)\wedge\varkappa -\frac{1}{8\pi G}\int_{\mathcal{N}}(\xi\hook\varkappa) \di\varepsilon - 
\frac{\I}{8\pi G}\int_{\mathcal{N}}\Big(F_{AB}\wedge \ou{e}{B}{A'}\xi^{AA'}-\CC\Big).\label{Cxi2}
\end{equation}
The first two terms cancel against each other, the third term is the integral of the vector-valued Einstein three-form \eref{curvt} over the null boundary. If the bulk plus boundary field equations are satisfied, the generator $C_\xi$ vanishes. Therefore, the fibre-preserving diffeomorphisms of $\mathcal{N}$ represent gauge transformation that map a given configuration of boundary fields $(\varkappa_a,\ell^a,m_a)$ into a gauge equivalent configuration $(\varphi^\ast\varkappa_a,\varphi^{-1}_\ast\ell^a,\varphi^\ast m_a)$.  
\paragraph{- Dilations} Next, we consider the dilations that send the triple $(\varkappa_a,\ell^a,m_a)$ into $(\varkappa_a+\partial_af,\E^f\ell^a,m_a)$ for gauge parameters $f:\mathcal{N}\rightarrow\R$ that vanish at the two ends of the null surface, i.e.\ $f\big|_{\partial \mathcal{N}}=0$. The corresponding vector field $\delta^{\mtext{dilat}}_f\in T\mathcal{H}_{\mtext{kin}}$ is given in \eref{dilat}. 
Consider then the one-form $k_a\in T^\ast\mathcal{N}$, which is dual to $\ell^a: k_a\ell^a=-1$, and the one-form $\ell_AD\ell^A$, see \eref{thetashear}. Going back to our definitions \eref{dilat} and \eref{sheardef} and \eref{exdef}, we immediately find that fields transform homogeneously, i.e.\
\begin{align}
\delta^{\mtext{dilat}}_f[\ell_AD\ell^A] & =  f\ell_AD\ell^A,\label{dilat3b}\\
\varepsilon\wedge\delta^{\mtext{dilat}}_f[k] &= \delta^{\mtext{dilat}}_f[\varepsilon\wedge k] = -f\varepsilon\wedge k.\label{dilat4b}
\end{align}
Let then $\delta\in T\mathcal{H}_{\rm{kin}}$ be a second linearly independent vector field. Inserting \eref{dilat3b} and \eref{dilat4b} back into the definition of the pre-symplectic two-form \eref{OmN}, we obtain a total derivative, namely
\begin{align}\nonumber
\Omega_{\mathcal{N}}(\delta^{\mtext{dilat}}_f,\delta)&=+\frac{1}{8\pi G}\int_{\mathcal{N}}\delta[\varepsilon]\wedge \di f+\frac{\I}{8\pi G}\int_{\mathcal{N}}f\delta\big[\ell_AD\ell^A\wedge k\wedge\bar{m}-\CC\big]=\\
&=+\frac{1}{8\pi G}\int_{\mathcal{N}}\delta[\varepsilon]\wedge \di f+\frac{1}{8\pi G}\int_{\mathcal{N}}f\delta[\di\varepsilon]=\frac{1}{8\pi G}\int_{\partial\mathcal{N}}f\delta[\varepsilon]=0.\label{dilatedge}
\end{align}
On shell, i.e.\ provided the bulk plus boundary field equations are satisfied, $\delta^{\mtext{dilat}}_f$ is a degenerate direction of the pre-symplectic two-form \eref{OmN}. Therefore, the vector field $\delta^{\mtext{dilat}}_f$ defines an unphysical gauge direction.

\paragraph{- $U(1)$ frame rotations} On the null surface $\mathcal{N}$, the complex-valued one-form $m_a\in T^\ast\mathcal{N}_\C$ param\-etrises the degenerate signature $(0$$+$$+)$ metric $q_{ab}=2m_{(a}\bar{m}_{b)}$. For any such metric, the dyad is unique modulo $U(1)$ gauge transformations. Given a $U(1)$ gauge parameter $\varphi:\mathcal{N}\rightarrow\R$, we define the corresponding infinitesimal vector field
\begin{equation}
\delta^{U(1)}_\varphi[m_a] = \I\varphi\, m_a.\label{rot1}
\end{equation}
Going back to the definition of shear and expansion, see \eref{sheardef} and \eref{exdef}, we also see that the one-form $\ell_AD\ell^A$ transforms homogeneously under such an infinitesimal $U(1)$ frame rotation, i.e.\
\begin{align}
\delta^{U(1)}_\varphi[\ell_AD\ell^A] & = \I\varphi\,\ell_AD\ell^A.\label{rot2}
\end{align}
On the other hand, the non-affinity one-form $\varkappa_a$ and the area two-form $\varepsilon=-\I m\wedge\bar{m}$ are uncharged and so is the one-form $k_a$, which is dual to the null vector $\ell^a:k_a\ell^a=-1$. Therefore, $\delta^{U(1)}_\varphi[\varepsilon_{ab}]=0$, etc. If we then insert such an infinitesimal $U(1)$ rotation back into the pre-symplectic two-form \eref{OmN}, we obtain
\begin{align}\nonumber
\Omega_{\mathcal{N}}(\delta^{U(1)}_\varphi,\delta) & = \frac{\I}{8\pi G}\int_{\mathcal{N}}\Big(\I\varphi\,\delta\big(\ell_AD\ell^A\wedge k\wedge \bar{m}\big)-\CC\Big)=\\
&=-\frac{1}{16\pi G}\int_{\mathcal{N}}\Big(\varphi\,\delta\big(\vartheta_{(\ell)}k\wedge m\wedge\bar{m}\big)+\CC\Big)=0,\label{rotnull}
\end{align}
which is the same as to say that the infinitesimal $U(1)$ transformations are unphysical null directions of the pre-symplectic potential. 
\paragraph{- Shifts of $\varkappa_a$} Finally, we consider the shift symmetry \eref{gauge3}. The corresponding vector field $\delta_\zeta^{\mtext{shift}}$ generates spacelike shifts of the one-form $\varkappa_a$: $\delta_\zeta^{\mtext{shift}}[\varkappa_a]=\zeta\bar{m}_a+\bar{\zeta}m_a$. All other boundary variables are annihilated by the vector field $\delta_\zeta^{\mtext{shift}}\in\mathcal{H}_{\mtext{kin}}$, i.e.\ $\delta_\zeta^{\mtext{shift}}[m_a]=0$, $\delta_\zeta^{\mtext{shift}}[\ell^a]=0$, etc. Let then $\delta\in\mathcal{H}_{\mtext{kin}}$ be a second linearly independent vector field on the space of kinematical histories. If we contract both such vector fields with the pre-symplectic two-form $\Omega_{\mathcal{N}}$, we find 
\begin{align}
\Omega_{\mathcal{N}}(\delta_\zeta^{\mtext{shift}},\delta)&=\frac{1}{8\pi G}\int_{\mathcal{N}}\delta[\varepsilon]\wedge\delta_\zeta^{\mtext{shift}}[\varkappa]=
\frac{1}{8\pi G}\int_{\mathcal{N}}\delta[\varepsilon]\wedge(\bar{\zeta}m+\CC)=0.\label{shiftnull}
\end{align}
The last term vanishes identically since any such vector field $\delta\in\mathcal{H}_{\mtext{kin}}$ preserves the direction of the null generators of $\mathcal{N}$, which implies $\ell^a\delta[m]_a=0$, which is the same as to say $\varepsilon\wedge\delta[m]= 0$, i.e.\ $\delta[\varepsilon]\wedge m= 0$, since $\varepsilon=-\I m\wedge\bar{m}$. We thus see that the shift transformations define yet another degenerate direction $\delta_\zeta^{\mtext{shift}}\in T\mathcal{H}_{\mtext{kin}}$ of the pre-symplectic two-form \eref{OmN}.

\paragraph{- Summary} In this section, we identified the degenerate gauge directions of the pre-symplectic two-form on the null hypersurface $\mathcal{N}$. First of all, there are the fibre preserving diffeomorphisms \eref{gauge1}. Such diffeomorphisms are  generated by a Hamiltonian functional  $C_\xi$, which is a smeared version of the Raychaudhuri equation. The Hamiltonian $C_\xi$ is a constraint that vanishes on the space of physical histories, where the fibre preserving diffeomorphisms turn into degenerated gauge directions of the pre-symplectic two-form \eref{OmN}.  On the space of physical histories (on-shell), the fibre-preserving diffeomorphisms remove, therefore, two dimensions from phase space (the gauge orbit plus the constraint $C_\xi=0$). On the other hand, the dilations \eref{dilat3b} and \eref{dilat4b} and $U(1)$ frame rotations \eref{rot1} remove one phase space dimension each. Finally, there is the shift symmetry \eref{gauge3} that removes another two dimensions from phase space, see \eref{shiftnull}. The triple of kinematical boundary fields $(\varkappa_a,\ell^a,m_a)$, whose functional differential determines the pre-symplectic two-form (\ref{OmN}, \ref{NTheta}) is characterised by eight local degrees of freedom along $\mathcal{N}$. Removing the fibre-preserving diffeomorphisms and imposing the Raychaudhuri constraint \eref{Cxi} brings this down to $8-2 = 6$ local degrees of freedom per point. The dilatations and $U(1)$ gauge rotations remove another two phase space dimensions per point. The shift symmetry removes yet another directions from phase space, which brings us down to $6-4=2$ physical degrees of freedom along $\mathcal{N}$, which are the two degrees of freedom of gravitational radiation at the full non-perturbative level. The resulting physical phase space is co-ordinatized by the quasi-local graviton \eref{gdef}, which represents the free initial data along the null hypersurface.

{We defined the quasi-local graviton as an equivalence class \eref{gdef}. In defining this equivalence class, we also removed the orbits of conformal transformations \eref{gauge4} from $(\varkappa_a,\ell^a,m_a)$. From a Hamiltonian perspective, this may seem odd, since such conformal transformations are \emph{not} gauge directions (it is easy to check that they do not define degenerate directions of the pre-symplectic two-form). However, no mistake is being made, because on every such orbit there is only one value of the conformal factor (the gauge parameter $\lambda$) that is compatible with the Raychaudhuri equation. In fact, the Rachaudhuri equation turns into a constraint \eref{Cxi} on the kinematical phase space, and it selects a unique value of $\lambda$ on every such gauge orbit for given initial conditions on a cross-section $\mathcal{C}_o$ of $\mathcal{N}$ (the initial values are $\lambda_o=\lambda|_{\mathcal{C}_o}$ and $\dot{\lambda}_o=\mathcal{L}_\ell\lambda|_{\mathcal{C}_o}$). The construction is reminiscent of conformal methods on a spacelike hypersurface, where the orbits of three-dimensional conformal transformations are used often to determine a local gauge-fixing for the Hamiltonian constraint, see \cite{Isenberg:2013iva,shapebook,Gomes:2011zi}. }
\section{Quasi-local boost and angular momentum charges}\label{sec4}
\paragraph{- Horizontal diffeomorphisms} 
On the space of kinematical histories $\mathcal{H}_{\mtext{kin}}$, the light rays $\pi^{-1}(z)$ are shared among different spacetime geometries (histories), but their parametrisation is not. There is therefore a preferred class of bulk diffeomorphisms $\hat{\varphi}\in \mathrm{Diff}(\mathcal{M})$, namely those, whose restriction to the light-like portion of the boundary generate \emph{horizontal diffeomorphisms},  
\begin{equation}
\operatorname{HDiff}(\mathcal{N})=\big\{\varphi\in\operatorname{Diff}(\mathcal{N}):\pi\circ\varphi\circ\pi^{-1}\in \operatorname{Diff}(S^2) \big\}.\label{HDiff}
\end{equation}
Any such horizontal diffeomorphism $\varphi$ maps fibres onto (possibly different) fibres, hence $\varphi_\ast\ell^a \in[\ell^a]$. If an element $\varphi\in\operatorname{HDiff}(\mathcal{N})$ is smoothly connected to the identity, it is generated by a vector field $\xi^a\in T\mathcal{N}:\varphi=\exp(\xi)$ that projects into a unique vector field $\xi^a_\downarrow =\pi_\ast\xi^a\in TS^2$ on the base manifold ($\xi^a$ is a horizontal lift of $\xi^a_\downarrow$). Since the null surface $\mathcal{N}\subset P(S^2,\pi,\R)$ has itself a boundary, which consists of two successive horizontal sections $\mathcal{C}_o$ and $\mathcal{C}_1$ of $P$, i.e.\ $\partial\mathcal{N}=\mathcal{C}_o\cup\mathcal{C}_1^{-1}$, and since the exponential $\exp(\xi)\in\operatorname{HDiff}(\mathcal{N})$ maps $\mathcal{N}$ onto itself, the vector fields $\xi^a\in T\mathcal{N}$ must be tangential to the two cross sections, i.e. $\xi^a\big|_{\mathcal{C}_o}\in T\mathcal{C}_o$ and equally $\xi^a\big|_{\mathcal{C}_1}\in T\mathcal{C}_1$. 

To  lift such a vector field $\xi^a$ into a vector field on the infinite-dimensional space of kinematical histories, we first need to extend it into a bulk vector field $\hat{\xi}^a\in T\mathcal{M}$ such that $\hat{\xi}^a\big|_{\mathcal{N}}=\xi^a$. There are infinitely many ways to do so and we will see in a moment that they are all  gauge equivalent. Given such an extension of $\xi^a\in T\mathcal{N}$ into the interior of the manifold, we may now define the $SL(2,\C)$ gauge covariant Lie derivative of the bulk plus boundary fields,
\begin{subalign}
\mathcal{L}_{\hat{\xi}} \ou{A}{A}{B} & = \hat{\xi}\hook\ou{F}{A}{B},\label{Liea} \\
\mathcal{L}_{\hat{\xi}} e_{AA'} & = \hat{\xi}\hook\nabla e_{AA'} + \nabla(\xi\hook e_{AA'}),\label{Lieb}\\
\mathcal{L}_{\hat{\xi}}\eta_A\equiv\mathcal{L}_\xi \eta_A & = {\xi}\hook D \eta_A + D(\xi\hook\eta_A),\label{Liec}\\
\mathcal{L}_{\hat{\xi}}\ell^A\equiv \mathcal{L}_\xi \ell^A & = {\xi}\hook D\ell^A,\label{Lied}
\end{subalign}
where $D=\varphi^\ast_{\mathcal{N}}\nabla$ is the induced $SL(2,\C)$ gauge covariant exterior derivative at the null boundary. The Lie derivative is a vector field $\mathcal{L}_\xi[\cdot]\in T\mathcal{H}_{\mtext{kin}}$ on the space of kinematical histories. Since a diffeomorphisms maps solutions of Einstein's equations onto themselves, it is clear that the Lie derivative also defines a tangent vector to the space of physical histories, i.e.\ $\mathcal{L}_\xi[\cdot]\big|_{\mathcal{H}_{\mtext{phys}}}\in T\mathcal{H}_{\mtext{phys}}$. 

\paragraph{- Quasi-local angular momentum} To identify the Hamiltonian generator of such horizontal diffeomorphisms, consider first the pre-symplectic two-form on the partial Cauchy hypersurfcae $M$, whose boundary intersects the null surface in a horizontal section $\mathcal{C}:\partial{M}=\mathcal{C}\subset\mathcal{N}$. Given the pre-symplectic potential \eref{thetamdef}, the pre-symplectic two-form is obtained by the exterior derivative,   
\begin{equation}
\Omega_M =\bbvar{d}\Theta_M = \frac{\I}{8\pi G}\left[\int_M\bbvar{d}\Sigma_{AB}\bbwedge\bbvar{d}A^{AB}-\oint_{\partial{M}}\bbvar{d}\eta_A\bbwedge\bbvar{d}\ell^A\right]+\CC,\label{OmMdef}
\end{equation}
where $\bbwedge$ is the wedge product of differential forms in $T^\ast\mathcal{H}_{\mtext{kin}}$. Let now $\delta\in T\mathcal{H}_{\mtext{phys}}$ be a second linearly independent vector field on the space of physical histories  (i.e.\ a linearised solution of the bulk plus boundary field equations). Contracting both $\mathcal{L}_\xi[\cdot]\in T\mathcal{H}_{kin}$ and $\delta$ with the pre-symplectic two-form \eref{OmMdef} and restricting our results to the space of physical histories (i.e.\ going on-shell), we obtain 
\begin{align}
\Omega_M(\mathcal{L}_{\hat{\xi}},\delta)\Big|_{\mathcal{H}_{\mtext{phys}}} =& \frac{\I}{8\pi G}\left[\int_M\Big(\nabla(\hat{\xi}\hook \Sigma_{AB})\wedge\delta A^{AB}-\delta\Sigma_{AB}\wedge\hat{\xi}\hook F^{AB}\Big)\right.+\nonumber\\
&\qquad\left.-\oint_{\partial{M}}\big(\mathcal{L}_\xi\eta_A\delta\ell^A-\delta\eta_A\mathcal{L}_\xi\ell^A\big)\right]\Big|_{\mathcal{H}_{\mtext{phys}}}+\CC=\nonumber\\
=& \frac{\I}{8\pi G}\left[\int_M\Big((\hat{\xi}\hook \Sigma_{AB})\wedge\delta F^{AB}-\delta\Sigma_{AB}\wedge\hat{\xi}\hook F^{AB}\Big)\right.+\nonumber\\
&\qquad\left.+\oint_{\partial{M}}\big(+{\xi}\hook\Sigma_{AB}\wedge\delta A^{AB}-\mathcal{L}_\xi\eta_A\delta\ell^A+\delta\eta_A\mathcal{L}_\xi\ell^A\big)\right]\Big|_{\mathcal{H}_{\mtext{phys}}}+\CC,
\end{align}
where we used Stokes's theorem to go from the first to the second line. On shell, the bulk integral vanishes: setting $\hat{\xi}_{AA'}=\hat{\xi}\hook e_{AA'}$, and taking into account the self-dual decomposition $e_{AA'}\wedge e_{BB'} =-\bar{\epsilon}_{AA'}\Sigma_{AB}+\CC$, see \eref{Pauliident}, we find
\begin{align}\nonumber
\int_M\Big((\hat{\xi}\hook \Sigma_{AB})\wedge\delta F^{AB}-\delta\Sigma_{AB}\wedge&\hat{\xi}\hook F^{AB}\Big) =  
-\int_M\Big(\hat{\xi}_{AC'}\uo{e}{B}{C'}\wedge\delta F^{AB}+\\\nonumber
&\hspace{10em} -\delta\uo{e}{B}{C'}\wedge e_{AC'}\wedge\hat{\xi}\hook F^{AB}\Big)=\\
&=-\int_M\Big(\hat{\xi}_{AC'}\delta\big[\uo{e}{B}{C'}\wedge F^{AB}\big]-\delta\uo{e}{B}{C'}\wedge\hat{\xi}\hook(e_{AC'}\wedge F^{AB})\Big).\label{Jdef1}
\end{align}
The two terms in the last line vanish thanks to the Einstein equations \eref{curvt}. We are thus left with a boundary term and this boundary term is a total derivative on field space. Inserting the gluing conditions \eref{signull} back into \eref{Jdef1}, we obtain
\begin{align}\nonumber
\Omega_M&(\mathcal{L}_{\hat{\xi}},\delta)\Big|_{\mathcal{H}_{\mtext{phys}}} = \frac{\I}{8\pi G}\left[\oint_{\partial{M}}\big({\xi}\hook\Sigma_{AB}\wedge\delta A^{AB}-\mathcal{L}_\xi\eta_A\delta\ell^A+\delta[\eta_A]\mathcal{L}_\xi\ell^A\big)\right]\Big|_{\mathcal{H}_{\mtext{phys}}}+\CC=\\
&= \frac{\I}{8\pi G}\left[\oint_{\partial{M}}\big(\eta_A{\xi}\hook\delta[\ou{A}{A}{B}]\ell^B+\eta_A\xi\hook D[\delta\ell^A]+\delta[\eta_A]\xi\hook D\ell^A\big)\right]\Big|_{\mathcal{H}_{\mtext{phys}}}+\CC=-\delta[J_\xi[\partial{M}]],
\end{align}
We have thus shown that any horizontal diffeomorphism that is generated by a vector field $\xi^a\in T\mathcal{N}:\pi_\ast\xi^a\in S^2$ can be lifted into a vector field $\mathcal{L}_{\hat{\xi}}\big|_{\mathcal{H}_{\mtext{phys}}}\in T\mathcal{H}_{\mtext{phys}}$ on the space of physical histories, which is Hamiltonian: the Lie derivative $\mathcal{L}_{\hat{\xi}}\big|_{\mathcal{H}_{\mtext{phys}}}\in T\mathcal{H}_{\mtext{phys}}$ is the Hamiltonian vector field of the quasi-local angular momentum,
\begin{equation}
J_\xi[\mathcal{C}]=-\frac{\I}{8\pi G}\oint_{\mathcal{C}}\Big(\eta_A\mathcal{L}_\xi\ell^A-\CC\Big).\label{Jdef}
\end{equation}

\paragraph{- Comparison with Komar charge} The more familiar Komar charge for tangential diffeomrophisms is given by the integral
\begin{equation}
J^{\mtext{Komar}}_\xi[\mathcal{C}] = -\frac{1}{32\pi G}\oint_{\mathcal{C}}\tilde{\eta}^{ab}\uo{\varepsilon}{ab}{cd}\nabla_c\xi_d,\label{Komardef}
\end{equation}
where $\tilde{\eta}^{ab}$ is the two-dimensional Levi-Civita tensor density on the cross-section $\mathcal{C}\subset\mathcal{N}$ of the null boundary. The difference between the two charges \eref{Jdef} and \eref{Komardef} results from the fact that the light-like normal to the null boundary $\mathcal{N}$ has no canonical normalisation. {If, in fact, the outer boundary is time-like rather than null, the pre-symplectic potential \eref{thetanull} on the partial Cauchy hypersurface $M$ will be replaced by $\Theta_M =\frac{1}{16\pi G}\int_M\ast\Sigma_{\alpha\beta}\wedge\bbvar{d}A^{\alpha\beta}-\oint_{\partial M}\ast\Sigma_{\alpha\beta}z^\alpha\bbvar{d}z^\beta$, where $\alpha,\beta,\dots$ are internal Lorentz indices, $\ast$ denotes the Hodge dual on internal indices, $\ou{A}{\alpha}{\beta}$ is the spin connection, $\Sigma_{\alpha\beta}=e_\alpha\wedge e_\beta$ is the Pleba\'{n}ski two-form and $z^\alpha:z_\alpha z^\alpha=1$ is the internal and spacelike normal to the boundary, i.e.\ $\varphi^\ast_{\partial M}z^\alpha e_\alpha=0$. The resulting Hamiltonian  for tangential diffeomorphisms is $J_\xi = -1/(8\pi G)\oint_{\partial M}\ast\Sigma_{\alpha\beta}z^\alpha \xi^a D_az^\beta$, which is nothing but the Komar charge \eref{Komardef} for tangential diffeomorphisms written in terms of first-order spin-connection variables \cite{Freidel:2020xyx}.     } 

Although the two charges differ for generic configurations on $\mathcal{H}_{\mtext{phys}}$ (the space of physical histories), they agree on those configurations that admit Killing symmetries: if the vector field $\hat{\xi}^a\in T\mathcal{M}$ is Killing, it will Lie drag the configuration variables in the bulk up to an internal Lorentz transformation,
\begin{align}
\mathcal{L}_{\hat{\xi}}e^{AA'}&= \hat{\xi}\hook\nabla e^{AA'}=\ou{\Lambda}{A}{B}e^{BA'}+\ou{\bar{\Lambda}}{A'}{B'}e^{AB'},\label{killng1}\\
\mathcal{L}_{\hat{\xi}}\ou{A}{A}{B}&=\hat{\xi}\hook\ou{F}{A}{B}=-\nabla\ou{\Lambda}{A}{B}.\label{killng2}
\end{align}
The gauge element $\ou{\Lambda}{A}{B}:\mathcal{M}\rightarrow\mathfrak{sl}(2,\C)$ is determined by the first derivative of the Killing vector field: the Killing equation implies that $\Lambda_{ab}=\nabla_b\hat{\xi}_a$ is anti-symmetric, its self-dual component\footnote{The soldering forms $e_{AA'a}$ allow to identify spacetime indices $a,b,\dots$ with pairs of spinor indices $AA',BB',\dots$.} is $\Lambda_{AB}$, thus $\Lambda_{ab}\equiv-\bar{\epsilon}_{A'B'}{\Lambda}_{AB}+\mathrm{cc}$. 
If such a vector field is a Killing field that lies tangential to $\mathcal{C}$, it will act onto the boundary fields via the internal Lorentz transformation $\ou{\Lambda}{A}{B}$ and an overall rescaling.\footnote{This can be proven by considering the finite diffeomorphism $\varphi_\varepsilon =\exp(\varepsilon\mathcal{L}_{\hat{\xi}})$. Since $\hat{\xi}^a\in T\mathcal{M}$ is Killing, equation \eref{killng1} will be satisfied. Thus $\varphi^\ast_\varepsilon\Sigma^{AB}=\ou{[\exp(\varepsilon\Lambda]}{A}{C}\ou{[\exp(\varepsilon\Lambda]}{B}{D}\Sigma^{CD}$, where $\Sigma_{AB}$ is the self-dual Pleba\'{n}ski two-form and $\varphi^\ast_\varepsilon\Sigma^{AB}$ is the solution to the differential equation $\frac{\di}{\di\varepsilon}\varphi^\ast_\varepsilon\Sigma^{AB}=\mathcal{L}_\xi\Sigma^{AB}$ to the initial condition $\varphi^\ast_{\varepsilon=0}\Sigma^{AB}=\Sigma^{AB}$ (i.e.\ a combination of the ordinary pull-back of differential forms and the spinor parallel transport along the integral curves of $\xi^a)$. Consider then the pull-back of the Pleba\'{n}ski two-form to the null boundary, see \eref{signull}. For given $\Sigma_{AB}$ in the bulk, the boundary fields $\eta_A$ and $\ell^A$ are unique up to an overall rescaling. We thus also know that there must be a function $\lambda_\varepsilon:\mathcal{N}\rightarrow\C$ such that $\varphi^\ast_\varepsilon\eta^A = \ou{[\exp(\varepsilon\Lambda]}{A}{B}\eta^B+\E^{-\frac{\lambda_\varepsilon}{2}}\eta^A$ and $\varphi^\ast_\varepsilon\ell^A = \ou{[\exp(\varepsilon\Lambda]}{A}{B}\ell^B+\E^{+\frac{\lambda_\varepsilon}{2}}\ell^A$. Taking the derivative with respect to $\varepsilon$, we obtain \eref{killng3} and \eref{killng4}.} We will thus have
\begin{align}
\mathcal{L}_\xi\ell^A &= \xi\hook D\ell^A =\ou{\Lambda}{A}{B}\ell^B+\frac{\lambda}{2}\ell^A,\label{killng3}\\
\mathcal{L}_\xi\eta_A &= \xi\hook D\eta_A+D(\xi\hook\eta_A) =-\ou{\Lambda}{B}{A}\eta_B-\frac{\lambda}{2}\eta_A,\label{killng4}
\end{align}
where $\lambda:\mathcal{N}\rightarrow\C$ defines an overall rescaling of the boundary fields. For a given solution of Einstein's equations with Killing vector $\xi^a$, we can then always find a rescaling of the boundary fields such that $\lambda=0$, and the boundary charge \eref{Jdef} will return the Komar integral,
\begin{align}\nonumber
J_\xi[\mathcal{C}]&=-\frac{\I}{8\pi G}\oint_{\mathcal{C}}\Big(\eta_A\ou{\Lambda}{A}{B}\ell^B-\CC\Big)=
\frac{\I}{8\pi G}\oint_{\mathcal{C}}\Big(\Lambda^{AB}\Sigma_{AB}-\CC\Big)=
\frac{1}{8\pi G}\oint_{\mathcal{C}}\ast\Lambda=\\
&=\frac{1}{16\pi G}\oint_{\mathcal{C}}\tilde{\eta}^{ab}(\ast\Lambda)_{ab}=-\frac{1}{32\pi G}\oint_{\mathcal{C}}\tilde{\eta}^{ab}\uo{\varepsilon}{ab}{cd}\nabla_c\xi_d.
\end{align}

\paragraph{- Helmholtz decomposition of angular moments} The quasi-local angular momentum \eref{Jdef} is evaluated against a  vector field $\xi^a$ that lies tangential to the two-dimensional cross section $\mathcal{C}$ of the null surface $\mathcal{N}$. The cross section is equipped with a Riemannian metric $q^{\mathcal{C}}_{ab}$, which is induced from the bulk, $q^{\mathcal{C}}_{ab}=\varphi^\ast_{\mathcal{C}}g_{ab}$. The corresponding Levi-Civita tensor is $\varepsilon_{ab}^{\mathcal{C}}$, and the dual tensors are $\varepsilon^{ab}_{\mathcal{C}}$ and $q^{ab}_{\mathcal{C}}$, such that e.g.\ $\varepsilon^{ac}_{\mathcal{C}}\varepsilon_{bc}^{\mathcal{C}}=q^{ac}_{\mathcal{C}}q_{bc}^{\mathcal{C}}=[\mathrm{id}_{\mathcal{C}}]^a_b$. Given a metric, we also have the metric compatible and torsionless covariant derivative  $\mathcal{D}_a$ on $\mathcal{C}$. This derivative can be  extended naturally to spinor-valued fields, where it acts via the pull-back of the spin connection, e.g.\
\begin{equation}
\mathcal{D}_a\ell^A =\varphi^\ast_{\mathcal{C}}D_a\ell^A.
\end{equation}

Since $\mathcal{C}$ is equipped with a Riemannian structure, we can  use the Helmholtz\,--\,Hodge decomposition of $\xi^a\in T\mathcal{C}$ to split the angular momentum $J_\xi[\mathcal{C}]$ into area-preserving and rotation-free parts. For a given vector field along $\mathcal{C}$, the Hodge\,--\,Helmholtz decomposition reads
\begin{equation}
T\mathcal{C}\ni\xi^a = \varepsilon^{ab}_{\mathcal{C}}\partial_b f + q^{ab}_{\mathcal{C}}\partial_b\tilde{f},
\end{equation}
where $f$ and $\tilde{f}$ are functions on the cross section $\mathcal{C}$. The first and second terms are the area-preserving and curl-free contributions respectively. To insert this decomposition back into the quasi-local angular momentum, it is useful to introduce 
 a dual and normalised spinor $k_A^{\mathcal{C}}:k_A^{\mathcal{C}}\ell^A=1$, which is defined as follows
\begin{equation}
k^{\mathcal{C}}_A\varepsilon^{\mathcal{C}}_{ab}:=\I\varphi^\ast_{\mathcal{C}}\eta_{Aab}.
\end{equation}
Going back to the definition of the quasi-local angular momentum \eref{Jdef}, we obtain
\begin{align}
J_\xi[\mathcal{C}]=-\frac{1}{8\pi G}\oint_{\mathcal{C}}\varepsilon\big(\xi^aA^{\mathcal{C}}_a+\CC)=\frac{1}{8\pi G}\oint_{\mathcal{C}}\Big(\di f\wedge A^{\mathcal{C}}-q^{ab}_{\mathcal{C}}\partial_a\tilde{f}A_b^{\mathcal{C}}+\CC\Big),
\end{align}
where we introduced the complexified $U(1)$ connection\footnote{N.b.\ the abelian connection $A^{\mathcal{C}}$ transforms as $A^{\mathcal{C}}\rightarrow A^{\mathcal{C}}+\tfrac{1}{2}\di\lambda$ under \eref{conf1} and \eref{conf2}.}
\begin{equation}
k^{\mathcal{C}}_A\mathcal{D}\ell^A=A^{\mathcal{C}}\label{Abeldef}
\end{equation}
on the cross section $\mathcal{C}$. The curvature of this connection is related to the self-dual part of the curvature of the spin connection,
\begin{align}
F^{\mathcal{C}}=\di A^{\mathcal{C}} &  = \mathcal{D}k^{\mathcal{C}}_A\wedge\mathcal{D}\ell^A +k^{\mathcal{C}}_A\mathcal{D}^2\ell^A=\nonumber\\
&=+k^{\mathcal{C}}_A\mathcal{D}k_{\mathcal{C}}^A\wedge\ell_B\mathcal{D}\ell^B - (\varphi^\ast_{\mathcal{C}}F_{AB})k_{\mathcal{C}}^A\ell^B.\label{abelianF}
\end{align}
The first term is a functional of the shear and expansion of the two null directions that span the plane orthogonal to $T\mathcal{C}$ (i.e.\ the extrinsic curvature of $\mathcal{C}$). As we have seen in \eref{thetashear}, the one-form $\ell_AD\ell^A$, which is intrinsic to the null boundary, encodes the shear and expansion of $\ell^a$. In the same way, the components of the one-form $k_A\mathcal{D}k^A$  determine shear and expansion of the transversal null direction $\tensor[^4]{k}{^a}=\I\uo{e}{AA'}{a}k^A\bar{k}^{A'}$ such that
\begin{equation}
k^{\mathcal{C}}_A\mathcal{D}k_{\mathcal{C}}^A = +\varphi^\ast_{\mathcal{C}}\Big(\frac{1}{2}\vartheta_{(k)}\bar{m}+\bar{\sigma}_{(k)}m\Big),\label{thetashear2}
\end{equation}
where shear and expansion are defined as in \eref{sheardef} and \eref{exdef} above, i.e.\ $\sigma_{(k)}=[\tensor[^4]{m}{}]^a[\tensor[^4]{m}{}]^b\nabla_a\tensor[^4]{k}{_b}$ and $\vartheta_{(k)}=2[\tensor[^4]{m}{}]^{(a}[\tensor[^4]{\bar{m}}{}]^{b)} \nabla_a\tensor[^4]{k}{_b}$. 
If the vacuum Einstein equations are satisfied, only the spin $(2,0)$ Weyl curvature component will be excited, i.e.\
\begin{equation}
F^{AB}=\ou{\Psi}{AB}{CD}\Sigma^{CD},\quad\Psi_{ABCD}=\Psi_{(ABCD)},\label{selfdualEEQ}
\end{equation}
where $\Psi_{ABCD}$ is the spin $(2,0)$ Weyl spinor. For a given spin frame $(k_A,\ell_A)$, its $(2\times 2+ 1)=5$ algebraically independent components are
\begin{equation}
\Psi_s = \Psi_{A_1\dots A_4}\ell^{A_1}\cdots\ell^{A_s}k^{A_{s+1}}\cdots k^{A_4}.\label{Psis}
\end{equation}
If we align the spin frame to the cross section, i.e.\ if we set $(k_A,\ell_A)\equiv (k_A^{\mathcal{C}},\ell_A)$, the curvature of the complexified $U(1)$ connection on $\mathcal{C}$ will depend only on $\Psi_2$. Restricting equation \eref{signull} to $\mathcal{C}$, we obtain
\begin{equation}
(\varphi^\ast_{\mathcal{C}}F_{AB})_{ab}k_{\mathcal{C}}^A\ell^B=\Psi_{ABCD}k_{\mathcal{C}}^A\ell^B(\varphi^\ast_{\mathcal{C}}\Sigma^{CD})_{ab}=
-\I\Psi_{ABCD}k_{\mathcal{C}}^A\ell^Bk_{\mathcal{C}}^C\ell^D\varepsilon^{\mathcal{C}}_{ab}=-\I\Psi_2^{\mathcal{C}}\varepsilon^{\mathcal{C}}_{ab}.\label{abelianF2}
\end{equation}
Combining \eref{abelianF2} with the expressions for shear and expansion along the two null directions, i.e.\ \eref{thetashear} and \eref{thetashear2}, we obtain the curvature of the complexified $U(1)$ connection \eref{Abeldef}, 
\begin{equation}
F^{\mathcal{C}} = \I\Big(\Psi_2^{\mathcal{C}}+\frac{1}{4}\vartheta_{(k)}\vartheta_{(\ell)}-\bar{\sigma}_{(k)}\sigma_{(\ell)}\Big)\varepsilon_{\mathcal{C}}\label{abelianF3}
\end{equation}
The cross section $\mathcal{C}$ has no boundary. Using Stokes's theorem, we obtain
\begin{equation}
J_\xi[\mathcal{C}] = \frac{1}{4\pi G}\oint_{\mathcal{C}}d^2v\,f\operatorname{\mathfrak{Im}}\left(\Psi_2^{\mathcal{C}}-\bar{\sigma}_{(k)}\sigma_{(\ell)}\right)+\frac{1}{4\pi G}\oint_{\mathcal{C}}d^2 v\,\tilde{f}\operatorname{\mathfrak{Re}}\left(\mathcal{D}_aA_{\mathcal{C}}^a\right),\label{Jdecomp}
\end{equation}
where $\mathcal{D}_aA_{\mathcal{C}}^a= q^{ab}_{\mathcal{C}}\mathcal{D}_aA^{\mathcal{C}}_b$ is the two-dimensional vector divergence of the abelian connection \eref{Abeldef} with respect to the induced metric on $\mathcal{C}$ and $d^2v=\tfrac{1}{2}\tilde{\eta}^{ab}\varepsilon_{ab}$ is the induced volume element. The shear of the two null directions is $\sigma_{(k)}$ and $\sigma_{(\ell)}$, and $\vartheta_{(k)}$ and $\vartheta_{(\ell)}$ denote their expansion respectively. The first term of \eref{Jdecomp} is the contribution to the quasi-local angular momentum \eref{Jdef} from area-preserving diffeomorphisms, the second term corresponds to curl-free vector fields on the two-dimensional cross section. For these charges to have a finite limit at $\mathcal{I}^+$, we must impose falloff conditions $f=\mathcal{O}(r)$ and $\tilde{f}=\mathcal{O}(r)$.

\paragraph{- Boost angular momentum} The null generators $\ell^a$ of $\mathcal{N}$ and therefore also the null flag $\ell^A:\ell^a = \I\uo{e}{AA'}{a}\ell^A\bar{\ell}^{A'}$ have no preferred  normalisation. Different normalisations are connected via a complexified scaling transformation,
\begin{align}
\delta_\lambda^{\mtext{boost}}[\ell^A] &= +\frac{\lambda}{2}\ell^A,\label{conf1}\\
\delta_\lambda^{\mtext{boost}}[\eta_A] & = -\frac{\lambda}{2}\eta_A,\label{conf2}\\
\delta_\lambda^{\mtext{boost}}[\ell^a] & = \frac{\lambda+\bar{\lambda}}{2}\ell^a,\label{conf3}
\end{align}
for $\lambda:\mathcal{N}\rightarrow\C$. All other variation of the fundamental bulk and boundary variables vanish under $\delta_\lambda$, e.g.\ $\delta_\lambda[\ou{A}{A}{Ba}]=0$ and $\delta_\lambda[\Sigma_{AB}]=0$. We thus have  a vector field  $\delta_\lambda\in T\mathcal{H}_{\mtext{kin}}$, and it is easy to check that this vector field is Hamiltonian. Going back to the definition of the pre-symplectic two-form \eref{OmMdef}, and restricting ourselves to the space of physical histories, we find
\begin{align}
\Omega_M(\delta^{\mtext{boost}}_\lambda,\delta)&=+\frac{\I}{8\pi G}\oint_{\mathcal{C}}\Big(\frac{\lambda}{2}\eta_A\delta[\ell^A]+\frac{\lambda}{2}\delta[\eta_A]\ell^A\Big)+\CC = \frac{\I}{16\pi G}\oint_{\mathcal{C}}\Big(\lambda\delta[\eta_A\ell^A]-\CC\Big)=\nonumber\\
&=\frac{1}{8\pi G}\oint_{\mathcal{C}}\operatorname{\mathfrak{Re}}(\lambda)\delta[\varepsilon]=-\delta\big[Q_\lambda[\mathcal{C}]\big],
\end{align}
where we introduced the boost generator,
\begin{equation}
Q_\lambda[\mathcal{C}]=-\frac{1}{8\pi G}\oint_{\mathcal{C}}\operatorname{\mathfrak{Re}}(\lambda) \varepsilon.\label{boostgen}
\end{equation}
The zero mode $\lambda=1$, which is the generator of global dilations of the null normal, returns the total area of the cross section. On a black hole horizon, this charge provides a quasi-local Hamiltonian for locally non-rotating observers \cite{FGPfirstlaw}.

\section{Radial regularization}\label{sec5}
\subsection{Peeling for a double null foliation of spacetime}
\paragraph{- Double null foliation} In the previous sections, we considered the gravitational phase space for a fixed bounded region $\mathcal{M}$ in spacetime. The boundary $\partial\mathcal{M}$ consist of two partial Cauchy hypersurfaces $M_o$ and $M_1$  and a null surface $\mathcal{N}$, i.e.\ $\partial\mathcal{M}=M_1\cup\mathcal{N}\cup M_o^{-1}$. So far, we have left the location of the null boundary undetermined. 
Natural choices will restrict it to a portion of an isolated horizon \cite{Isohorizonreview,Ashtekar:aa,Ashtekar:2001is,Ashtekar:2004gp,Bodendorfer:2013jba,Pranzetti:2014tla} or a cosmological horizon or an asymptotic boundary. In the following, we consider only the case of an asymptotic boundary, namely future null infinity. The limit to the asymptotic boundary will be obtained by introducing an auxiliary parameter $\rho$ and sending $\rho\rightarrow\infty$. This limit can be understood both as a limit in spacetime and a limit within the infinite-dimensional quasi-local phase space. In the first case, $\rho$ is simply an advanced time coordinate on a given solution to Einstein's equations, in the latter case it is to be treated as one of the canonical variables on phase space.\footnote{The asymptotic limit $\rho\rightarrow\infty$ removes the radial coordinate from the quasi-local phase space. To introduce a symplectic structure and obtain a phase space, we will also have to impose a gauge-fixing condition on the conjugate momentum $p_\rho$ (upon choosing a polarization). The asymptotic $\rho\rightarrow\infty$ limit will remove, therefore, both $\rho$ and $p_\rho$ from the quasi-local phase space on a partial Cauchy hypersurface $\Theta_{M_\rho}$.} 

Instead of working on a fixed region as in above, we consider thus a one-parameter family of such regions $\{\mathcal{M}_\rho\}_{\rho\in\R_>}:\mathcal{M}_\rho\subset\mathcal{M}_{\rho'}$ for all $\rho<\rho'$, which are embedded into an asymptotically flat spacetime, with conformal completion $(\tilde{\mathcal{M}},\tilde{g}_{ab})$. The physical metric is $g_{ab}=\Omega^{-2}\tilde{g}_{ab}$, and we choose the conformal factor $\Omega:\tilde{\mathcal{M}}\rightarrow\R_>$ in such a way that the $\Omega=\mathrm{const}.$  hypersurfaces

\begin{equation}
\mathcal{N}_\rho=\big\{p\in\overline{\mathcal{M}}_\rho:\Omega(p)=\rho^{-1}\big\}
\end{equation}
are light-like (null). This condition is useful for us, since it allows us to match the regions $\{\mathcal{M}_\rho\}_{\rho\in\R_>}$ with the level sets of $\Omega$. In fact, for every $\rho\in\R_>$, we choose these regions in such a way that the  boundary  $\partial\mathcal{M}_\rho$ consists of two partial Cauchy surfaces $M_o^\rho$ and $M_1^\rho$ that are joined together via the null surface $\mathcal{N}_\rho$. 
Notice that every $\mathcal{N}_\rho$ has a boundary:  $\partial\mathcal{N}_\rho=\mathcal{C}^\rho_o\cup[\mathcal{C}^\rho_1]^{-1}$, which are the corners of the partial Cauchy surfaces, i.e.\ $\partial M_o^\rho=\mathcal{C}_o^\rho$ and $\partial M^\rho_1=\mathcal{C}^\rho_1$. 

Since the family of null surfaces $\{\mathcal{N}_\rho\}_{\rho\in\R_>}$ defines a foliation, the normal vector $\ell^a$ to every such null surface $\mathcal{N}_\rho$ defines a one-form $\ell_a=g_{ab}\ell^b$ that satisfies the Frobenius integrability condition 
%
%
\begin{equation}
\di \ell=-\psi_{(\ell)}\wedge\ell.\label{diEll}
\end{equation}
The one-form $\psi_{(\ell)}$ determines the non-affinity $\kappa=-[\psi_{(\ell)}]_b\ell^b$ of the null generators: $\ell^b\nabla_b\ell^a = \kappa\ell^a$. The light-like normal vector $\ell^a$ to the boundary is unique up to overall dilations sending $\ell^a$ into $\E^\lambda\ell^a$. This rescaling freedom allows us to choose the normal vectors $\ell^a$ such that they are all geodesic at $\Omega=0$, i.e.\
\begin{equation}
\kappa\big|_{\Omega=0}=0.\label{geodsc}
\end{equation}

Next, we extend the null vector $\ell^a$ into a null tetrad. We do this by introducing a transveral foliation, which is defined via a time coordinate $u:\tilde{\mathcal{M}}\rightarrow \R$ that foliates the region $\bigcup_{\rho\in\R_>}\mathcal{N}_\rho$ into transversal null hypersurfaces that intersect future null infinity in such a way that the two-dimensional (spherical) intersections are Lie dragged along the null generators. In other words,
\begin{equation}
\ell^a\nabla_a u\big|_{\Omega=0}=1.\label{timefnctn}
\end{equation}
We thus have a double null foliation, which is defined by two scalar functions $\rho$ and $u$ on $\tilde{\mathcal{M}}$. The next step ahead is to introduce and adapted NP null tetrad   and compute the falloff conditions of the various spin coefficients for such a particular gauge choice.\footnote{The falloff conditions are usually given for different gauge conditions, where only the $u=\mathrm{const}.$ surfaces are null, whereas the $\Omega=\rho^{-1}=\mathrm{const}.$ surfaces become null only asymptotically, i.e.\ for $\Omega\rightarrow 0$.} 
Let then $k_a=-\nabla_a u$ be the light-like normal to the transversal $u=\mathrm{const}.$ null surfaces. The relative normalisation between the two null vectors is $N_{(k,\ell)}:=-k_a\ell^a$. Since equation \eref{timefnctn} is satisfied at the asymptotic boundary, the inner product $N_{(k,\ell)}$ admits the $\Omega$-expansion  $N_{(k,\ell)}=1+\mathcal{O}(\Omega)$. The $\Omega=\mathrm{const}.$ surfaces are null, with null normals $\ell^a$. Therefore, $\ell^a\nabla_a\Omega=0$ and the gradient $\ell^a\nabla_aN_{(k,\ell)}$ vanishes at the asymptotic boundary, i.e.\ $\ell^a\nabla_aN_{(k,\ell)}=\mathcal{O}(\Omega)$. By rescaling  $\ell^a$ via $\ell^a\rightarrow N_{(k,\ell)}^{-1}\ell^a$, we obtain a null vector field, whose non-affinity $\kappa:\ell^b\nabla_b\ell^a =\kappa\ell^a$ still vanishes at $\Omega=0$. We can assume, therefore, without loss of generality that the two null normals $k^a$  and $\ell^a$ satisfy 
\begin{equation}
k_a=-\partial_a u,\quad k_a\ell^a=-1,\quad\ell^b\nabla_b\ell^a\big|_{\Omega=0}=0,\label{gaugecond1}
\end{equation}
where $u:\tilde{\mathcal{M}}\rightarrow\R$ is a retarded time function, which is constant along the transversal null surfaces. Since $\nabla_{[a}k_{b]}=0$, it immediately follows that $k^b\nabla_bk^a=0$. This implies that there exists an affine coordinate $r$ in $\tilde{\mathcal{M}}$ such that
\begin{equation}
k^a=\left[\frac{\di}{\di r}\right]^a.\label{rdef}
\end{equation}

\paragraph{- Relation between the radial coordinate and advanced time} There are now two natural radial coordinates, namely the affine parameter $r$, as introduced in \eref{rdef}, and the inverse conformal factor $\Omega^{-1}=:\rho$. What is the relation between the two? Since the null vector $\ell^a$ lies tangential to the $\rho=\mathrm{const}.$ light-like hypersurfaces, there exists a lapse function $N_{(\ell)}>0$ such that
\begin{equation}
\ell_a=N_{(\ell)}\nabla_a\Omega\label{lapsedef}
\end{equation}
Consider now the physical metric,
\begin{equation}
g_{ab}=-k_a\ell_b-\ell_ak_b+q_{ab}=\Omega^{-2}\tilde{g}_{ab}.\label{metric2}
\end{equation}
where $\tilde{g}_{ab}$ is the conformally rescaled metric and $q_{ab}$ is the two-dimensional Riemannian metric on the $u=\mathrm{const}.$ cross sections $\mathcal{C}_{\rho,u}\subset\mathcal{N}_\rho$. Since the metric is asymptotically flat, and both $\nabla_au$ and $\nabla_a\Omega$ do not vanish on $\mathcal{I}^+$, we can infer from \eref{lapsedef}  and \eref{metric2} the following falloff condition for the lapse function,
\begin{equation}
N_{(\ell)}=\mathcal{O}(\Omega^{-2}).
\end{equation}
Taking into account that $k^a=\partial^a_r$ and $k^a\ell_a=-1$, we infer the expansion of the gradient $k^a\nabla_a\Omega$
\begin{equation}
k^a\nabla_a\Omega=- \rho^{-2}\frac{\di\rho}{\di r}=-N^{-1}_{(\ell)}=\mathcal{O}(\rho^{-2}).
\end{equation}
If we integrate this equation along the outgoing null geodesics, we obtain
\begin{equation}
\rho = \rho^{(0)}(u,z,\bar{z})r+\mathcal{O}(r^0),\label{lapse}
\end{equation}
where the complex coordinates $(z,\bar{z})$ parametrise the two-dimensional surfaces, where the  $u=\mathrm{const}.$ surfaces, which are null, and the $r=\mathrm{const}.$  hypersurfaces intersect.\footnote{In general, the intersection of an $u=\mathrm{const}$ and an $r=\mathrm{const}$ surface will \emph{not} be a cross section of $\mathcal{N}_\rho$.} Equation \eref{lapse} implies then the falloff conditions 
\begin{equation}
\mathcal{O}(\rho^{n})=\mathcal{O}(r^{n}),\label{rhoexpans}
\end{equation}
which means that we can realise the asymptotic limit either as an $\rho\rightarrow\infty$ or an $r\rightarrow\infty$ limit.

\paragraph{- Tetrad and connection} The $u=\mathrm{const}.$ surfaces and the $\Omega=\mathrm{const}.$ surfaces each define a  foliation, with corresponding null co-normals $k_a=-\nabla_au$ and $\ell_a:k_a\ell^a=-1$. Any two such $\Omega=\mathrm{const}.$ and $u=\mathrm{const}.$ surfaces intersect each other at two-dimensional surfaces $\mathcal{C}_{\rho,u}$, which have the topology of a two-sphere. Let $(m^a,\bar{m}^a)$ be a normalised dyad in the complexified tangent space to every such cross section, i.e.\ $m_am^a=0$, $\bar{m}_am^a=1$, and $k_am^a=\ell_am^a=0$. By choosing an associate spin dyad $(k^A,\ell^A):k_A\ell^A=1$, we introduce the soldering form 
\begin{equation}
e^{AA'}= -\I\ell^{A}\bar{\ell}^{A'} k - \I k^A\bar{k}^{A'} \ell +\I k^A\bar{\ell}^{A'}m+\I\ell^A\bar{k}^{A'}\bar{m}.\label{nulltetra2}
\end{equation}
Given the metric $g_{ab}$ and the co-vector fields $k_a$ and $\ell_a$, the one-form $m_a$ is unique up to residual $U(1)$ gauge transformations, $m_a\rightarrow\E^{\I\varphi}m_a$. Using this gauge freedom, we can always require that
\begin{equation}
\bar{m}_ak^b\nabla_bm^a =-{m}_ak^b\nabla_b\bar{m}^a= 0.\label{U1gauge}
\end{equation}

The exterior derivative of the one-forms $(k_a,\ell_a,m_a)$ defines the anholonomy coefficients that determine the various spin coefficients. Consider first the exterior derivative of the one-form $m_a$, which admits the following decomposition,
\begin{align}
\di m = &-\frac{1}{2}\vartheta_{(k)}\ell\wedge m -\frac{1}{2}\big(\vartheta_{(\ell)}+2\I\phi\big)k\wedge m+\nonumber\\
&+\I\gamma\, m\wedge\bar{m}-\sigma_{(\ell)}k\wedge\bar{m}-\sigma_{(k)}\ell\wedge\bar{m}+(\alpha-\beta)k\wedge\ell.\label{anhol}
\end{align}
The various components have an immediate physical interpretation: the pair $(\sigma_{(\ell)},\sigma_{(k)})$ denotes the shear of the two null congruences $(\ell^a,k^a)$, and $(\vartheta_{(\ell)},\vartheta_{(k)})$ denotes their expansion. The spin coefficient $\gamma$ defines an abelian $U(1)$ spin connection $\gamma \bar{m}_a+\bar{\gamma}m_a$ on the two-dimensional cross sections $\mathcal{C}_{\rho,u}$, and $\phi$ is the time component of this abelian connection. That the $\ell\wedge m$ component of the exterior derivative $\di m$ has no imaginary part is a consequence of the gauge condition \eref{U1gauge}. The $k\wedge\ell$ component, on the other hand, 
measures the failure of the transversal null directions $k^a$ and $\ell^a$ to commute, i.e.\ $[\ell,k]^am_a\equiv m_a(\ell^b\nabla_bk^a-k^b\nabla_b\ell^a)=-(\alpha-\beta)$.

The remaining spin coefficients are given by the exterior derivative of $\ell_a$, which is determined by the one-form 
\begin{equation}
\psi_{(\ell)} =\kappa\, k +(\bar\alpha+\bar\beta)m+(\alpha+\beta)\bar{m},
\end{equation}
where $\kappa$ denotes the non-affinity of $\ell^a:\ell^b\nabla_b\ell^a = \kappa \ell^a$ and $(\alpha+\beta)$ determines the radial component (i.e.\ $k^a$-component) of the Lie bracket  $[k,m]^a=k^b\nabla_bm^a-m^b\nabla_bk^a$.

Given the adapted null tetrad \eref{nulltetra2}, the torsionless condition \eref{tors} determines the components of the spin connection $\ou{\Gamma}{A}{Ba}$,
\begin{equation}
\nabla_ak^A =\ou{\Gamma}{A}{Ba}k^B,\quad\nabla_a\ell^A =\ou{\Gamma}{A}{Ba}\ell^B,
\end{equation}
where $\ou{\Gamma}{A}{Ba}$ are the spin coefficients with respect to the spin basis $(k^A,\ell^A)$. A short calculation yields
\begin{align}
\nonumber\ou{\Gamma}{AB}{a} = &-\Big((\kappa+\I\phi)k_a+\I(\bar\gamma-\I\bar\alpha)m_a+\I(\gamma-\I\alpha)\bar{m}_a\Big)k^{(A}\ell^{B)}+\\
\nonumber&-\Big(\frac{1}{2}\vartheta_{(k)}\bar{m}_a+\bar{\sigma}_{(k)}m_a+\bar{\alpha}k_a\Big)\ell^A\ell^B+\\
&+\Big(\frac{1}{2}\vartheta_{(\ell)}m_a+{\sigma}_{(\ell)}\bar{m}_a+\beta \ell_a\Big)k^Ak^B.\label{connctn}
\end{align}
\paragraph{- Radial renormalisation and  evolution equations} For given boundary and $U(1)$ gauge fixing conditions \eref{gaugecond1} and  \eref{U1gauge}, we evaluate the Einstein equations for the null tetrad \eref{nulltetra2} and determine the components of the self-dual curvature two-form,
\begin{equation}
\ou{F}{A}{B}=\nabla\ou{\Gamma}{A}{B}-\ou{\Gamma}{A}{C}\wedge\ou{\Gamma}{C}{B}=\ou{\Psi}{A}{BCD}\Sigma^{CD},
\end{equation}
where $\Psi_{ABCD}\Sigma^{CD}$ is the self-dual part of the Weyl tensor. All other curvature components vanish thanks to the Einstein equations.

An advantage of the double null foliation is that the components of the Weyl tensor neatly split into three types of equations: first of all, there are the radial evolution equations (containing radial $k^a$-derivatives, but no $\ell^a$-derivatives), next there are the boundary evolution equations (containing $\ell^a$-derivatives, but no radial $k^a$-derivatives), and finally there are  constraint equations that contain only $m^a$-derivatives, which are intrinsic to the two-dimensional cross sections $\mathcal{C}_{\rho,u}$. 

Consider first the radial evolution equations, which determine the evolution away from the $\Omega=\rho^{-1}=\mathrm{const}.$ null hypersurfaces $\{\mathcal{N}_\rho\}_{\rho\in\R_>}$,
\begin{subequations}
\begin{align}
\frac{\di}{\di r}\vartheta_{(k)}+\frac{1}{2}\vartheta_{(k)}^2+2\sigma_{(k)}\bar{\sigma}_{(k)}&=0,\label{rada}\\
\frac{\di}{\di r}\vartheta_{(\ell)}+\frac{1}{2}\vartheta_{(k)}\vartheta_{(\ell)}+2\sigma_{(\ell)}\bar{\sigma}_{(k)}+2\big(\mathcal{L}_{\bar{m}}[\beta]+\I\bar{\gamma}\beta-\beta\bar{\beta}\big)&=2\Psi_2\label{radb}\\
\frac{\di}{\di r}\bar{\sigma}_{(k)}+\vartheta_{(k)}\bar{\sigma}_{(k)}&=-\Psi_0\\
\frac{\di}{\di r}\sigma_{(\ell)}+\frac{1}{2}\vartheta_{(k)}\sigma_{(\ell)}+\frac{1}{2}\vartheta_{(\ell)}\sigma_{(k)}+\mathcal{L}_m[\beta]+\I\gamma\beta-\beta^2&=0,\label{radd}\\
\I\frac{\di}{\di r}(\gamma-\I\alpha)+\frac{\I}{2}\vartheta_{(k)}(\gamma-\I\alpha)+\I\sigma_{(k)}(\bar{\gamma}-\I\bar{\alpha})-\beta\vartheta_{(k)}&=0\\
\I\frac{\di}{\di r}(\bar{\gamma}-\I\bar{\alpha})+\frac{\I}{2}\vartheta_{(k)}(\bar{\gamma}-\I\bar{\alpha})+\I\bar{\sigma}_{(k)}(\gamma-\I\alpha)-2\beta\bar{\sigma}_{(k)}&=2\Psi_1
\end{align}
\begin{align}
\frac{\di}{\di r}(\kappa+\I\varphi)+\I(\alpha-\beta)(\bar{\gamma}-\I\bar{\alpha})+\I(\bar{\alpha}-\bar{\beta})(\gamma-\I\alpha)-2\bar{\alpha}\beta&=-2\Psi_2\label{radg}\\
\frac{\di}{\di r}\bar{\alpha}+\frac{1}{2}(\bar\alpha-\bar\beta)\vartheta_{(k)}+(\alpha-\beta)\bar{\sigma}_{(k)}&=\Psi_1,\label{radf}
\end{align}
\end{subequations}
where $\mathcal{L}_{\bar{m}}[\beta]$ denotes the ordinary derivative $\mathcal{L}_{\bar{m}}[\beta]=\bar{m}^a\nabla_a\beta$. Next, we have the evolution equations that are intrinsic to the null surfaces $\{\mathcal{N}_\rho\}$, 
\begin{subalign}
\mathcal{L}_\ell[\vartheta_{(\ell)}]-\vartheta_{(\ell)}(\kappa-\frac{1}{2}\vartheta_{(\ell)})+2\sigma_{(\ell)}\bar{\sigma}_{(\ell)}&=0\label{evolva}\\
\mathcal{L}_\ell[\vartheta_{(k)}]+\vartheta_{(k)}(\kappa+\frac{1}{2}\vartheta_{(\ell)})+2\sigma_{(\ell)}\bar{\sigma}_{(k)}+2(\mathcal{L}_m[\bar{\alpha}]-(\I\gamma+\alpha)\bar{\alpha})&=2\Psi_2\label{evolvb}\\
\mathcal{L}_\ell[{\sigma}_{(\ell)}]-(\kappa-\vartheta_{(\ell)}+2\I\phi)\sigma_{(\ell)}&=-\Psi_4\label{evolvc}\\
\mathcal{L}_\ell[\bar{\sigma}_{(k)}]+(\kappa+\frac{1}{2}\vartheta_{(\ell)}+2\I\phi)\bar{\sigma}_{(k)}+
\mathcal{L}_{\bar{m}}[\bar{\alpha}]-(\I\bar{\gamma}+\bar{\alpha})\bar{\alpha}+\frac{1}{2}\bar{\sigma}_{(\ell)}\vartheta_{(k)}&=0\label{evolvd}\\
\mathcal{L}_\ell[\beta]-\I\phi\beta-\frac{1}{2}(\alpha-\beta)\vartheta_{(\ell)}-(\bar{\alpha}-\bar{\beta})\sigma_{(\ell)}&=\Psi_3\\
\I\mathcal{L}_\ell[\bar{\gamma}-\I\bar{\alpha}]+\mathcal{L}_{\bar{m}}[\kappa+\I\phi]+\frac{\I}{2}(\vartheta_{(\ell)}+2\I\phi)(\bar{\gamma}-\I\bar{\alpha})+\I\bar{\sigma}_{(\ell)}(\gamma-\I\alpha)+\vartheta_{(\ell)}\bar{\alpha}&=0\label{evolvf}\\
\I\mathcal{L}_\ell[\gamma-\I\alpha]+\mathcal{L}_m[\kappa+\I\phi]+\frac{\I}{2}(\vartheta_{(\ell)}-2\I\phi)(\gamma-\I\alpha)+\I\sigma_{(\ell)}(\bar{\gamma}-\I\bar{\alpha})+2\bar{\alpha}\sigma_{(\ell)}&=-2\Psi_3,\label{evolvg}
\end{subalign}
where e.g.\ $\mathcal{L}_\ell[\vartheta_{(\ell)}]=\ell^a\nabla_a\vartheta_{(\ell)}$ is the time derivative along the null generators of the null surface $\mathcal{N}_\rho$. Finally, there are the constraint equations
\begin{subalign}
-\frac{1}{2}(\mathcal{L}_{\bar{m}}[\vartheta_{(k)}]-\bar{\alpha}\vartheta_{(k)})+\mathcal{L}_m[\bar{\sigma}_{(k)}]-(2\I\gamma+\alpha)\bar{\sigma}_{(k)}&=-\Psi_1\label{consa}\\
-\frac{1}{2}(\mathcal{L}_m[\vartheta_{(\ell)}]+\alpha\vartheta_{(\ell)})+\mathcal{L}_{\bar{m}}[\sigma_{(\ell)}]+(2\I\bar{\gamma}+\bar{\alpha})\sigma_{(\ell)}&=-\Psi_3\\\nonumber
\I(\mathcal{L}_{\bar{m}}[\gamma]-\mathcal{L}_m[\bar{\gamma}])-2\gamma\bar{\gamma}+(\mathcal{L}_{\bar{m}}[\alpha]+\I\bar{\gamma}\alpha)-(\mathcal{L}_m[\bar{\alpha}]-\I\gamma\bar{\alpha})+\hspace{2em}&\\
+\frac{1}{2}\vartheta_{(k)}\vartheta_{(\ell)}-2\bar{\sigma}_{(k)}\sigma_{(\ell)}&=-2\Psi_2\label{consc}
\end{subalign}
For a given metric, the system of equations (\ref{rada}--\ref{consc}) is redundant.\footnote{There are 16 complex-valued equations and two real-valued equations, which are the Raychaudhuri equations \eref{rada} and \eref{evolva}. Of these 17 complex-valued equations, five of them define the components of the Weyl spinor. In addition, there are the ten components of the Einstein equations. The remaining 14 real-valued equations are redundant thanks to the Bianchi identities. } The ten components of the Einstein equations are $0=\Phi_{00'}=-\eref{rada}$, $\Phi_{11'}=\eref{consc}-\eref{radg}$, $\Phi_{22'}=-1/2\times\eref{evolva}$, $\Phi_{01'}=\eref{consa}+\eref{radf}$, $\Phi_{02'}=\eref{evolvd}$, $\Phi_{12'}=-\eref{evolvf}$ and real part of \eref{consc}+\eref{evolvb}.
\paragraph{- Radial renormalisation and peeling of the Weyl spinor} As mentioned before, we may extend the radial $r$ coordinate and the $u$ coordinate, see \eref{gaugecond1} and \eref{rdef}, \eref{lapse}, into a four-dimensional coordinate system $(r,u,z,\bar{z})$ in the vicinity of $\mathcal{I}^+$. The complex coordinates $(z,\bar{z})$  parametrise the two-dimensional surfaces, where the $r=\mathrm{const}.$ and $u=\mathrm{const}.$ surfaces intersect. We can then always find a coordinate transformation $z\rightarrow\tilde{z}=\tilde{z}(u,z,\bar{z})$ such that the co-tetrad $(k_a,\ell_a,m_a,\bar{m}_a)$ admits the following asymptotic expansion,\footnote{The coordinate transformation $z\rightarrow\tilde{z}=\tilde{z}(u,z,\bar{z})$ is merely used to guarantee that the $\ou{J}{m}{\bar{z}}$ off-diagonal entries of the matrix $J$ are $\mathcal{O}(1)$ rather than $\mathcal{O}(r)$. In addition, we  assume a polynomial expansion, i.e. $a=\mathcal{O}(r^n)$ means $a= a_n r^{n}+a_{n+1}r^{n-1}+\dots$ is convergent in a neighbourhood of $\mathcal{I}^+$. }
\begin{equation}
\begin{pmatrix}k\\\ell\\m\\\bar{m}\end{pmatrix}=\begin{pmatrix}
-1&0&0&0\\
\mathcal{O}(r)&-1&\mathcal{O}(r)&\mathcal{O}(r)\\
\mathcal{O}(r)&0&\mathcal{O}(r)&\mathcal{O}(1)\\
\mathcal{O}(r)&0&\mathcal{O}(1)&\mathcal{O}(r)
\end{pmatrix}\begin{pmatrix}\di u\\\di r\\\di z\\\di\bar{z}\end{pmatrix}\equiv J\cdot\begin{pmatrix}\di u\\\di r\\\di z\\\di\bar{z}\end{pmatrix}.\label{Jacobi}
\end{equation}
The inverse transformation $J^{-1}$ maps the basis vectors $(\partial^a_u,\partial^a_r,\partial^a_{z},\partial^a_{\bar{z}})$ back into the null tetrad $(-\ell^a,-k^a,$ $\bar{m}^a,m^a)=(\partial^a_u,\partial^a_r,\partial^a_{z},\partial^a_{\bar{z}})\cdot J^{-1}$. We introduce the decomposition,
\begin{align}
\ell^a&=\partial^a_u+N\partial^a_r+\bar{\zeta}m^a+{\zeta}\bar{m}^a,\label{ellvec}\\
m^a&= w \partial^a_r+\mu\partial^a_{\bar{z}}+\nu\partial^a_z\label{emvec}.
\end{align}
The falloff conditions for $(N,\zeta,w,\mu,\nu)$ can be inferred algebraically from $\ou{[J^{-1}]}{m}{n}=\frac{\partial}{\partial \ou{J}{n}{m}}\ln\det(J)$ and \eref{Jacobi}. We obtain,
\begin{align}
N&=\mathcal{O}(r),\zeta=\mathcal{O}(r^{-1}),w=\mathcal{O}(1),\mu=\mathcal{O}(r^{-1)},\nu=\mathcal{O}(r^{-2})\label{falloff0}.
\end{align}
In the same way, the falloff conditions for the components of the matrix-valued one-form $\di JJ^{-1}$ determine the falloff conditions of the spin coefficients,
\begin{align}
&\vartheta_{(\ell)}=\mathcal{O}(1),\phi=\mathcal{O}(1),\kappa=\mathcal{O}(1),\alpha-\beta =\mathcal{O}(1),\label{falloff1}\\
&\vartheta_{(k)}=\mathcal{O}(r^{-1}),\sigma_{(\ell)}=\mathcal{O}(r^{-1}),\sigma_{(k)}=\mathcal{O}(r^{-2}),\gamma=\mathcal{O}(r^{-1}),\alpha+\beta=\mathcal{O}(r^{-1})\label{falloff2}.
\end{align}
Notice that the falloff conditions \eref{falloff1} and \eref{falloff2} are a consequence of \eref{Jacobi} alone. In particular, we have not yet employed the equations of motion \eref{rada}--\eref{consc} nor the gauge fixing conditions \eref{geodsc} and \eref{U1gauge}. So far, we only have a rough estimate and some of the spin coefficients will fall off faster than \eref{falloff1} and \eref{falloff2} would suggest. For example, we know from the boundary conditions \eref{gaugecond1} that the non-affinity $\kappa$ will admit the expansion
\begin{equation}
\kappa=\mathcal{O}(r^{-1}).
\end{equation}
In addition, we can always find a gauge parameter $\varphi=\mathcal{O}(1)$ such that the $U(1)$ gauge transformation $m_a\rightarrow\E^{\I\varphi}m_a$ maps the spin coefficient  $\phi$ into $\phi+\ell^a\nabla_a\varphi$ such that $\phi+\ell^a\nabla_a\varphi=\mathcal{O}(r^{-1})$. Notice that we may always choose $\varphi$ such that  the gauge fixing condition \eref{U1gauge} is still satisfied. Without loss of generality we can thus always assume that
\begin{equation}
\kappa+\I\phi=\mathcal{O}(r^{-1}).\label{falloff3}
\end{equation}
Inserting the falloff condition \eref{falloff0}, \eref{falloff1}, \eref{falloff2} and \eref{falloff3}, back into the constraint equation \eref{consc}, we can see then also that $\Psi_2=\mathcal{O}(r^{-1})$ or faster. Going back to the radial evolution equation for the tangential expansion $\vartheta_{(\ell)}$, i.e.\ going back to equation  \eref{radb}, and again using the falloff conditions, i.e.\ \eref{falloff0}, \eref{falloff1} \eref{falloff2}, we infer $\beta=\mathcal{O}(r^{-1})$ rather than $\beta=\mathcal{O}(1)$. Since, however, $\alpha+\beta=\mathcal{O}(r^{-1})$, we thus also know $\alpha=\mathcal{O}(r^{-1})$. Taking the sum of equations \eref{radb} and \eref{consc}, and solving the resulting equation to leading order in $r$, we find that $\vartheta_{(\ell)}=\mathcal{O}(r^{-1})$. Therefore, 
\begin{equation}
\alpha=\mathcal{O}(r^{-1}),\beta=\mathcal{O}(r^{-1}),\vartheta_{(\ell)}=\mathcal{O}(r^{-1}).
\end{equation}
Next, we solve the radial evolution equations \eref{rada} to leading order in $\mathcal{O}(r^{-n})$. Going back to \eref{falloff2}, we have $\vartheta_{(k)}=\mathcal{O}(r^{-1})$ and $\sigma_{(k)}=\mathcal{O}(r^{-2})$, which implies that $\vartheta_{(k)}$ admits the $1/r$-expansion
\begin{equation}
\vartheta_{(k)}=\frac{2}{r}+\mathcal{O}(r^{-2}).\label{falloff4}
\end{equation}

We have now all parts together to determine the $\mathcal{O}(r^{-n})$ expansion of the components of the Weyl spinor, which can be derived from the Bianchi identities \cite{penroserindler}. If the Einstein equations are satisfied, the first Bianchi identity reads
\begin{equation}
\uo{e}{AA'}{a}\nabla_a\Psi^{AB_1B_2B_3}=0.
\end{equation}
If we contract this equation with $\bar{k}^{A'}$ and various powers of $k^B$ and $\ell^B$, we obtain the radial evolution equations for the components of the Weyl spinor. A short calculation gives,
\begin{align}\nonumber
\frac{\di}{\di r}[\Psi_s]=&-\frac{1}{2}(5-s)\vartheta_{(k)}\Psi_s+\\
&+\mathcal{L}_{m}\big[\Psi_{s-1}]+\big[\I(s-3)({\gamma}-\I\alpha)-s{\beta}\big]\Psi_{s-1}-(s-1){\sigma}_{(\ell)}\Psi_{s-2},\label{radPsi}
\end{align}
where $\Psi_s\equiv 0$ for $s<0$. To solve these equations to leading order in $r$, we will consider $\vartheta^{(0)}_{(k)}=2/r$  as the \emph{free radial Hamiltonian}, while all other terms represent the interaction term . Working in the interaction picture, we introduce the rescaled components of the Weyl spinor
\begin{equation}
\tilde{\Psi}_s = r^{5-s}\Psi_s.\label{intpic}
\end{equation}
Using the falloff conditions for the metric and spin coefficients, i.e.\ \eref{falloff1}, \eref{falloff2}, \eref{falloff3} and \eref{falloff4}, we obtain the falloff conditions of the radial evolution equations,
\begin{equation}
\frac{\di}{\di r}
\begin{pmatrix}\tilde{\Psi}_4\\\tilde{\Psi}_3\\\tilde{\Psi}_2\\\tilde{\Psi}_1\\\tilde{\Psi}_0\end{pmatrix}=\begin{pmatrix}
\mathcal{O}(r^{-2})&\mathcal{O}(r^{-2})&\mathcal{O}(r^{-3})&0&0\\
0&\mathcal{O}(r^{-2})&\mathcal{O}(r^{-2})&\mathcal{O}(r^{-3})&0\\
0&0&\mathcal{O}(r^{-2})&\mathcal{O}(r^{-2})&\mathcal{O}(r^{-3})\\
0&0&0&\mathcal{O}(r^{-2})&\mathcal{O}(r^{-2})\\
0&0&0&0&\mathcal{O}(r^{-2})
\end{pmatrix}
\begin{pmatrix}\tilde{\Psi}_4\\\tilde{\Psi}_3\\\tilde{\Psi}_2\\\tilde{\Psi}_1\\\tilde{\Psi}_0\end{pmatrix}=:-\I \bbvar{H}\begin{pmatrix}\tilde{\Psi}_4\\\tilde{\Psi}_3\\\tilde{\Psi}_2\\\tilde{\Psi}_1\\\tilde{\Psi}_0\end{pmatrix},\label{nullevolv}
\end{equation}
where we introduced a radial interaction Hamiltonian $\bbvar{H}$. Next, we formally integrate these equations along the outgoing null rays $\gamma_{(u,z,\bar{z})}(r)$ that generate a given $u=\mathrm{const}.$ null hypersurface, with $(z,\bar{z})$ denoting the angular coordinates on the $r=\mathrm{const.}$, $u=\mathrm{const}.$ cross sections of the double null foliation. Using the radially ordered exponential, i.e.\ the path ordered exponential along the outgoing null generators, we obtain
\begin{equation}
\tilde{\Psi}(u,r_1,z,\bar{z})=\mathrm{Rexp}\Big(-\I\int_{r_o}^{r_1}\di r\,\gamma_{(u,z,\bar{z})}^\ast\bbvar{H}\Big)\Psi(u,r_o,z,\bar{z})\equiv \bbvar{U}(r_o\rightarrow r_1|u,z,\bar{z})\Psi(u,r_o,z,\bar{z}).
\end{equation}
The falloff conditions of the components of the transfer matrix $\bbvar{U}(r_o\rightarrow r_1|u,z,\bar{z})$ can be inferred directly from \eref{nullevolv},
\begin{equation}
\bbvar{U}(r_o\rightarrow r_1|u,z,\bar{z})= \bbvar{1} + \mathcal{O}(r^{-1}).
\end{equation}
Going back to the physical components of the Weyl spinor \eref{intpic}, we obtain the familiar falloff conditions for the components of the Weyl spinor,%
\begin{subalign}
\Psi_4&=\mathcal{O}(r^{-1}),\label{psi4}\\
\Psi_3&=\mathcal{O}(r^{-2}),\label{psi3}\\
\Psi_2&=\mathcal{O}(r^{-3}),\label{psi2}\\
\Psi_1&=\mathcal{O}(r^{-4}),\label{psi1}\\
\Psi_0&=\mathcal{O}(r^{-5}),\label{psi0}
\end{subalign}
where the components are now computed with respect to the  null tetrad $(k^a,\ell^a,m^a,\bar{m^a})$, which is adapted to the double null foliation (i.e.\ both $k^a$ and $\ell^a$ are surface orthogonal).

\paragraph{- Falloff conditions for the metric coefficients $N$ and $w$} To calculate physical observables, we also need to understand the subleading terms in the $1/r$ expansion of the metric coefficients, in particular $N$ and $w$, as defined in \eref{ellvec}. To infer the subleading terms of  the $1/r$ expansion, consider the radial and tangential evolution equations
\begin{subalign}
\frac{\di}{\di r}&N-(\alpha+\beta)\bar{\zeta}-(\bar{\alpha}+\bar{\beta})\zeta =\kappa,\label{Nevolv}\\
\mathcal{L}_\ell[w]&-\mathcal{L}_m[N+\bar{\zeta}w+\zeta\bar{w}]=-\frac{1}{2}\big(\vartheta_{(\ell)}-2\I\phi\big)w-\sigma_{(\ell)}\bar{w},\label{wevolv}
\end{subalign}
which are a consequence of the Lie brackets $[k,\ell]^a=k^b\nabla_b\ell^a-\ell^b\nabla_bk^a=\kappa k^a+(\alpha-\beta)\bar{m}^a+(\bar{\alpha}-\bar{\beta})m^a$ and $[\ell,m]^a=\ell^b\nabla_bm^a-m^b\nabla_b\ell^a=-\frac{1}{2}(\vartheta_{(\ell)}-2\I\phi)m^a-\sigma_{(\ell)}\bar{m}^a$. We have built the double-null foliation in such a way that $\kappa=\mathcal{O}(r^{-1})$ and we also saw that for an asymptotically flat spacetime the falloff conditions $N=\mathcal{O}(r)$, $w=\mathcal{O}(1)$, $\alpha=\mathcal{O}(r^{-1})=\beta$ and $\zeta=\mathcal{O}(r^{-1})$ will be satisfied, see (\ref{falloff0}--\ref{falloff3}). These falloff conditions are compatible with equation \eref{Nevolv}, only if $N=\mathcal{O}(1)$ rather than $N=\mathcal{O}(r)$, see \eref{falloff0}. 

Before further expanding on $N$, let us now consider the $\mathcal{O}(r^{-1})$ expansion of $w=m^a\nabla_ar$. Going back to \eref{ellvec}, which provides the vector field $\ell^a$ in terms of the coordinate basis $(\partial^a_u,\partial^a_r,\partial^a_z,\partial^a_{\bar{z}})$, and taking into account the falloff conditions \eref{falloff0}, we obtain the evolution equations
\begin{align}
\frac{\di}{\di u}w^{(0)}(u,z,\bar{z})&=0,\label{wevolv2}\\
\frac{\di}{\di u}w^{(1)}(u,z,\bar{z})&=r\mathcal{L}_m[N^{(0)}]+\mathcal{O}(r^{-1}),
\end{align}
where $w=w^{(0)}(u,z,\bar{z})+w^{(1)}(u,z,\bar{z})r^{-1}+\mathcal{O}(r^{-2})$ and $N=N^{(0)}(u,z,\bar{z})+\mathcal{O}(r^{-1})$. We may now always choose initial conditions on an $u=u_o=\mathrm{const}.$ initial null hypersurface such that $w^{(0)}=0$. The easiest way to impose such initial conditions is to choose a specific foliation of $\rho=\mathrm{const}.$ surfaces, where the three-dimensional $\rho=\mathrm{const}.$ null surfaces $\mathcal{N}_\rho$ intersect a fixed $u=u_o=\mathrm{const}.$ initial null hypersurface, from where the construction of $\mathcal{N}_\rho$ starts, at constant values of $r$. In other words,
\begin{equation}
\rho(u_o,r,z,\bar{z}) = \Omega^{-1}(u_o,r,z,\bar{z}) = r.\label{intersect}
\end{equation}
This equation \eref{intersect} implies $k^a\nabla_a\rho|_{u_o}=1$ such that the pull-back of $\ell_a\propto\nabla_a\rho$ to the $u=u_o$ hypersurface is simply $-\di r$. Since $m^a$ lies tangential to the $u=u_o$ surface ($m^ak_a=-m^a\nabla_au=0$), we also have
\begin{equation}
0=m^a\ell_a\big|_{u=u_o}= - m^a\nabla_ar\big|_{u=u_o} = - w\big|_{u=u_o},\label{wintl}
\end{equation}
which implies the desired initial condition $w|_{u_o}=0$. If we insert these initial conditions back into \eref{wevolv2}, we obtain a double null foliation, i.e.\ a local foliation into $\rho=\mathrm{const}.$ and $u=\mathrm{const}.$ null hypersurfaces, such that
\begin{align}
w&=m^a\nabla_a r =\int_{u_o}^u\di u' \mathcal{L}_m[N^{(0)}](r,u',z,\bar{z})+\mathcal{O}(r^{-2}).\label{falloff5}
\end{align}

Let us now return to the expansion of $N$. The leading order coefficient $N^{(0)}(u,z,\bar{z})$ of the $1/r$ expansion of the metric component $N=N^{(0)}(u,z,\bar{z})+\mathcal{O}(r^{-1})$ can be set to zero via an affine transformation of the outgoing null generators that sends the radial $r$ coordinate into $r-\int_{u_o}^u\di u'N^{(0)}(u',z,\bar{z})$. We may therefore assume, without loss of generality, that 
\begin{subalign}
N&=\frac{N^{(1)}(u,z,\bar{z})}{r}+\mathcal{O}(r^{-2}),\label{falloff6a}\\
w&=m^a\nabla_a r =\mathcal{O}(r^{-2}).\label{falloff5a}
\end{subalign}
\paragraph{- Falloff conditions for  the inverse conformal factor} Recall that the inverse conformal factor defines the radial coordinate $\rho=\Omega^{-1}$. The $\rho=\mathrm{const}.$ surfaces are null, with null normal $\ell^a:\ell^a\nabla_a\rho=0$. From \eref{lapse}, we know that $\rho$ admits the expansion $\rho = \rho^{(0)}(u,z,\bar{z})r+\rho^{(1)}(u,z,\bar{z})+\mathcal{O}(r^{-1})$ and $\ell^a=\partial^a_u+N\partial^a_r+\bar{\zeta}m^a+\zeta\bar{m}^a$. Given the various falloff conditions and the initial condition \eref{intersect}, we obtain
\begin{equation}
\rho(u,r,z,\bar{z})=r+\mathcal{O}(r^{-1}).\label{rhoexpand}
\end{equation}

\paragraph{- Falloff conditions for  $\alpha$ and $\beta$} The Lie bracket $[k,m]^a=k^b\nabla_bm^a-m^b\nabla_bk^a=-(\alpha+\beta)k^a-\frac{1}{2}\vartheta_{(k)}m^a-\sigma_{(k)}\bar{m}^a$ implies the radial evolution equations
\begin{equation}
\frac{\di}{\di r} w+\frac{1}{2}\vartheta_{(k)}w+\sigma_{(k)}\bar{w}=-(\alpha+\beta).
\end{equation}
Since $w=\mathcal{O}(r^{-1})$ and $\vartheta_{(k)}=2/r+\mathcal{O}(r^{-2})$, see \eref{falloff4}, and $\sigma_{(k)}=\mathcal{O}(r^{-2})$ as inferred from \eref{falloff2}, we find
\begin{equation}
\alpha+\beta=\mathcal{O}(r^{-3}).\label{falloff6b}
\end{equation}
Going back to the radial evolution equations \eref{radf} for $\bar{\alpha}$ and taking into account the falloff conditions \eref{falloff4}, \eref{psi1}, \eref{falloff6b} and \eref{falloff2}, we obtain
\begin{equation}
\alpha=\frac{\alpha^{(0)}(u,z,\bar{z})}{r^2}+\mathcal{O}(r^{-3}),\quad \beta=-\frac{\alpha^{(0)}(u,z,\bar{z})}{r^2}+\mathcal{O}(r^{-3}).\label{falloff6c}
\end{equation}
 
\paragraph{- Falloff conditions for non-affinity $\kappa$} The $1/r$ expansion of the non-affinity $\kappa$ can be inferred from the radial evolution equation \eref{radg}. Solving this equation to leading order in $1/r$ and taking the falloff conditions \eref{falloff2}, \eref{falloff3} \eref{psi2}, and \eref{falloff6c} into account, we find
\begin{equation}
\kappa=+\frac{\operatorname{\mathfrak{Re}}(\Psi^{(0)}_2)}{r^2}+\mathcal{O}(r^{-3}),
\end{equation}
and $\varphi=\mathcal{O}(r^{-2})$. Given the leading order of the $1/r$ expansion of $\kappa$, we solve the radial evolution equation for the metric coefficient $N=N^{(1)}r^{-1}+\mathcal{O}(r^{-2})$ such that
\begin{equation}
N=-\frac{\operatorname{\mathfrak{Re}}(\Psi^{(0)}_2)}{r}+\mathcal{O}(r^{-2})\label{Nlead},
\end{equation}
where
\begin{equation}
\Psi_2 = \frac{\Psi_2^{(0)}(u,z,\bar{z})}{r}+\mathcal{O}(r^{-2}).
\end{equation}

\paragraph{- Falloff conditions for $\sigma_{(k)}$ and $\sigma_{(\ell)}$} Next, we consider the $1/r$ expansion of the transversal and tangential shear $\sigma_{(k)}$ and $\sigma_{(\ell)}$. Given the falloff condition \eref{falloff2}, the transversal goes like $1/r^{2}$ such that we may write
\begin{equation}
\sigma_{(k)} = \frac{\sigma(u,z,\bar{z})}{r^2} + \mathcal{O}(r^{-3}).\label{falloff7}
\end{equation}
The evolution of the transversal shear $\sigma_{(k)}$ along the null generators $\ell^a$ of the null surfaces $\mathcal{N}_\rho$ is determined by the evolution equation \eref{evolvd}. If the falloff conditions are satisfied, we can solve this equation perturbatively in $r^{-1}$. We obtain
\begin{equation}
\sigma_{(\ell)} = - \frac{\dot{\sigma}(u,z,\bar{z})}{r}+\mathcal{O}(r^{-2}),\label{falloff8a}
\end{equation}
where $\dot{\sigma}(u,z,\bar{z}):=\frac{\di}{\di u}\sigma(u,z,\bar{z})$.

\paragraph{- Falloff conditions for $\vartheta_{(k)}$ and $\vartheta_{(\ell)}$}The $1/r$ expansion of $\vartheta_{(k)}$ can be inferred directly from the radial Raychaudhuri equation \eref{rada}. For given asymptotic shear \eref{falloff7}, the first three terms in the $1/r$ expansion are given by
\begin{equation}
\vartheta_{(k)}=\frac{2}{r}+\frac{2U}{r^2}+\frac{2(\sigma\bar{\sigma}+U^2)}{r^3}+\mathcal{O}(r^{-4}),\label{falloff9}
\end{equation}
where $U\equiv U(u,z,\bar{z})$ characterises the next to leading term of $\vartheta_{(k)}=2/r+\mathcal{O}(r^{-2})$. Consider then the first two terms of the tangential expansion
\begin{equation}
\vartheta_{(\ell)} = \frac{\vartheta_{(\ell)}^{(1)}(u,z,\bar{z})}{r}+\frac{\vartheta_{(\ell)}^{(2)}(u,z,\bar{z})}{r^2}+\mathcal{O}(r^{-3}).\label{falloff10}
\end{equation}
The transversal expansion satisfies the evolution equation \eref{evolvb} along the null generators of $\mathcal{N}_\rho$. We  solve this evolution equation order by order in $1/r$, from which we obtain the coefficients of the $1/r$ expansion of $\vartheta_{(\ell)}$. To leading order, we obtain
\begin{equation}
2\dot{U}(u,z,\bar{z}) + \vartheta_{(\ell)}^{(1)}(u,z,\bar{z}) = 0,\label{order1}
\end{equation}
where $\dot{U}=\frac{\di}{\di u} U$. We will see below that $\dot{U}$ has a geometric interpretation: it is simply the Ricci scalar of the $u=\mathrm{const}.$ cross sections of the ${\rho}=\mathrm{const}.$ null hypersurfaces. 

To compute the next to leading order of $\vartheta_{(\ell)}$ from the evolution equation \eref{evolvb}, let us first note that the radial $r$ has a derivative along $\ell^a$. We have, in fact,
\begin{equation}
\mathcal{L}_\ell[r^{-1}]=-r^{-2}\ell^a\nabla_ar = -r^{-2}(N+\bar{\zeta}w+\zeta\bar{w})=\frac{\operatorname{\mathfrak{Re}}(\Psi_2^{(0)})}{r^3}+\mathcal{O}(r^{-4}),
\end{equation}
which is a consequence of \eref{ellvec} and \eref{Nlead}. Taking into account the falloff conditions, and the leading order equations \eref{order1}, we obtain the next to leading order  for $\mathcal{L}_\ell[\vartheta_{(k)}$], namely,
\begin{align}
\mathcal{L}_\ell[\vartheta_{(k)}]=\frac{2\dot{U}}{r^2}+\frac{2}{r^3}\Big(\operatorname{\mathfrak{Re}}(\Psi_2^{(0)})+\frac{\di}{\di u}|\sigma|^2+2U\dot{U}\Big)+\mathcal{O}(r^{-4})\label{order2}\end{align}
If we insert equation \eref{order2} back into the tangential evolution equation \eref{evolvb}, we obtain
\begin{align}\nonumber
\vartheta_{(\ell)}^{(2)} &  = -4\operatorname{\mathfrak{Re}}(\Psi_2^{(0)})+2\Psi_2^{(0)}-2\sigma\dot{\bar{\sigma}}-2U\dot{U}-2r^3\big(\mathcal{L}_m[\bar{\alpha}]-\I\gamma\bar{\alpha}\big)+\mathcal{O}(r^{-1}) = \\
& = -2\bar{\Psi}_2^{(0)}-2\sigma\dot{\bar{\sigma}}-2U\dot{U}-2\bar{\mathcal{D}}\bar{\alpha}^{(0)}+\mathcal{O}(r^{-1}),\label{falloff11a}
\end{align}
where $\mathcal{D}$ is the $U(1)$ covariant derivative on the two-dimensional cross sections $\mathcal{C}_{\rho,u}$. If we perform a $U(1)$ transformation $m_a\rightarrow\E^{\I\varphi}m_a$ for a gauge parameter $\varphi$, the corresponding $U(1)$ component of the spin connection transforms as $\gamma\rightarrow\E^{\I\varphi}\gamma-\mathcal{L}_m[\lambda]$. If, in addition, there is a spin coefficient $X$, with spin weight $s$, that transforms as $X\rightarrow \E^{\I s\lambda}X$ under such a $U(1)$ transformation and admits the $1/r$ expansion $X = X^{(0)}+\mathcal{O}(r^{-1})$, the leading order of the $U(1)$ covariant derivative will be defined by $\bar{\mathcal{D}}X^{(0)}:=\lim_{r\rightarrow\infty}r(\mathcal{L}_m X+\I s\gamma X )$ and  ${\mathcal{D}}X^{(0)}:=\lim_{r\rightarrow\infty}r(\mathcal{L}_{\bar{m}} X+\I s\bar{\gamma} X )$. 

We thus see from \eref{falloff11a} that the next to leading order $\vartheta_{(\ell)}^{(2)}$ of the tangential expansion $\vartheta_{(\ell)}$ depends on the asymptotic shear $\sigma(u,z,\bar{z})$, on the next to leading order of $\vartheta_{(k)}=2/r(1+U/r+\mathcal{O}(r^{-2}))$ and on the leading order $\alpha^{(0)}(u,z,\bar{z})$ of the spin coefficient $\bar{\alpha}=k_A\ell^a\nabla_ak^A$, see \eref{falloff7}. The spin coefficient $\alpha$ can be eliminated from this equation. The dependence can be inferred from the constraint equation for $\Psi_1$, i.e.\ \eref{consa}. Taking into account the various falloff conditions, in particular \eref{psi1}, \eref{falloff9} and \eref{falloff7}, we obtain
\begin{equation}
\bar{\alpha}^{(0)}=\mathcal{D}U-\bar{\mathcal{D}}\bar{\sigma}.\label{falloff11b}
\end{equation}
This in turn allows us tow write the next to leading order of the tangential expansion in terms of $\Psi_2^{(0)}$, and in terms of the asymptotic shear and the next to leading term of the outgoing expansion,
\begin{equation}
\vartheta_{(\ell)}^{(2)}(u,z,\bar{z})=-2\Big(\bar{\Psi}_2^{(0)}+\sigma\dot{\bar{\sigma}}-\bar{\mathcal{D}}\bar{\mathcal{D}}\bar{\sigma}+U\dot{U}+\bar{\mathcal{D}}\mathcal{D}U\Big).\label{falloff12}
\end{equation}

The tangential Raychaudhuri equation \eref{evolva} determines the time evolution of the various coefficients of the $1/r$ expansion \eref{falloff10}. Inserting \eref{order1} and \eref{order2} back into \eref{falloff10} and \eref{evolva}, we obtain the evolution equations
\begin{subalign}
\frac{\di^2}{\di u^2}&U(u,z,\bar{z})=0,\label{Uevolv}\\
\frac{\di}{\di u}&\vartheta_{(\ell)}^{(2)}(u,z,\bar{z})+2\dot{U}^2(z,\bar{z})=-2|\dot{\sigma}(u,z,\bar{z})|,\label{Uevolv2}
\end{subalign}

Finally, let us explain how $\dot{U}$, which is constant in $u$, is related to the Ricci curvature of the two-dimensional $u=\mathrm{const}.$ cross sections of $\mathcal{N}_\rho$. In equation \eref{falloff11a}, we introduced the two-dimensional $U(1)$ covariant derivative $\mathcal{D}$. The $1/r$ expansion of the curvature of the abelian connection $\Gamma=\gamma\bar{m}+\bar{\gamma}m$ can be inferred  from the constraint equation \eref{consc} and the various falloff conditions. A short calculation gives $[{\mathcal{D}},\bar{\mathcal{D}}]X^{(0)}=2s \dot{U}X^{(0)}$, such that
\begin{equation}
\mathcal{R}[q^o_{ab}]= {4\dot{U}}+\mathcal{O}(r^{-1})\label{twoRicci}
\end{equation}
is the Ricci scalar of the conformally rescaled metric $q_{ab}^o:q_{ab}=r^2q^o_{ab}+\mathcal{O}(r)$, where  $q_{ab}=\varphi^\ast_{\mathcal{C_{\rho,u}}}g_{ab}$ is the pull-back of the physical metric $g_{ab}$ to the two-dimensional $\rho=\mathrm{const}.$ and $u=\mathrm{const}.$ spherical cross sections.
\section{Bondi energy and radiative phase space}\label{sec6}
\subsection{Radiative phase space from radial renormalisation}
\noindent 
Our first task in this section is to explain how to recover the radiative phase space on $\mathcal{I}^+$ via an asymptotic $\rho\rightarrow\infty$ limit of the quasi-local radiative phase space that we introduced in \hyperref[sec3.4]{section 3.4}. For each null hypersurface $\mathcal{N}_\rho$ of the foliation $\{\mathcal{N}_\rho\}_{\rho>0}$, we introduce the quasi-local symplectic potential \eref{thetanull}. Using the definition of the one-form $\ell_AD\ell^A$, see \eref{thetashear}, we find  
\begin{align}\nonumber
\Theta_{\mathcal{N}_\rho} =  - \frac{1}{8\pi G}\int_{\mathcal{N}_\rho}\big(\varepsilon\wedge\bbvar{d}\varkappa-&k_a\bbvar{d}\ell^a\di\varepsilon+\frac{1}{2}\vartheta_{(\ell)}k\wedge\bbvar{d}\varepsilon\big)+\\
&+\frac{\I}{8\pi G}\int_{\mathcal{N}_\rho}\big(\sigma_{(\ell)}k\wedge\bar{m}\wedge\bbvar{d}\bar{m}-\CC\big)\equiv\int_{\mathcal{N}_\rho}\theta_{\mathcal{N}_\rho}.\label{nullsympl}
\end{align}

To evaluate $\Theta_{\mathcal{N}_\rho}$ for our falloff and gauge fixing conditions in the asymptotic $\rho\rightarrow\infty$ limit to future null infinity, we need to know the falloff conditions for a linearised solution $\delta[\cdot]$ of the bulk plus boundary field equations (as summarised in \hyperref[tab1]{table 1}). The $\rho\rightarrow\infty$ limit removes the $\rho$-coordinate, i.e.\ the inverse conformal factor, from the quasi-local phase space, and we may treat, therefore, the foliation as a fiducial background structure, such that the surfaces $\mathcal{N}_\rho$ are locked into the abstract manifold $\tilde{\mathcal{M}}$. In other words, $\delta[\rho]=0$. Going back to the $1/r$-expansion of the radial $\rho$ coordinate as a function of $(u,r,z,\bar{z})$, see \eref{rhoexpand}, and solving the equation $\delta[\rho]=0$ order by order in $r$, we obtain
\begin{equation}
\delta[r]=\mathcal{O}(r^{-1}).\label{deltafall2}
\end{equation}
From $\ell_a=N_{(\ell)}\nabla_a\Omega$ and $N_{(\ell)}=\mathcal{O}(\Omega^{-2})$, and $\delta\Omega=\delta\rho^{-1}=0$, we infer the  falloff conditions
\begin{equation}
\delta\ell_a = \lambda\ell_a,\qquad\lambda=\mathcal{O}(r^0).\label{deltafall1}
\end{equation}
In a neighbourhood of null infinity, the  $(u, z, \bar{z}$) coordinates complete the radial $r$ coordinate into a four-dimensional coordinate system $(u,r,z,\bar{z})$. To guarantee that these coordinates are regular for $r\rightarrow\infty$, we impose the boundary conditions
\begin{align}
\delta[u]&=\mathcal{O}(r^0),\label{deltafall3}\\
\delta[z]&=\mathcal{O}(r^0).\label{deltafall4}
\end{align}
Since $k_a=-\nabla_au$, and $\delta[u]=\mathcal{O}(r^0)$ we may now also infer the falloff conditions for the components of the one-form $\delta[k_a]$. Going back to \eref{ellvec} and \eref{emvec}, we obtain, schematically,
\begin{equation}
\delta[k_a] = \mathcal{O}(r^{-2})\ell_a+\mathcal{O}(r^0)k_a+\mathcal{O}(r^{-1})m_a+\mathcal{O}(r^{-1})\bar{m}_a.\label{deltafall5}
\end{equation}
To recover the symplectic structure on the radiative phase space in terms of the asymptotic shear, we express the $1/r$ expansion of $\delta[m_a]$ in terms of $\delta[\sigma_{(k)}]=\delta\sigma/r^2+\mathcal{O}(r^{-3})$ and the variation of $\delta[\vartheta_{(k)}]=2\delta[U]/r^2+\mathcal{O}(r^{-3})$. For every value of $\rho$, the null surface $\mathcal{N}_\rho$ is equipped with a universal ruling, which determines the direction of the null generators, i.e.\ the equivalence class $[\ell^a]$. This ruling is a universal background structure that we consider to be fixed on the covariant phase space, hence $\delta\ell^a=\tilde{\lambda}\ell^a$. The falloff conditions for $\tilde{\lambda}$ can be inferred directly from $k_a\ell^a=-1$ and \eref{deltafall5}, which implies $\tilde{\lambda}=\mathcal{O}(r^0)$. The existence of such a fixed ruling of $\mathcal{N}_\rho$ also implies that the variation of the complex-valued one-form $m_a$ will admit the expansion
\begin{equation}
\delta[m_a] = \text{\sl{f}}\,\ell_a + \text{\sl{g}}\,m_a +\text{\sl{h}}\,\bar{m}_a.
\end{equation}
Where $\text{\sl{f}}=\mathcal{O}(r^{-1})$ and $\text{\sl{g}}=\mathcal{O}(r^0)$, $\text{\sl{h}}=\mathcal{O}(r^0)$, which is a consequence of (\ref{deltafall2}, \ref{deltafall3}, \ref{deltafall4}) and the falloff conditions for the metric coefficients $\zeta$, $\mu$, $\nu$ that define the one-form $m_{a}=-\zeta\partial_au+(\mu\bar{\mu}-\nu\bar{\nu})^{-1}[\mu\partial_az-\nu\partial_a{\bar{z}}]$, see (\ref{ellvec}, \ref{emvec}). We consider thus the ansatz,
\begin{align}
\text{\sl{f}} & = \frac{\text{\sl{f}\hspace{0.2em}}^{(0)}}{r}+\mathcal{O}(r^{-2}),\\
\text{\sl{g}} & = \text{\sl{g}}^{(0)} + \frac{\text{\sl{g}}^{(1)}}{r}+\mathcal{O}(r^{-2}),\\
\text{\sl{h}} & = \text{\sl{h}}^{(0)} + \frac{\text{\sl{h}}^{(1)}}{r}+\mathcal{O}(r^{-2}).
\end{align}
To evaluate the symplectic potential \eref{nullsympl} at future null infinity, we now want to express the subleading terms of this expansion in terms of variations of the asymptotic shear $\delta\sigma_{(k)}=\delta\sigma/r^2+\mathcal{O}(r^{-3})$ and the variation of the asymptotic expansion $\delta\vartheta_{(k)}=2\delta[U]/r^2+\mathcal{O}(r^{-3})$. Consider then the radial and tangential evolution equations for the pull-back of $m_a$ to the null hypersurface,
\begin{align}
{\varphi}^\ast_{\mathcal{N}_\rho}\big[\mathcal{L}_km\big]_a&=\frac{1}{2}\vartheta_{(k)}{\varphi}^\ast_{\mathcal{N}_\rho}m_a+\sigma_{(k)}{\varphi}^\ast_{\mathcal{N}_\rho}\bar{m}_a+(\alpha-\beta){\varphi}^\ast_{\mathcal{N}_\rho}k_a,\\
{\varphi}^\ast_{\mathcal{N}_\rho}\big[\mathcal{L}_\ell m\big]_a&=\frac{1}{2}\big(\vartheta_{(\ell)}+2\I\phi\big){\varphi}^\ast_{\mathcal{N}_\rho}m_a+\sigma_{(\ell)}{\varphi}^\ast_{\mathcal{N}_\rho}\bar{m}_a,
\end{align}
which follow directly from \eref{anhol}. Taking into account that $0=\delta[k^a\nabla_a r]=\delta[k^a]\nabla_a r+\frac{\di}{\di r}\delta[r]=\delta[k^a]\nabla_a r+\mathcal{O}(r^{-2})$, and $0=\delta[k^a\nabla_a u]=\delta[k^a]\nabla_a u+\frac{\di}{\di r}\delta[u]=\delta[k^a]\nabla_a u+\mathcal{O}(r^{-2})$, we obtain from $\ell_a=-\partial_ar +N\nabla_au +w\bar{m}_a+\bar{w}m_a$ and the falloff conditions for $N$ and $w$ that
\begin{align}
\delta[k^a]&=\delta\left[\frac{\di}{\di r}\right]^a = \mathcal{O}(r^{-2})k^a + \mathcal{O}(r^{-2})\ell^a + \hspace{0.3em}{\bar{\text{\hspace{-0.3em}\sl{f}}}}\,m^a  + {{\text{\sl{f}}}}\,\bar{m}^a,\\
\delta[\ell^a]&=\mathcal{O}(r^0)\ell^a.
\end{align}
Since $\delta\rho=0$, the pull-back to $\mathcal{N}_\rho$ commutes with the variation, and we obtain from $\delta\big[{\varphi}^\ast_{\mathcal{N}_\rho}[\mathcal{L}_\ell m]\big]={\varphi}^\ast_{\mathcal{N}_\rho}[\mathcal{L}_\ell \delta m]-{\varphi}^\ast_{\mathcal{N}_\rho}[\mathcal{L}_{\delta \ell} m]$  that
\begin{align}
\frac{\di}{\di u}{\text{\sl{f}}\hspace{0.2em}}^{(0)} &=0, \quad \text{\sl{f}\hspace{0.2em}}^{(0)}=\text{\sl{f}\hspace{0.2em}}^{(0)}(z,\bar{z}),\label{deltamdot1}\\
\frac{\di}{\di u}{\text{\sl{g}}}^{(0)} &=0, \quad \text{\sl{g}}^{(0)}=\text{\sl{g}}^{(0)}(z,\bar{z}),\label{deltamdot2}\\
\frac{\di}{\di u}{\text{\sl{h}}}^{(0)} &=0, \quad \text{\sl{h}}^{(0)}=\text{\sl{h}}^{(0)}(z,\bar{z})\label{deltamdot3}.
\end{align}
The next to leading order perturbations $\text{\sl{g}}^{(1)}$ and $\text{\sl{h}}^{(1)}$ can be obtained from the variation of the radial evolution equation, i.e.\ $\delta\big[{\varphi}^\ast_{\mathcal{N}_\rho}[\mathcal{L}_k m]\big]={\varphi}^\ast_{\mathcal{N}_\rho}[\mathcal{L}_k \delta m]-{\varphi}^\ast_{\mathcal{N}_\rho}[\mathcal{L}_{\delta k} m]$. Taking into account the various falloff conditions, we obtain
\begin{align}
{\text{\sl{g}}}^{(1)} & = -\delta U + {\text{\sl{h}}}^{(0)}\bar{\sigma}-{\bar{\text{\sl{h}}}}^{(0)}\sigma-\I\gamma^{(0)}\hspace{0.3em}{\bar{\text{\hspace{-0.3em}\sl{f}}}\hspace{0.2em}}^{(0)},\\
{\text{\sl{h}}}^{(1)} & = -\delta\sigma + ({\text{\sl{g}}}^{(0)} - {\bar{\text{\sl{g}}}}^{(0)})\sigma+\I\gamma^{(0)}\text{\sl{f}\hspace{0.2em}}^{(0)},
\end{align}
where $\gamma^{(0)}$ is the $1/r$ leading term of the expansion $\gamma=\gamma^{(0)}/r+\mathcal{O}(r^{-2})$. The next to leading order of ${\text{\sl{h}}}$ and ${\text{\sl{g}}}$ is thus sourced by the variation of the radial shear and expansion,
\begin{align}
\delta[\vartheta_{(k)}] & = \frac{2\delta[U]}{r^2}+\mathcal{O}(r^{-3}),\\
\delta[\sigma_{(k)}] & = \frac{\delta[\sigma]}{r^2}+\mathcal{O}(r^{-3}).
\end{align}

We have seen in \hyperref[sec3.4]{section 3.4} that those bulk diffeomorphism, whose pullback to the null surface $\mathcal{N}_\rho$  map every light ray onto itself are unphysical gauge directions on the covariant phase space. We remove this gauge freedom by imposing the following boundary conditions on the field variation of the retarded $u$ time coordinate,\footnote{Notice that the radial coordinate $r$ and the retarded time $u$ depend via the gauge and falloff conditions \eref{geodsc}, \eref{rdef}, \eref{lapse} implicitly on the gravitational variables, hence $\delta[r]\neq 0$ and $\delta[u]=0$.} 
\begin{equation}
\delta[u] = \text{\sl{s}}(z,\bar{z})  + \mathcal{O}(r^{-1}).\label{deltafall6}
\end{equation}

 We have now everything at hand to recover the radiative symplectic potential on future null infinity. We have seen in (\ref{deltamdot1}, \ref{deltamdot2}, \ref{deltamdot3}) that the leading coefficients ${\text{\sl{f}}\hspace{0.2em}}^{(0)}$, ${\text{\sl{g}}}^{(0)}$, ${\text{\sl{h}}}^{(0)}$ that determine $\delta[\varphi^\ast_{\mathcal{N}_\rho} m_a]$ are constant along the null generators of $\mathcal{J}^+$. Therefore, they cannot represent radiative modes, which  characterise \emph{local} degrees of freedom of the gravitational field at $\mathcal{I}^+$. To infer the \emph{radiative} symplectic structure on $\mathcal{I}^+$ from the $\rho\rightarrow\infty$ limit of the quasi-local symplectic potential, we set those variations to zero, otherwise we would be left with an IR divergent integral along the null generators (the range of the $u$-coordinate is the entire real line). For the same reason, we set $\delta[U]=0$ such that $\delta[\vartheta_{(k)}]=\mathcal{O}(r^{-3})$. In fact, we have seen in \eref{Uevolv} that $U(u,z,\bar{z})$ is linear in the affine parameter: the derivative $\dot{U}>0$ is constant along the null generators and determines the Ricci curvature \eref{twoRicci} of the $u=\mathrm{const}.$ cross sections of $\mathcal{I}^+$. If we restrict ourselves to cross sections, where the  two-dimensional metric $q^o_{ab}$ is simply the metric of the round two-sphere, we immediately have $\delta[U]=0$.
 
 If we then remove such IR divergent terms, i.e.\ after imposing that the $u$ independent terms ${\text{\sl{g}}}^{(0)}$, ${\text{\sl{h}}}^{(0)}$ and $\delta[\dot{U}]$ vanish, the $(m_a,\bar{m}_a)$-components of a tangent vector $\delta_{\mtext{rad}}$ to the radiative phase space will satisfy the falloff conditions
\begin{equation}
\delta_{\mtext{rad}}[\varphi^\ast_{\mathcal{N}_\rho}m]_a = \mathcal{O}(r^{-2})\,[\varphi^\ast_{\mathcal{N}_\rho}m]_a-\left(\frac{\delta_{\mtext{rad}}[\sigma]}{r}+\mathcal{O}(r^{-2})\right)[\varphi^\ast_{\mathcal{N}_\rho}\bar{m}]_a\label{deltafall7}.
\end{equation}

To obtain the radiative phase space, we insert both \eref{deltafall2} and \eref{deltafall6} together with \eref{deltafall7} back into the pre-symplectic potential \eref{nullsympl}, and evaluate the integral as $\rho=r+\mathcal{O}(r^{-1})\rightarrow\infty$, such that
\begin{equation}
\Omega_{\mathcal{I}^+}\big(\delta_1^{\mtext{rad}},\delta_2^{\mtext{rad}}\big):=\lim_{\rho\rightarrow\infty}\Omega_{\mathcal{N}_\rho}\big(\delta_1^{\mtext{rad}},\delta_2^{\mtext{rad}}\big) = \frac{1}{4\pi G}\int_{\mathcal{I}^+}k\wedge d^2\Omega\,\big(\delta^{\mtext{rad}}_{[1}\dot{\sigma}\,\delta^{\mtext{rad}}_{2]}\bar{\sigma}+\CC\big).\label{radSympl}
\end{equation}
where the family of bounded null surfaces $\{\mathcal{N}_\rho\}_{\rho>0}$ is chosen such that $\lim_{\rho\rightarrow\infty}\mathcal{N}_\rho=\mathcal{I}^+$ and  $\Omega_{\mathcal{N}_\rho}=\bbvar{d}\Theta_{\mathcal{N}_\rho}$ is the pre-symplectic two-form that we introduced in \hyperref[sec3.4]{section 3.4}. In addition, $d^2\Omega$ is the fiducial area element on two-dimensional cross sections of $\mathcal{I}^+$, which can be inferred from the $1/r$ expansion of the physical area two-form $\varepsilon=-\I m\wedge\bar{m}$,
\begin{equation}
-\I \varphi^\ast_{\mathcal{N}_\rho}\big(m\wedge\bar{m}\big) = d^2\Omega\, (r^2-2Ur+\mathcal{O}(r^0)),\label{falloff8b}
\end{equation}
and $k_a$ is a one-form such that $k_a\partial^a_u=-1$. Given the symplectic two-form \eref{radSympl}, it is also useful to introduce the corresponding symplectic current. Choosing the same polarisation as in \eref{nullsympl}, we obtain
\begin{equation}
\theta_{\mathcal{I}^+}(\delta_{\mtext{rad}}) = \frac{1}{8\pi G} k\wedge d^2\Omega\big(\dot{\sigma}\delta\bar{\sigma}+\CC).\label{thetainf}
\end{equation}

\subsection{Bondi energy and Helmholtz free energy of gravitational edge modes}
\noindent It is now possible to identify the Hamiltonian on a partial Cauchy hypersurface $M_u$ that intersects future null infinity at constant values of $u$ (the boundary $\partial M_u = \mathcal{C}_u$ will be a $u=\mathrm{const}.$ cross section of $\mathcal{I}^+$). We call this Hamiltonian $H_\xi[\mathcal{C}_u]$ and it will  generate time translations along the vector field $\xi^a\in T\mathcal{M}$, which is null and lies tangential to the generators of the null foliation,
\begin{equation}
\xi^a\big|_{\mathcal{N}}\in[\ell^a].
\end{equation}

Following what we said in equation \eref{Hvar2} above, we {define} the generator as a functional $H_\xi$ on the space of physical histories, which is larger than phase space,\footnote{The space of physical histories is larger than phase space, because (i) it contains configurations that would be gauge equivalent on phase space (ii) includes the boundary data on $\mathcal{I}^+$, which is otherwise fixed by the boundary and gauge fixing conditions, i.e. $\delta\sigma(u,z,\bar{z})=0$.} such that
\begin{equation}
\delta\big[H_\xi[\mathcal{C}]\big] = - \Omega_M(\mathcal{L}_\xi,\delta) + \int_{\partial M}\xi\hook\theta_{\mathcal{N}}(\delta),\label{deltaHxi}
\end{equation}
where $\theta_{\mathcal{N}}$ is the symplectic current, i.e.\ the integrand of \eref{nullsympl}, and $\delta[\cdot]\in T\mathcal{H}_{\mtext{phys}}$ denotes a linearised solution of the bulk plus boundary field equations, see \hyperref[tab1]{table 1}. {The relative minus sign between \eref{deltaHxi} and \eref{Hvar2} results from a change of orientation on $\mathcal{C}=\partial M$, which is equipped with the induced orientation from $M$ rather than $\mathcal{N}$.}

The entire calculation of \eref{deltaHxi} is valid only on-shell, which is to say provided the bulk plus boundary field equations are satisfied. As in \hyperref[sec4]{section 4} above, the first term is a total boundary term. Using the definition of the $SL(2,\C)$ gauge covariant Lie derivative, see (\ref{Liea}--\ref{Lied}), we obtain, in fact
\begin{align}\nonumber
\Omega_M(\mathcal{L}_\xi,\delta) & =  \frac{\I}{8\pi G}\bigg[\int_M\Big(\nabla(\xi\hook\Sigma_{AB})\wedge\delta[A^{AB}]-\delta[\Sigma_{AB}]\wedge\xi\hook F^{AB}\Big)+\\\nonumber
&\hspace{1.5em}-\oint_{\partial M}\Big(\xi\hook D\eta_A\delta\ell^A+D(\xi\hook\eta_A)\delta\ell^A-\delta\eta_A\xi\hook D\ell^A\Big)\bigg]+\CC=\\\nonumber
& =  \frac{\I}{8\pi G}\bigg[\int_M\Big((\xi\hook\Sigma_{AB})\wedge\delta[F^{AB}]-\delta[\Sigma_{AB}]\wedge\xi\hook F^{AB}\Big)+\\
&\hspace{1.5em}+\oint_{\partial M}\Big(\xi\hook\Sigma_{AB}\wedge\delta[A^{AB}]-\xi\hook D\eta_A\delta\ell^A-D(\xi\hook\eta_A)\delta\ell^A+\delta\eta_A\xi\hook D\ell^A\Big)\bigg]+\CC\label{workterm}
\end{align}
Since $\delta[\cdot]$ is a linearised solution of the bulk plus boundary field equations, the three-dimensional bulk integral vanishes,
\begin{align}\nonumber
\int_M&\Big(\xi\hook\Sigma_{AB}\wedge\delta F^{AB}-\delta\Sigma_{AB}\wedge\xi\hook F^{AB}\Big) = \\\nonumber
&=-\int_M\Big(\xi_{AA'}\uo{e}{B}{A'}\wedge\delta F^{AB}-e_{AA'}\wedge\delta\uo{e}{B}{A'}\wedge\xi\hook F^{AB}\Big)=\\
&=-\int_M\Big(\xi_{AA'}\uo{e}{B}{A'}\wedge\delta F^{AB}+\xi_{AA'}\wedge\delta\uo{e}{B}{A'}\wedge F^{AB}\Big)=-\int_M\xi_{AA'}\delta[\uo{e}{B}{A'}\wedge F^{AB}]=0,\label{bulkterm}
\end{align}
where $\xi_{AA'}=\xi\hook e_{AA'}$ and $\Sigma_{AB}=-\frac{1}{2}e_{AC'}\wedge\uo{e}{B}{C'}$ and the field equations in the bulk, i.e.\ $e_{BA'}\wedge \ou{F}{B}{A}=0$,$\nabla e_{AA'}=0$ are satisfied. If we then also take into account the boundary field equations (\ref{bndryeq1}, \ref{bndryeq2}), we can further simplify the various contributions to \eref{deltaHxi}. Going back to \eref{bndryeq1} and \eref{bndryeq2}, we find 
\begin{align}\nonumber
\xi\hook  D\eta_A\delta\ell^A & = -\xi\hook\Big((\omega+\frac{1}{2}\varkappa)\wedge\eta_A\Big)\delta\ell^A-\xi\hook N_A\wedge\bar{m}\delta\ell^A=\\\nonumber
& = -\xi\hook\big(\omega+\frac{1}{2}\varkappa\wedge\eta_A\big)\delta\ell^A+\xi^ak_a \ell\hook N_A\wedge\bar{m}\delta\ell^A=\\
& = -\xi\hook\big(\omega+\frac{1}{2}\varkappa\wedge\eta_A\big)\delta\ell^A+\xi^ak_a \bar{m}\wedge\delta\ell_A D\ell^A.\label{bndryterm}
\end{align}
If we insert \eref{bndryterm} back into \eref{workterm}, two terms appear: the first term is linear in the variation of the two-dimensional area two-form $\varepsilon=-\I m\wedge\bar{m}$ on $\mathcal{N}$, and the other term only contains variations of the one-form $\ell_AD\ell^A$ that determines shear and expansion of $\mathcal{N}$, see \eref{thetashear}. More precisely,
\begin{align}\nonumber
\Omega_M(\mathcal{L}_\xi,\delta) &= -\frac{\I}{8\pi G}\oint_{\partial M}\Big(\xi\hook\eta_A\delta[\ou{A}{A}{B}]\ell^B+\xi^ak_a\bar{m}\wedge\delta[\ell_A]D\ell^A+\xi\hook\eta_AD\delta\ell^A+\\\nonumber
&\hspace{5em}-\xi\hook(\omega+\frac{1}{2}\varkappa)\eta_A\delta\ell^A-\xi\hook(\omega+\frac{1}{2}\kappa)\delta[\eta_A]\ell^A\Big)+\CC=\\
&=\frac{1}{8\pi G}\oint_{\partial M}(\xi\hook\varkappa)\,\delta[\varepsilon]-\frac{\I}{8\pi G}\oint_{\partial M}\xi^ak_a\big(\bar{m}\wedge\delta(\ell_AD\ell^A)-\CC\big).
\end{align}
We assume $\delta[\xi^a]=0$, and insert the radiative symplectic potential \eref{thetanull} back into the definition of the generator \eref{deltaHxi}. This leads us to
\begin{align}\nonumber
\delta\big[H_\xi[\partial M]\big] & = -\Omega_M(\mathcal{L}_\xi,\delta)+\oint_{\partial M}\xi\hook\theta_{\mathcal{N}}(\delta)=\\\nonumber
& = -\frac{1}{8\pi G}\oint_{\partial M}\delta\big[\varepsilon\,\xi\hook\varkappa\big]+\frac{\I}{8\pi G}\oint_{\partial M}\delta\big[\xi^ak_a\bar{m}\wedge\ell_AD\ell^A-\CC\big]=\\
& = -\frac{1}{8\pi G}\oint_{\partial M}\delta\big[\varepsilon\,\xi\hook\varkappa\big]-\frac{1}{8\pi G}\oint_{\partial M}\delta\big[\varepsilon\,\xi^ak_a\,\vartheta_{(\ell)}\big],\label{Fenergy}
\end{align}
which is a total derivative \emph{on the space of physical histories}.

To extract the Bondi mass from the time-dependent Hamiltonian \eref{Fenergy}, we impose the various gauge fixing and falloff conditions for the double null foliation that we defined in the last section and evaluate the integral in the limit $\rho\rightarrow\infty$. In addition, and to guarantee that $H_\xi$ vanishes in Minkowski space, we impose the following falloff conditions,
\begin{equation}
\xi^a = \Big(1+\frac{U}{r}+\mathcal{O}(r^{-2})\Big)\ell^a,\label{xifalloff}
\end{equation}
hence $\xi^a$ is a field-dependent vector field.\footnote{Notice that $\xi^a$ depends on $U$ and $r$ and $\ell^a$, hence $\delta[\xi^a]\neq 0$ or more precisely $\delta[\xi^a]=\mathcal{O}(r^{-2})\ell^a$ provided \eref{deltafall2} and \eref{deltafall6} are satisfied. For a vector field that depends itself on the configuration variables, equation \eref{deltaHxi} gets replaced by $\delta[H_\xi]-H_{\delta[\xi]}=-\Omega(\mathcal{L}_\xi,\delta)+\oint_{\partial M}\xi\hook\theta_{\mathcal{N}}(\delta)$. If, however, $\delta\xi^a=\mathcal{O}(r^{-2})\ell^a$ this subtlety can be ignored since the integrand $\delta[\xi^a]k_a\vartheta_{(\ell)}\varepsilon$ will vanish as $r\rightarrow\infty$.} The falloff conditions for the tangential expansion \eref{falloff10}, \eref{order1}, for the area element \eref{falloff8b}, and for the vector field $\xi^a$ imply now the following $1/r$ expansion of the variation
\begin{align}
\frac{1}{8\pi G}\oint_{\partial M}\delta\big[\xi^ak_a\,\varepsilon\,\vartheta_{(\ell)}\big] & =-\frac{1}{8\pi G}\delta\oint_{\partial M}d^2\Omega\Big[-2r\dot{U}+\vartheta_{(\ell)}^{(2)}+2U\dot{U}+\mathcal{O}(r^{-1})\Big].\label{Bondimass1}
\end{align}
The first term, which is linear in $r$, is a potential source of an IR divergence, but this term is harmless, since its variation vanishes for the gauge fixing and falloff conditions that we have chosen above. Since, in fact, $\delta\rho=0$, which is to say that wee keep the $\{\mathcal{N}_\rho\}_{\rho>0}$ foliation fixed, and $\rho=r+\mathcal{O}(r^{-1})$, see \eref{rhoexpand} and \eref{deltafall2}, we obtain 
\begin{equation}
\lim_{\rho\rightarrow\infty}\delta\bigg[\oint_{\partial M_{\rho,u}}d^2\Omega\,r\,\dot{U}\bigg]=\lim_{\rho\rightarrow\infty}\bigg(\rho\,\delta\bigg[\oint_{\partial M_{\rho,u}}d^2\Omega\,\dot{U}\bigg]\bigg).\label{Eulercharact1}
\end{equation}
On the other hand,  $\dot{U}$ is proportional to the two-dimensional Ricci scalar \eref{twoRicci}. The resulting integral  $(4\pi)^{-1}\int_{\partial M_{\rho,u}}d^2\Omega \,\mathcal{R}$ is the Euler characteristic $\chi[\partial M_{\rho,u}]$ of the two-dimensional boundary $\partial M_{\rho,u}$, which has the topology of a two sphere.  Since $\chi[S^2]=2$, the variation vanishes,
\begin{equation}
\lim_{\rho\rightarrow\infty}\delta\bigg[\oint_{\partial M_{\rho,u}}d^2\Omega\,r\,\dot{U}\bigg]=\pi\lim_{\rho\rightarrow\infty}\big(\rho\,\delta\big[\chi[\partial M_{\rho,u}]\big]\big)=0.\label{Eulercharact}
\end{equation}

It is now possible to insert \eref{Bondimass1} and \eref{Eulercharact}  back into \eref{Fenergy}. Taking the asymptotic $\rho\rightarrow\infty$ limit, we obtain
\begin{equation}
\lim_{\rho\rightarrow\infty}\delta\big[H_\xi[\mathcal{C}_{\rho,u}]\big] = -\frac{1}{8\pi G}\delta\bigg[\oint_{\mathcal{C}_u}\varepsilon\,\kappa\bigg]+\delta\big[M_{\mathrm{B}}(u)\big],\label{varHxi}
\end{equation}
where $M_{\mathrm{B}}(u)$ is the $\mathcal{O}(r^0)$ contribution to \eref{Bondimass1}, i.e.\ the integral of $\vartheta_{(\ell)}^{(2)}+2U\dot{U}$ over the $u=\mathrm{const}.$ cross section of $\mathcal{I}^+$. Going back to equation \eref{falloff11a}, we have
\begin{equation}
M_{\mathrm{B}}(u) = -\frac{1}{4\pi G}\oint_{\mathcal{C}_u}d^2\Omega\,\Big(\bar{\Psi}_2^{(0)}+\sigma\dot{\bar{\sigma}}+\bar{\mathcal{D}}\bar{\alpha}^{(0)}\Big),
\end{equation}
which is the Bondi mass. The mass loss formula follows from the next to leading order of the tangential Raychaudhuri equation, i.e.\ \eref{Uevolv2}, such that
\begin{equation}
\frac{\di}{\di u} M_{\mathrm{B}}(u) = -\frac{1}{4\pi G}\oint_{\mathcal{C}_u}d^2\Omega\,|\dot{\sigma}|^2=\oint_{\mathcal{C}_u}\xi\hook\theta_{\mathcal{I}^+}(\tfrac{\di}{\di u})\leq 0.\label{possflux}
\end{equation}

\paragraph{- Bondi energy as free energy} On a black hole spacetime, the ADM mass at infinity, the ADM angular momentum and the area of the horizon all have a thermodynamical interpretation. The functional variation \eref{varHxi} and the mass loss formula \eref{possflux} suggest a similar understanding as well:
\begin{equation}
\underbrace{H_\xi[\mathcal{C}_{u}\subset\mathcal{I}^+]}_{\mtext{{internal energy}}}\quad+\quad\underbrace{\frac{1}{8\pi G}\oint_{\mathcal{C}_u} \varepsilon\,\kappa}_{-\mtext{{$S\times T\phantom{-}$}}}\quad = \underbrace{M_{\mathrm{B}}(u).}_{\mtext{{free energy}}}\label{fEnergydef}
\end{equation}

In other words, our suggestion is to identify Bondi's radiative energy with the Helmholtz \emph{free energy} of the system. The free energy of what \emph{system}? 
To answer this question, it seems crucial to understand how observables and phase space itself depends on the chosen boundary conditions. 
Different boundary conditions represent altogether different physical processes, i.e.\ different physical systems with different phase spaces and different Hamiltonians. In our case, this physical difference is realised mathematically by the difference between the space of \emph{physical histories} $\mathcal{H}_{\mtext{phys}}$, i.e.\ the solution space of the field equations for arbitrary boundary conditions, and the \emph{covariant phase space}, which is the space of solutions to the field equations for specific boundary and falloff conditions. 
The space of \emph{physical histories} is therefore larger than phase space, and depending on how restrictive the boundary conditions are, the size of the resulting phase space will be different.  A simple example was given in \hyperref[sec2]{section 2}, where we considered a time-dependent Hamiltonian $H[\vec{p},\vec{q},\omega(t)]$, whose time-dependence enters through a time-dependent background field $\omega(t)$. A generic tangent vector $\delta[\cdot]\in T\mathcal{H}_{\mtext{phys}}$ on the space of physical histories will generate infinitesimal changes of these parameters, hence $\delta[\omega]\neq0$. However, any actual physical trajectory\,---\,a point on phase space\,---\,is realised only for a particular choice of the background field $\omega(t)$. Since $\omega(t)$ can be tuned continuously, the space of histories foliates into a whole family of phase spaces $\bigsqcup_\omega\mathcal{P}_\omega =\mathcal{H}_{\mtext{phys}}$, and for every $\omega(t)$ there is a different phase space.\footnote{The background fields represent knobs and controls that allow us to manipulate the experiment. Such manipulations may happen directly by changing the controls of the experimental setup, or retroactively by post-selecting a subset of observations from an ensemble of similar observations with  different boundary conditions.} Each of these phase spaces is equipped with a symplectic two-form $\Omega$, which is obtained by the pull-back of the pre-symplectic two-form $\Omega_{\mtext{bulk}}\in\Omega^2(T^\ast\mathcal{H}_{\mtext{phys}})$ to $\mathcal{P}_\omega$ modulo gauge.

Thus, we need to identify the phase space, and hence the system, for which the functional $H_\xi[\mathcal{C}_u\subset\mathcal{I}^+]:\mathcal{H}_{\mtext{phys}}\rightarrow\R$, (\ref{deltaHxi}, \ref{fEnergydef}), is the Hamiltonian. Consider the problem at the linearised level. A tangent vector $\delta[\cdot]\in T\mathcal{H}_{\mtext{phys}}$ that lies tangential to this yet unspecified phase space $\mathcal{P}_{\mtext{edge}}\subset\mathcal{H}_{\mtext{phys}}$ will satisfy Hamilton's equations, $\delta[H_\xi]=-(\varphi^\ast_{\mathcal{P}}\Omega_{M_u})(\mathcal{L}_\xi,\delta)$, where $M_u$ is a partial Cauchy surface that intersects $\mathcal{I}^+$ at some $u=\mathrm{const}.$ cross section $\mathcal{C}_u$. On the other hand, a generic such vector field $\delta[\cdot]\in T\mathcal{H}_{\mtext{phys}}$ will satisfy \eref{deltaHxi}. Both equations can be true, only if the symplectic current vanishes, i.e.\ $\lim_{\rho\rightarrow\infty}\oint_{\mathcal{C}_{\rho,u}}\xi\hook\theta_{\mathcal{N}_\rho}(\delta)=0$. For a generic configuration on $\mathcal{H}_{\mtext{phys}}$, this implies that the variation of the asymptotic shear will vanish. Hence $\delta\sigma(u,z,\bar{z})=0$. If we insist to have a phase space for which $H_\xi[\mathcal{C}_u]$ is the Hamiltonian, we should add this condition to our boundary and falloff conditions on $\mathcal{I}^+$. Clearly, this is a very restrictive condition. Since the asymptotic shear characterises all the outgoing radiation, the boundary condition $\delta\sigma(u,z,\bar{z})=0$ removes  all radiative modes from the phase space that we would otherwise associate to the partial Cauchy hypersurface $M_u$.\footnote{Unless there are singularities or different asymptotic regions that could capture some of the radiative degrees of freedom.} The phase space $\mathcal{P}_{\mtext{edge}}$ contains, therefore, no local degrees of freedom from within the partial Cauchy hypersurface. Yet there should be still infinitely many boundary modes left that characterise e.g.\ large diffeomorphsims and boosts, see e.g.\ \eref{Jdef} and \eref{boostgen}.\footnote{The corresponding smearing functions will have the following falloff in a neighbourhood of $\mathcal{I}^+$, $\mathcal{\lambda}=\mathcal{O}(r^{-2})$ and $\xi^a =\xi\bar{m}^a+\bar{x}m^a$, $\xi=\mathcal{O}(r)$.} In fact, this is precisely what is suggested by general relativity in dimensions smaller than four. Consider, for example, three-dimensional gravity in the Chern\,--\,Simons or BF formulation. On a closed manifold $\mathcal{M}$, the resulting gravitational phase space is the moduli space of flat connections, which is finite-dimensional. On the other hand, if we break the manifold $\mathcal{M}$ into two parts $\mathcal{M}=\mathcal{M}^+\cup\mathcal{M}^-$, new degrees of freedom are excited along the boundary $\partial\mathcal{M}^+ =\mathcal{B}$. The splitting of the manifold into two parts destroys diffeomorphism invariance and directions  on field space that would have otherwise been considered unphysical represent now physical boundary modes (gravitational edge modes). The dynamics of these boundary modes along the two-dimensional boundary depends on the boundary conditions chosen. Different boundary conditions correspond to different boundary field theories with different phase spaces and different notions of energy. Perhaps the most important such example is the Wess\,--\,Zumino\,--\,Witten model, which provides a possible boundary field theory for gravitational edge modes in three-dimensional gravity, but there are many other  boundary field theories as well, both at the level of the discrete spin network representation and in the continuum, see e.g.\ \cite{Dittrich:2018xuk,Dittrich:2017hnl,Dittrich:2017rvb,Wieland:2017cmf,Wieland:2018ymr,Freidel:2020xyx}. By removing the radiative modes from the partial Cauchy surface $M_u$ and encoding them into auxiliary background fields on $\mathcal{I}^+$, we are in a very similar situation as well. The resulting phase on a partial Cauchy surface $M_u$ will be stripped off its radiative data, and can only consist of gravitational edge modes alone.


If we accept such a reasoning, which is supported by  recent results from various approaches \cite{Lashkari:2016idm,Chandrasekaran:2019ewn,DeLorenzo:2017tgx,PhysRevD.101.064023},  equation \eref{fEnergydef} suggests to identify Bondi's radiative energy with the free energy of gravitational edge modes. Accordingly, the mass loss formula \eref{possflux} turns into the gravitational equivalent of the statement that the free energy decreases towards thermal equilibrium, and thermal equilibrium is reached once $\dot{\sigma}=0$. In analogy to the first law of  black hole thermodynamics, we may then also identify the entropy density $s$ with the area density of the two-dimensional $u=\mathrm{const}.$ cross sections of $\mathcal{I}^+$, i.e.\ $s=\varepsilon/(4G) $. Clearly, the entropy diverges as $\rho\rightarrow\infty$, but the product $\varepsilon\,\kappa$ remains finite, because the non-affinity $\kappa:\ell^b\nabla_b\ell^a=\kappa\ell^a$ admits the $1/r$ expansion $\kappa = {\operatorname{\mathfrak{Re}}(\Psi_2^{(0)})}/{r^2}+\mathcal{O}(r^{-3})$. In equilibrium, $\dot\sigma=0$ and the Bondi mass is simply the integral $M_{\mathrm{B}}=-\frac{1}{4\pi G}\oint_{\mathcal{C}}d^2\Omega\,\bar{\Psi}^{(0)}_2$ such that $S\times T = M_{\mathrm{B}}/2\geq 0$, with the asymptotic equivalent of temperature given by  $T=-\frac{1}{2\pi}{\operatorname{\mathfrak{Re}}(\Psi_2^{(0)})}/{r^2}+\mathcal{O}(r^{-3})$, which vanishes as $r\rightarrow\infty$.

\section{Summary and discussion}\label{sec5}
\noindent We have developed, step by step, a representation of the radiative gravitational phase space at generic null boundaries in terms of an adapted Newman\,--\,Penrose tetrad. The starting point (\hyperref[sec3]{section 3}) was the introduction of the appropriate counter terms on the null boundary such that the action is stationary provided the  boundary conditions and equations of motion are satisfied. The boundary conditions are such that an equivalence class $\mathcal{g}=[\varkappa_a,\ell^a,m_a]/_\sim$ is kept fixed at the boundary, where $\ell^a$ is a representative of the null vectors (vertical vector fields) that generate the null boundary, $\varkappa_a$ is an abelian connection such that $\kappa=\ell^a\varkappa_a$ is the non-affinity of $\ell^a$, i.e.\ $\ell^b\nabla_b\ell^a=\kappa\ell^a$, and the co-dyad $m_a$ is a complex-valued one-form intrinsic to the null boundary such that $q_{ab}=2m_{(a}m_{b)}$ is the induced signature $(0$$+$$+)$ metric on the boundary. Two configurations $(\varkappa_a,\ell^a,m_a)$ and $(\tilde{\varkappa}_a,\tilde{\ell}^a,\tilde{m}_a)$ define the same equivalence class $\mathcal{g}=[\varkappa_a,\ell^a,m_a]/_\sim$ if they are related by a combination of (i) vertical diffeomorphisms \eref{gauge1} along the vector fields  $\xi^a\propto \ell^a$, (ii) shifts \eref{gauge3} of $\varkappa_a$, (iii) dilations \eref{gauge2} of $(\varkappa_a,\ell^a)$, and (iv) complexified conformal transformations of $(\ell^a,m_a)$, see \eref{gauge4}. The resulting equivalence class $[\varkappa_a,\ell^a,m_a]/_\sim$ characterises two local degrees of freedom on the null surface, which are the two radiative modes of the gravitational field at the full non-perturbative level.  \hyperref[sec3]{Section 3} provides the resulting radiative phase space at the quasi-local level.  Besides the two radiative modes, there are additional edge degrees of freedom. Such edge modes appear, because the null surface $\mathcal{N}$ has itself a boundary (two consecutive cross sections $\mathcal{C}_o$ and $\mathcal{C}_1$, $\partial\mathcal{N}=\mathcal{C}_o\cup\mathcal{C_1}$).  In \hyperref[sec4]{section 4}, we studied such edge modes from the perspective of the gravitational degrees of freedom in the bulk, i.e.\ on a partial Cauchy surface that is attached to the null boundary. We introduced the pre-symplectic two-form $\Omega_M$ on such a partial Cauchy surface and identified the Hamiltonian generators for tangential diffeomorphisms (generalised angular moments), see \eref{Jdef}, and dilations of the null normal \eref{boostgen}. The corresponding Hamiltonian that generates shifts along the null generators was introduced in \hyperref[sec6]{section 6}.

The second half of the paper was about the asymptotic $r\rightarrow\infty$ limit that sends the finite boundary $\mathcal{N}$ to future null infinity $\mathcal{I}^+$. We admit that our presentation was a bit involved, but we believe that this was crucial to obtain the limit to $\mathcal{I}^+$ from a quasi-local perspective. First of all, we introduced a Newman\,--\,Penrose (NP) tetrad adapted to a double null foliation. We then considered the $1/r$ expansion of perturbations of such an adapted null frame around a given solution of Einstein's equations. In a certain way, our gauge choices were dual to those that would be used normally in the Newman\,--\,Penrose formalism. In our case, the two NP null directions are \emph{both} surface forming, hence $\ell_{[a}\nabla_b\ell_{c]}=k_{[a}\nabla_bk_{c]}=0$. In the more standard NP gauge, only the outgoing radial null direction $k^a$ (what is called $l^a$ in the NP formalism) is surface forming, whereas $\ell^a$ (i.e.\ $n^a$ in the NP formalism) is not. Such a gauge choice would be inconvenient for us. To send $\mathcal{N}$ to $\mathcal{I}^+$, we found it necessary to work with a double null foliation, where the \emph{infalling} (collapsing) null surfaces $\{\mathcal{N}_\rho\}_{\rho>0}$ approach future null infinity as $\rho\rightarrow\infty$. Accordingly, we have to relax some other gauge fixing conditions that are otherwise frequently imposed in the NP formalism: in the NP formalism, the parallel transport propagates the null frame along the outgoing null direction ($k^a$ in our case). Such a gauge condition would be incompatible with the integrabilty conditions of the co-vectors $\ell_a$ and $k_a$. Therefore, we had to impose a weaker condition, namely \eref{U1gauge}, which can always be reached thanks to the gauge freedom $m_a\rightarrow \E^{\I\varphi}m_a$. Our discussion completes earlier results \cite{Hopfmuller:2016scf,Chandrasekaran:2018aop,Goldberg_1992,Goldberg:1995gb,Corichi:2016zac,DePaoli:2017sar,Frodden:2019ylc,BarberoG.:2020tmm} on the subject by clarifying the falloff conditions on the covariant phase space in terms of an adapted Newman\,--\,Penrose null tetrad on a generic double null foliation of spacetime.

Taking into account the falloff conditions and removing otherwise IR divergent terms, we obtained the well-known symplectic structure of the radiative modes at future null infinity from the radial renormalisation of the quasi-local symplectic potential, see \eref{radSympl}, \eref{thetanull}. Finally, we computed the time-dependent Hamiltonian, which generates translations along the null generators $\ell^a$. At finite distance, the Hamiltonian is simply the difference of the non-affinity $\kappa$ and the expansion $\vartheta_{(\ell)}$, which are integrated over a two-dimensional cross section of the boundary. This integral diverges in the limit $\rho\rightarrow\infty$, but this divergence is mild. In fact, what we obtain from the covariant phase space approach, is not directly the Hamiltonian, but rather its variation, see \eref{Fenergy}. The potentially IR divergent contribution to the Hamiltonian is proportional to the Euler characteristic of the cross section. This is a topological invariant, whose variation vanishes on the covariant phase space. The variation of the Hamiltonian $\delta[H_\xi]$ is finite \eref{varHxi} and returns the Bondi mass plus a term, which is given by the integral of the non-affinity $\kappa$ over the two-sphere at infinity. The Hamiltonian $H_\xi$ is explicitly time-dependent and integrable on a reduced phase space, where all the radiative modes that would otherwise exist on $M_u$ have been translated into fixed background fields at null infinity. The experience from gravity in dimensions $d<(3+1)$ suggests that the resulting reduced phase space is the phase space of gravitational edge modes alone. 

The article was about classical gravity, but the main motivation for this research has to do with quantum gravity. The quantum version of our approach will provide boundary transition amplitudes that are  conditioned on the asymptotic shear as a classical background field and are evaluated between quantum states at two consecutive cross sections of future null infinity, with the generator of asymptotic symmetries providing the time-dependent Hamiltonian. In \cite{Wieland:2019hkz}, we have given a proposal for how to construct such amplitudes  from a three-dimensional field theory on the null cone.\footnote{Although the programme is centred around boundary field theories, it would be misleading to call such an approach holographic: it is not that we try to translate the two radiative modes of the gravitational field in the bulk into the degrees of freedom of some dual field theory on the light cone. On dimensional grounds, this may very well be impossible: the radiative portion of the gravitational phase space in the bulk has $2\times2\times\infty^3$ dimensions (two polarisations of the graviton per point), the phase space of a boundary field theory has $2\times N\times\infty^2$ dimensions.} In two and three spacetime dimensions, the problem simplifies dramatically. There are no radiative modes to begin with, and the only physical degrees of freedom are the gravitational edge modes alone. Quasi-local realisations of the boundary dynamics for such gravitational edge modes have been explored recently from within loop quantum gravity and related approaches, see for instance \cite{Dittrich:2018xuk,Dittrich:2017hnl,Dittrich:2017rvb,Wieland:2017cmf,Wieland:2018ymr,Freidel:2020xyx}. 

The idea to treat the asymptotic shear as an auxiliary (classical) background field is reminiscent of developments in various other approaches. For example, there is a recent interest in describing quantum systems in relation to reference frames that are themselves quantum. It is then necessary to explain how to \emph{jump} from one such quantum reference system into another thereby creating quantum entanglement among the remaining constituents of the system \cite{Vanrietvelde:2018pgb,Hoehn:2019owq,Castro-Ruiz:2019nnl,Giacomini:2019fvi,Giacomini:2019aa,Krumm:2020fws}. In our case, the asymptotic shear $\sigma(u,z,\bar{z})$ for a given Bondi frame $(u,z,\bar{z})$ provides the classical frame of reference, the quantum variables are the gravitational edge modes, namely generators of horizontal diffeomorphisms \eref{Jdef} or boosts \eref{boostgen}. In addition, we would also like to stress that there seem to be recent developments from within the AdS/CFT community that supports our viewpoint as well, see \cite{Fiorucci:2020xto,Compere:2008us,Troessaert:2015nia}. \vspace{-0.5em}

\paragraph{- Acknowledgments} Discussions with Abhay Ashtekar, Laurent Freidel and Simone Speziale are gratefully acknowledged. This research was supported by the Austrian Academy of Science and the Institute for Quantum Optics and Quantum Information in Vienna.  I am very grateful for the support and I would like to thank \v{C}aslav Brukner in particular.
 This publication was supported by the ID 61466 grant from the John Templeton Foundation, as part of The Quantum Information Structure of Spacetime (QISS) Project (qiss.fr). The opinions expressed in this publication are those of the author and do not necessarily reflect the views of the John Templeton Foundation.

\section*{Appendix: NP formalism and double null foliation}\label{apex}
\phantom{x}\vspace{-2.5em}

\renewcommand{\arraystretch}{1.3}\setcounter{table}{0}
\begin{table}[h]{\small
\begin{tabularx}{\textwidth}[h]{p{0.9em}  p{7em} p{33.6em} X}\cline{1-4}
\multicolumn{1}{l}{}&\raisebox{0.1em}{\it NP Formalism}&\multicolumn{2}{l}{\raisebox{0.1em}{\it Double Null Foliation} }\phantom{x}\\\hline
\parbox[t]{1em}{\multirow{4}{*}{\rotatebox[origin=c]{90}{\it null tetrad}}}  & $l^a = o^A\bar{o}^{A'}$  & {$k^a =\I k^A\bar{k}^{A'}=\partial^a_r$, $k_a=-\nabla_au$} & \\[0.1em]

& $n^a = \iota^A\bar{\iota}^{A'}$ & $\ell^a=\I\ell^A\bar{\ell}^{A'}=\partial^a_u+N\partial^a_r+\bar{\zeta}m^a+\zeta m^a,$\quad  $N=-\frac{\operatorname{\mathfrak{Re}}(\Psi^{(0)}_2)}{r}+\mathcal{O}(r^{-2})$, $\zeta=\mathcal{O}(r^{-1})$ \\

& $m^a = o^A\bar{\iota}^{A'}$ & {$m^a=\I\ell^A\bar{k}^{A'}=w\partial^a_r+\mu\partial^a_{\bar{z}}+\bar{\nu}\partial^a_z$,}\quad $w=\mathcal{O}(r^{-2})$, $\mu=\frac{\mu^{(0)}(z,\bar{z})}{r}+\mathcal{O}(r^{-2})$, \\

& $\bar{m}^a = \iota^A\bar{o}^{A'}$ & {$\bar{m}^a=\I k^A\bar{\ell}^{A'}=\bar{w}\partial^a_r+\bar{\mu}\partial^a_{{z}}+{\nu}\partial^a_{\bar{z}}$},\quad $\nu=\mathcal{O}(r^{-2})$ &  \\[0.1em]\hline

\parbox[t]{1em}{\multirow{12}{*}{\rotatebox[origin=c]{90}{\it spin coefficients}}} & $\pi =\iota^A\nabla_{00'}\iota_A$ & {$ \bar{\ell}^{A'}\nabla_k\bar{\ell}_{A'}=-\bar{\beta}=\bar{\alpha}+\mathcal{O}(r^{-3})$}  \\

& $\lambda=\iota^A\nabla_{10'}\iota_A$ & {$\bar{\ell}^{A'}\nabla_{\bar{m}}\bar{\ell}_{A'}=\bar{\sigma}_{(\ell)}=-\frac{\dot{\bar\sigma}(u,z,\bar{z})}{r}+\mathcal{O}(r^{-2})$}   \\

& $\mu=\iota^A\nabla_{01'}\iota_A$ &  {$ \bar{\ell}^{A'}\nabla_{{m}}\bar{\ell}_{A'}=\frac{1}{2}\vartheta_{(\ell)}=-\frac{\dot{U}(z,\bar{z})}{r}+\mathcal{O}(r^{-2})$}  \\

& $\nu=\iota^A\nabla_{11'}\iota_A$ &  {$ \bar{\ell}^{A'}\nabla_{\ell}\bar{\ell}_{A'}=0$}   \\

& $\varepsilon=\iota^A\nabla_{00'}o_A$ &  {$ \bar{\ell}^{A'}\nabla_{k}\bar{k}_{A'}=0$}   \\

& $\alpha=\iota^A\nabla_{10'}o_A$ & {$ \bar{\ell}^{A'}\nabla_{\bar{m}}\bar{k}_{A'}=-\frac{\I}{2}\big(\bar{\gamma}+\I\bar{\alpha}\big)=\mathcal{O}(r^{-1})$},\quad $\gamma=-\frac{\I}{r}\frac{\mu^{(0)}}{\bar{\mu}^{(0)}}\partial_{\bar{z}}\bar{\mu}^{(0)}+\mathcal{O}(r^{-2})$ &  \\

& $\beta=\iota^A\nabla_{01'}o_A$ & {$ \bar{\ell}^{A'}\nabla_{{m}}\bar{k}_{A'}=-\frac{\I}{2}\big({\gamma}+\I{\alpha}\big)$}   \\

& $\gamma=\iota^A\nabla_{11'}o_A$ & {$ \bar{\ell}^{A'}\nabla_{\ell}\bar{k}_{A'}=-\frac{1}{2}\big(\kappa-\I\phi\big)=\mathcal{O}(r^{-1}),\quad\kappa=\frac{\operatorname{\mathfrak{Re}}(\Psi_2^{(0)}(u,z,\bar{z}))}{r^2}+\mathcal{O}(r^{-3})$}   \\

& $\kappa=o^A\nabla_{00'}o_A$ &  {$ \bar{k}^{A'}\nabla_{k}\bar{k}_{A'}=0$}   \\

& $\rho=o^A\nabla_{10'}o_A$ &  {$ \bar{k}^{A'}\nabla_{\bar{m}}\bar{k}_{A'}=-\frac{1}{2}\vartheta_{(k)}=-\frac{1}{r}-\frac{U}{r^2}-\frac{\sigma\bar\sigma+U^2}{r^3}+\mathcal{O}(r^4)$,\quad $\ddot{U}=0$}   \\

& $\sigma=o^A\nabla_{01'}o_A$ &  {$ \bar{k}^{A'}\nabla_{{m}}\bar{k}_{A'}=-\sigma_{(k)}=-\frac{\sigma(u,z,\bar{z})}{r^2}+\mathcal{O}(r^{-3})$} \\

& $\tau=o^A\nabla_{11'}o_A$ &  {$ \bar{k}^{A'}\nabla_{\ell}\bar{k}_{A'}=\alpha =\frac{\alpha^{(0)}}{r^2}+\mathcal{O}(r^{-3})$,\quad $\bar{\alpha}^{(0)}=\mathcal{D} U -\bar{\mathcal{D}}\bar{\sigma}$} \\\hline

& $\Psi_s$& $\bar{\Psi}_s = \bar{\Psi}_{A'_1\dots A'_{s}A'_{s+1}\dots A'_5}\bar{\ell}^{A'_1}\dots\bar{\ell}^{A'_s}\bar{k}^{A'_{s+1}}\dots\bar{k}^{A'_5}=\mathcal{O}(r^{s-5})$\\[0.1em]\hline

\end{tabularx}\vspace{1.2em}
}
\caption{Dictionary between the Newman\,--\,Penrose (NP) formalism and the conventions used in this paper. Our metric signature is $(-$$+$$+$$+)$. Spacetime vectors $V^a$ correspond to anti-hermitian $(1/2,1/2)$ spinors $V^{AA'}=-\bar{V}^{A'A}$. The role of primed and unprimed indices are switched by parity. What we call e.g.\ $\Psi_2$ corresponds, therefore, to $\bar{\Psi}_2$ in the NP formalism.  Notice that both $k^a$ and $\ell^a$ are surface orthogonal. }\label{tab2} 
\end{table}
\vspace{-1em}

\providecommand{\href}[2]{#2}\begingroup\raggedright\endgroup


\end{document}